\documentclass{aa}
\usepackage[normalem]{ulem}
\usepackage{graphicx}
\usepackage{subfigure}
\usepackage{amssymb}
\begin{document}

%       UNITS
%
\def\cm{\,{\rm cm}}
\def\cmc{\,{\rm cm^{-3}}}
\def\erg{\,{\rm erg}}
\def\kms{\,{\rm km\,s^{-1}}}
\def\degr{\hbox{$^\circ$}\,}
\def\farcm{\hbox{$.\mkern-4mu^\prime$}}
\def\farcs{\hbox{$.\!\!^{\prime\prime}$}}
\def\Jy{\,{\rm Jy}\,}
\def\Jyb{\,{\rm Jy/beam}\,}
\def\mJy{\,{\rm mJy}\,}
\def\muJy{\,\mu{\rm mJy}\,}
\def\mJyb{\,{\rm mJy/beam}\,}
\def\muJyb{\,\mu{\rm mJy/beam}\,}
\def\K{\,{\rm K}}
\def\mkG{\,\mu{\rm G}}
\def\MHz{\,{\rm MHz}\,}
\def\GHz{\,{\rm GHz}\,}
\def\p{\,{\rm pc}\,}
\def\kpc{\,{\rm kpc}\,}
\def\Mpc{\,\mathrm{Mpc}\,}
\def\radm{\,\mathrm{rad\,m^{-2}}\,}

\newcommand\HI{\ion{H}{i}}
\newcommand\HII{\ion{H}{ii}}
\def\ne{n_\mathrm{e}}
\def\DP{{\rm DP}}
\def\RM{{\rm RM}}
\def\EM{{\rm EM}}
\def\PI{\mbox{$P$}}
\def\s{\,{\rm s}}
\def\yr{\,{\rm yr}}
\def\pheins{\phantom{1}}
\newcommand{\ncr}{n_\mathrm{cr}}
\newcommand{\tot}{_I}
\newcommand{\an}{_\mathrm{a}}
\newcommand{\reg}{_P}
\newcommand\deriv[2]{ \displaystyle\frac{\partial #1}{\partial #2} }

%%%%%%%%%%%%%%%%%%%%%%%%%%%%%%%%%%%%
\newcommand\sfrac[2]{{\textstyle{\frac{#1}{#2}}}}
\newcommand\bra[1]{\langle #1\rangle}
\newcommand\cs{c_\mathrm{s}}
\newcommand\mean[1]{\overline{#1}}
\newcommand{\wav}[1]{$\lambda#1\,\mathrm{cm}$}  % one wavelength
\newcommand{\wwav}[2]{$\lambda\lambda#1,#2\,\mathrm{cm}$}  % 2  wavelengths
\newcommand{\wwwav}[3]{$\lambda\lambda#1,#2,#3\,\mathrm{cm}$}  % 3 wavelengths
\newcommand{\pa}{p_\mathrm{A}} %pitch angle

\title{Magnetic fields in barred galaxies.\\
        IV. NGC~1097 and NGC~1365}

\author{ R.~Beck\inst{1},
          A.~Fletcher\inst{1,2},
          A.~Shukurov\inst{2},
          A.~Snodin\inst{2},
          D.~D.~Sokoloff\inst{3},
          M.~Ehle\inst{4},
          D.~Moss\inst{5},
          V.~Shoutenkov\inst{6}
}

\institute{Max-Planck-Institut f\"ur Radioastronomie,
              Auf dem H\"ugel 69,
              53121 Bonn, Germany
\and     School of Mathematics and Statistics, University of Newcastle,
              Newcastle upon Tyne NE1 7RU, UK
\and     Department of Physics, Moscow State University,
              119992 Moscow, Russia
\and     XMM-Newton Science Operations Centre, European Space Astronomy Centre (ESAC),
    	   European Space Agency, P.O. Box 50727, 28080 Madrid, Spain
\and 	   School of Mathematics, University of Manchester,
	   Manchester M13 9PL, UK
\and     Pushchino Radioastronomy Observatory, Astro Space Center,
              142292 Pushchino, Russia}

\offprints{ R. Beck }

\date{Received 2 June 2005 / Accepted 2 August 2005}

\titlerunning{Magnetic fields in barred galaxies. IV.}
\authorrunning{R.~Beck et al.}

\abstract{We present $\lambda3.5$~cm and $\lambda6.2$~cm radio
continuum maps in total and polarized intensity of the barred
galaxies NGC~1097 (at 2\arcsec--15\arcsec\ resolution) and NGC~1365
(at 9\arcsec--25\arcsec\ resolution). A previously unknown
radio galaxy southwest of NGC~1097 is reported.
Apart from a smooth faint envelope and a bright central region,
both galaxies exhibit radio ridges roughly overlapping with the massive dust
lanes in the bar region. The contrast in total intensity across the radio ridges
is compatible with compression and shear of an isotropic random magnetic field,
where the gas density compression ratio is approximately equal to 4
and the cosmic ray density is constant across the ridges.
The contrast in polarized intensity is
significantly smaller than that expected from compression and shearing
of the regular magnetic field; this could be the result of decoupling
of the regular field from the dense molecular clouds. The regular field
in the ridge is probably strong enough to
reduce significantly shear in the diffuse gas (to which it is coupled)
and hence to reduce magnetic field amplification by shearing.
This contributes to the misalignment of the observed field orientation
with respect to the velocity vectors of the dense gas.
Our observations, for the first time, indicate that
\emph{magnetic forces can control the
flow of the diffuse interstellar gas at kiloparsec scales.}
The total radio intensity reaches its maximum in the circumnuclear starburst
regions, where the equipartition field strength is about $60\mkG$,
amongst the strongest fields detected in spiral galaxies so far. The regular field in
the inner region has a spiral shape with large pitch angle,
indicating the action of a dynamo.  Magnetic stress leads to mass
inflow towards the centre, sufficient to feed the active nucleus
in NGC~1097.  --
We detected diffuse X-ray emission, possibly forming a
halo of hot gas around NGC~1097.
 }

\maketitle

\section{Introduction}

Radio polarization observations have revealed basic properties of
interstellar
magnetic fields in galaxies of various morphological types (Beck et
al.\ \cite{BB96}; Beck\ \cite{beck05}).
Until recently magnetic fields in barred galaxies remained
relatively unexplored, although they can be expected to have
interesting properties.  Strong non-axisymmetric
gas flows and large-scale shocks will have a major effect on
interstellar magnetic fields;
velocity gradients may enhance the regular magnetic field (Chiba \&
Lesch\ \cite{CL94},
Otmianowska-Mazur et al.\ \cite{OL97}) and dynamo action should also be
strongly affected
by the presence of a bar (Moss et al.\ \cite{M98}).  According to
gas-dynamical simulations, a shear shock occurs in the bar where
the gas streamlines are deflected inwards (Athanassoula\ \cite{A92a},
\cite{A92b}; Piner et al.\ \cite{PS95}). To date, none of the numerical models
includes the magnetic field in a way that allows its influence on the shock to
be explored, although it is well known that magnetized shocks can develop
properties different from those of hydrodynamic shocks.

Radio observations of barred galaxies can then provide useful
insight into their interstellar gas dynamics. In order to clarify the
nonthermal properties of barred galaxies, we have performed a radio
survey
of a sample of barred galaxies (Beck et al.\ \cite{BS02}). Among the
twenty galaxies
studied, the prototypical examples NGC~1097 and NGC~1365 are most
spectacular.
Here we discuss in detail and interpret, in terms of gas flow
and magnetic field models, the nonthermal properties of these two galaxies.

The total [polarized + unpolarized] synchrotron radio intensity
depends on the strength of the total [regular + random] magnetic field
in the sky plane and the energy density of cosmic ray electrons.
Polarized synchrotron emission is produced
by cosmic ray electrons in the presence of either
a regular (coherent) magnetic field
\footnote{The regular field is also known as the mean, average, ordered,
or large-scale field, but these terms could be misleading and are
avoided in this paper.}
, or an anisotropic random (incoherent) field, or a combination of both.
Faraday rotation is produced by the component of the regular field
along the line of sight and thermal electrons.
Since the polarization angle is only sensitive to the orientation
of the magnetic field in the synchrotron source (rather than to its direction),
only Faraday rotation is sensitive to the distinction between regular and
random anisotropic magnetic fields. The anisotropy of the random interstellar
magnetic field can be significant in barred galaxies where it is produced by strong
shear and shocks. Therefore, the magnetic field strength
obtained from polarized intensity,
\footnote{In observational papers it is generally stated that the polarized emission
is a signature of the regular field, without defining the term `regular'.
In many papers the possible contribution of the anisotropic random field to the
polarized emission is neglected. Our definition is such that only \emph{coherent\/}
fields are called regular.}
denoted here $B\reg$, must be carefully distinguished
from that of of the regular magnetic field $\mean{B}$.
Denoting the strength of the anisotropic part of the
random magnetic field ($\vec{b}$)
by $b\an$, we can write $B\reg^2=\mean{B}^2+\mean{b\an^2}$.

The first high-resolution radio map of a barred galaxy, NGC~1097
(Ondrechen \& van der Hulst\ \cite{OH83}), showed narrow radio ridges
coinciding with the dust lanes, the tracers of compression regions along the
leading edge of the bar. A similar result was obtained for M83
(Ondrechen\ \cite{O85}), a galaxy with a smaller bar than NGC~1097.
The first detection of polarized radio emission from a bar was
reported by Ondrechen (\cite{O85}) for M83, with a mean fractional
polarization of 25\%.

The first high-resolution polarization observations of a galaxy with a
massive bar, NGC~1097, by Beck et al.\ (\cite{BE99}) had a resolution of
15\arcsec.
Magnetic field orientations in and around the bar were shown to
approximately follow the velocity field of the gas in the corotating frame,
while the outer field has a spiral pattern. A narrow ridge of vanishing polarized
intensity indicated deflection of the field lines in a shear shock, similar to
the deflection of the velocity vectors (Athanassoula\ \cite{A92b}).

Moss et al.\ (\cite{MS01}) presented a generic dynamo model,
based on the model velocity field of Athanassoula (\cite{A92b}), which
could explain the major magnetic features observed in radio polarization.
A similar dynamo model, but now based on the specific velocity and
gas density fields of NGC~1365, is discussed by Moss et al.\ (\cite{MS05}).
Beck et al.\ (\cite{BS02}) presented an
atlas of radio maps (in total and polarized intensity) of northern and
southern barred galaxies observed with the VLA and the ATCA at a
resolution of 30\arcsec. Harnett et al.\ (\cite{HE04}) discussed ATCA
radio observations of the peculiar barred galaxy NGC~2442.
The circumnuclear rings of the southern barred galaxies
NGC~1672 and NGC~7552 were discussed by Beck et al.\ (\cite{BE05}).

In this paper, we present VLA observations of NGC~1097 and NGC~1365,
with higher resolution than in Beck et al.\ (\cite{BE99}) and Beck et al.\
(\cite{BS02}), which resolve the structure of the magnetic field in the
ridges of NGC~1097 and NGC~1365 and in the circumnuclear starburst regions.
Furthermore, we present a ROSAT X-ray map of NGC~1097.

{\bf NGC~1097} is a barred galaxy of morphological type SBb(s) at
about 17~Mpc distance ($1\arcsec\approx83\,\p$), with a bar of about 20\kpc
in length continuing into two optical spiral arms. The galaxy plane is inclined by
about $45^\circ$ to the line of sight, and its line of nodes has a position
angle of about $-45^\circ$ with respect to the north-south direction
(Ondrechen et al.\ \cite{OHH89}), with the south-western side closer to
us. The bar has a similar position angle of about
$-32^\circ$ and thus lies almost in the plane of the sky. In the
galaxy's plane, the bar is located at $\simeq18^\circ$ azimuthal angle
from the major axis. Very little \HI\ gas has been found in the bar
(Ondrechen et al.\ \cite{OHH89}). CO observations by Gerin et al.\
(\cite{GN88}) were
not sensitive enough to detect cold gas in the bar. Crosthwaite
(\cite{C01}) detected extended CO emission in the (1--0) and (2--1)
transitions, but the spatial resolution was insufficient to see any
gas compression in the bar. Roussel et al.\ (\cite{RV01}) imaged
NGC~1097 and NGC~1365 in the mid-infrared. Dust emission is relatively weak in the
bar's ridge, and substantial diffuse emission was detected around the
bar.

The nuclear ring of 18\arcsec\ (1.5~kpc) diameter, formed by accreting
gas, shows enhanced star formation. This conspicuous feature is
visible in the optical, CO, radio continuum and X-ray spectral ranges.
Some spiral dust filaments were discovered interior to the ring (Barth
et al.\ \cite{BH95}, Prieto et al.\ \cite{P05}),
possibly indicating mass inflow towards the active nucleus.

{\bf NGC~1097A} is a companion dwarf galaxy at about 18~kpc projected
distance towards the north-west (see Fig.~\ref{n1097s15}), which causes
gravitational distortions in the northern half of NGC~1097 and a tidal
arm in the \HI\ gas (Ondrechen et al.\ \cite{OHH89}).

%%----------------------------------------

%FIGS NGC1097 6\arcsec\
\begin{figure*}[htbp]
\includegraphics[bb = 62 135 528 645,width=0.465\textwidth,clip=]{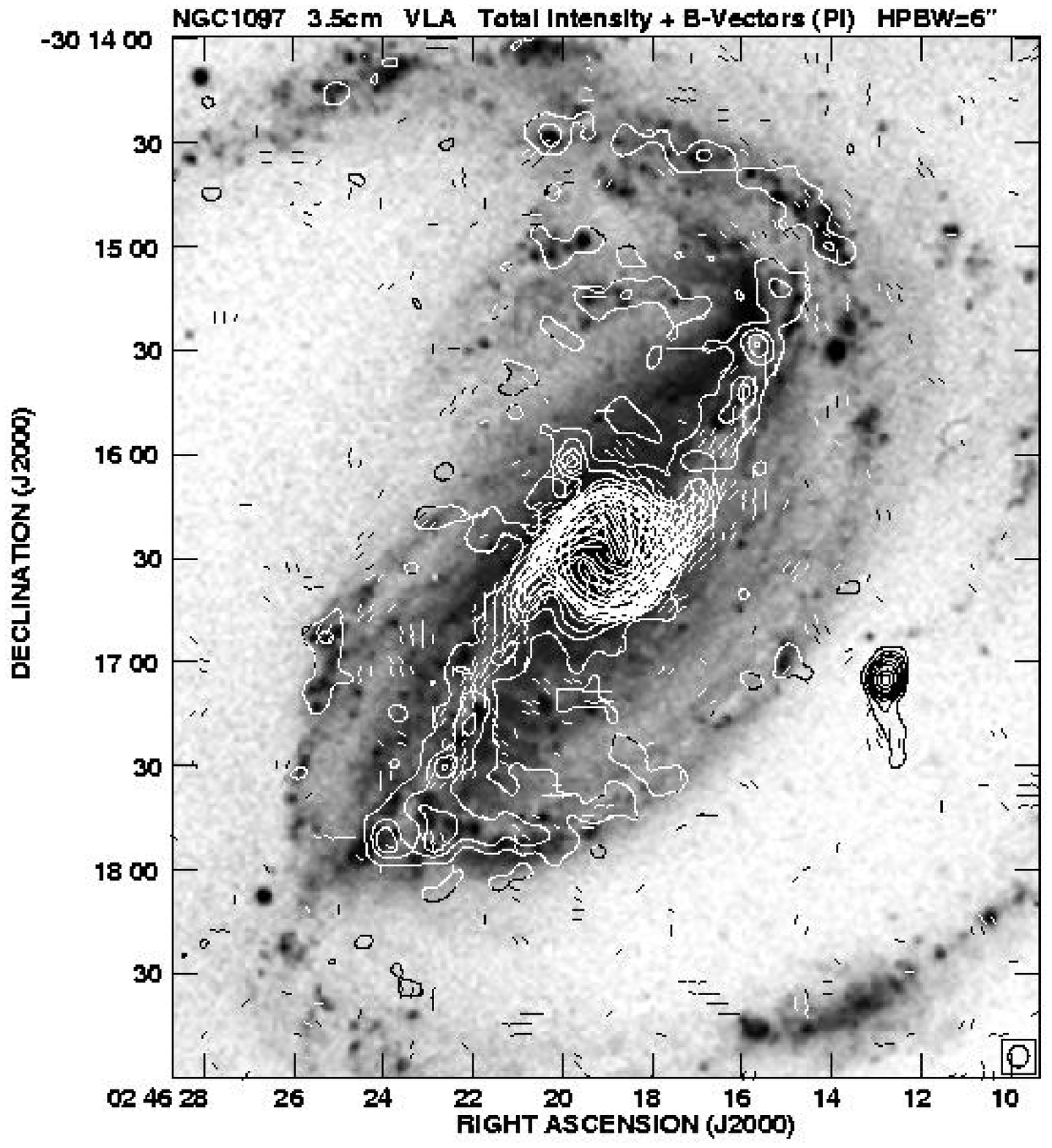}
\hfill
\includegraphics[bb = 62 135 528 645,width=0.465\textwidth,clip=]{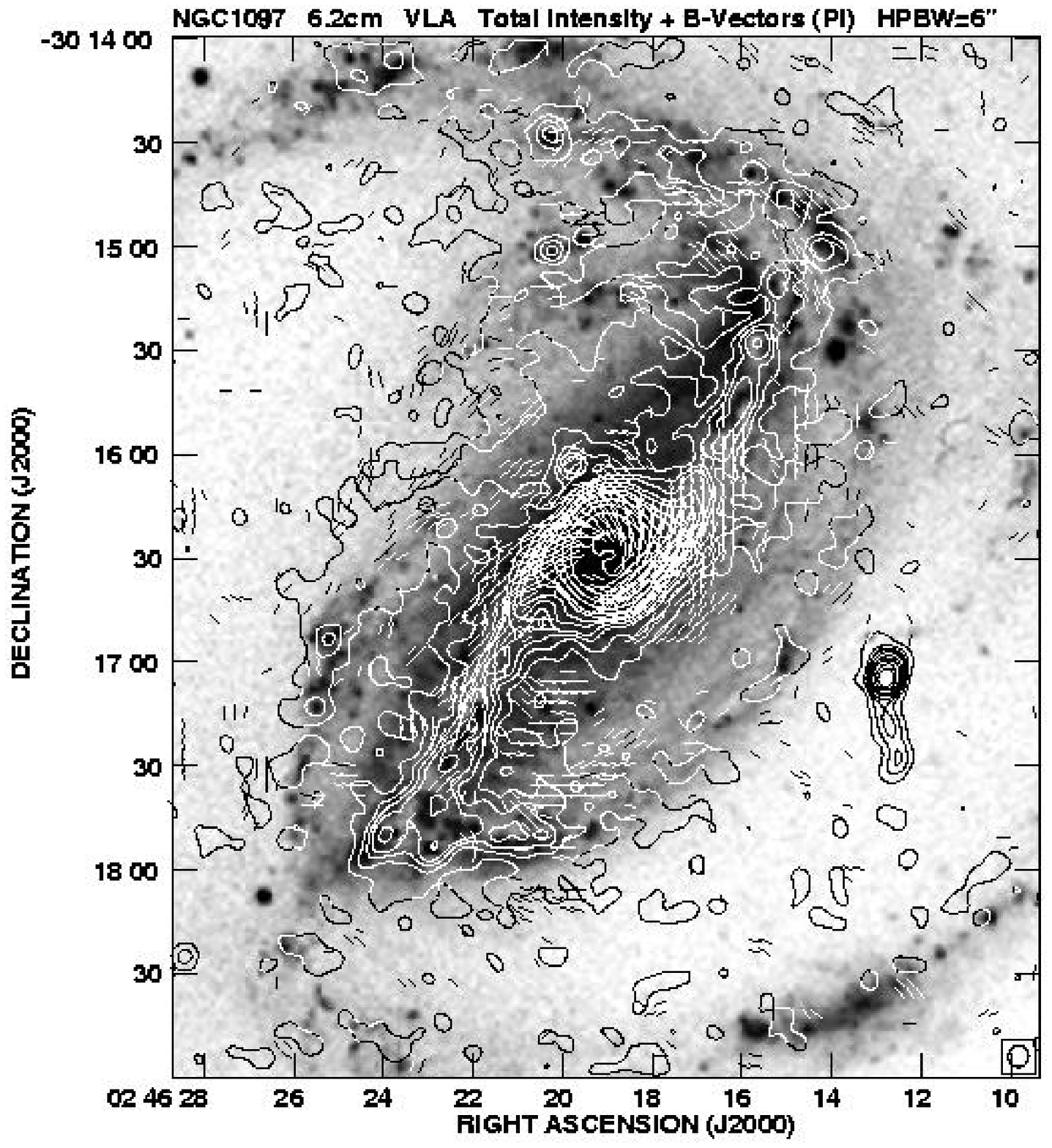}
\vspace*{0.25cm}
\includegraphics[bb = 62 135 536 645,width=0.465\textwidth,clip=]{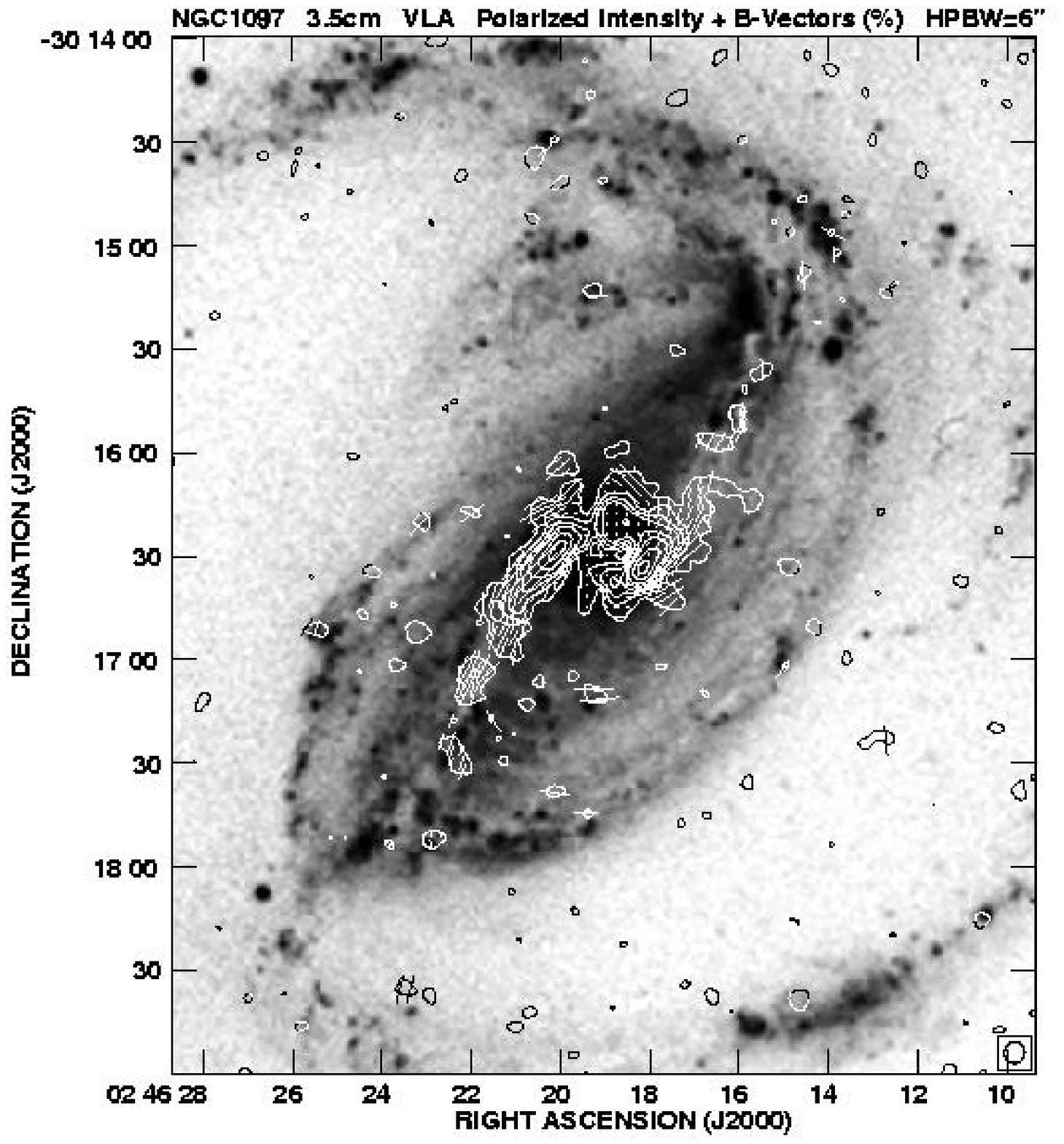}
\hfill
\includegraphics[bb = 62 135 536 645,width=0.465\textwidth,clip=]{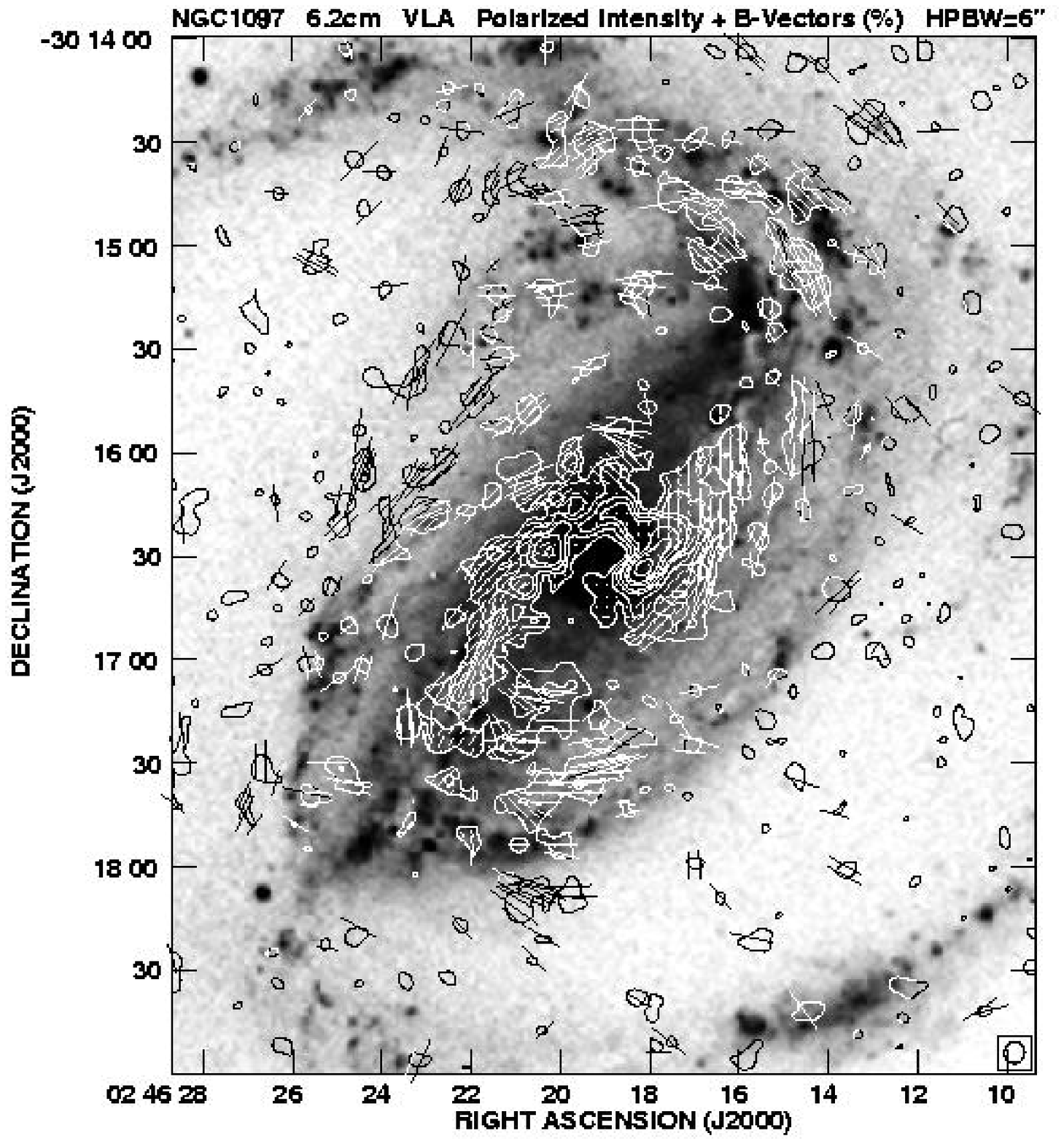}
\caption{
{\it Top row:\/} Total intensity contours and observed
$B$-vectors ($E+90^\circ$) of NGC~1097 at $\lambda3.5$~cm (left) and
$\lambda6.2$~cm (right) at 6\arcsec\ resolution.
The contour intervals are $1, 2, 3, 4, 6, 8, 12, 16, 32, 64, 128$
times 40~$\mu$Jy/beam area. The vector length
is proportional to polarized intensity, 3\arcsec\ length corresponds
to 20~$\mu$Jy/beam area.
The vector orientations are not corrected for Faraday
rotation. The background optical image was kindly provided by Halton Arp.
{\it Bottom row:\/}
Polarized intensity contours and observed $B$-vectors
at $\lambda3.5$~cm (left) and $\lambda6.2$~cm (right)
at 6\arcsec\ resolution. The vector length is proportional to fractional
polarization, 3\arcsec\ length corresponds to 20\%.
The contour intervals are
$1, 2, 3, 4, 6, 8, 12, 16$ times 20~$\mu$Jy/beam area.
The vector orientations are not corrected for Faraday rotation.
Here and in other maps, the beam size is shown in the bottom right
corner of each panel.}
\label{n1097s6}
\end{figure*}

%%------------------------------------------

%%------------------------------------------

%FIGS NGC1097 10\arcsec\
\begin{figure*}[htbp]
\includegraphics[bb = 62 135 536 645,width=0.465\textwidth,clip=]{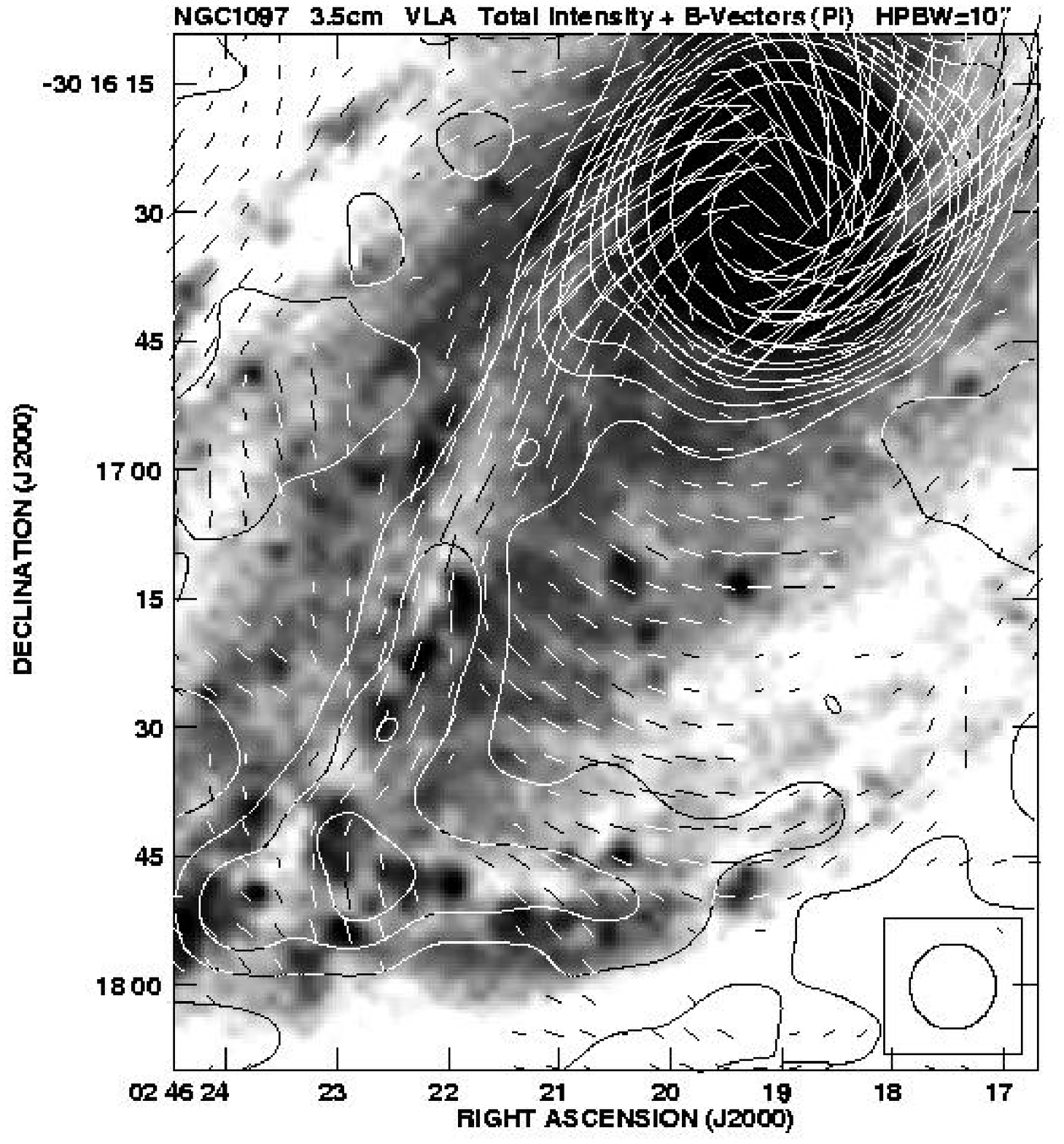}
\hfill
\includegraphics[bb = 62 135 536 652,width=0.465\textwidth,clip=]{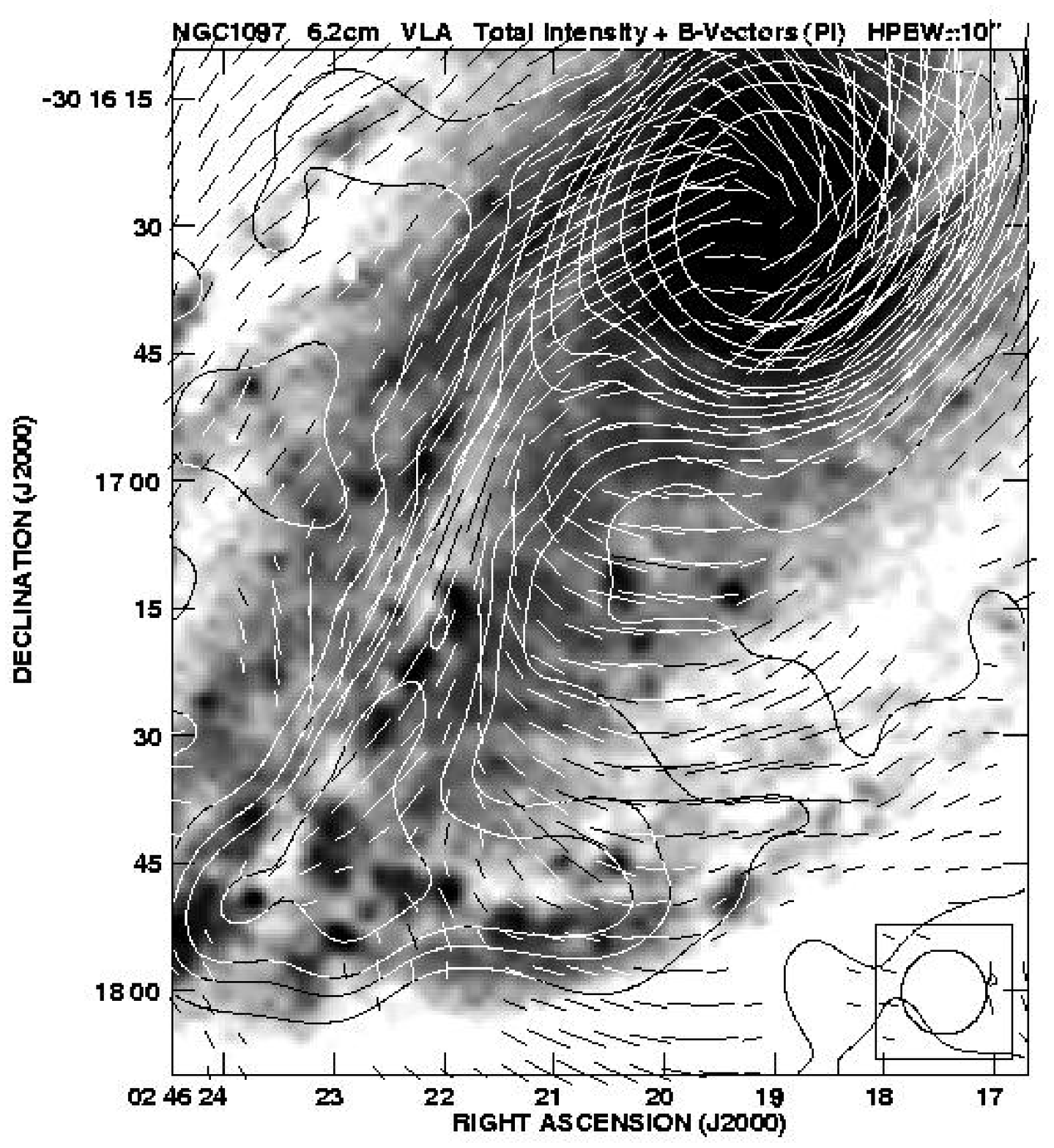}

\vspace{0.2cm}

\includegraphics[bb = 62 135 543 645,width=0.465\textwidth,clip=]{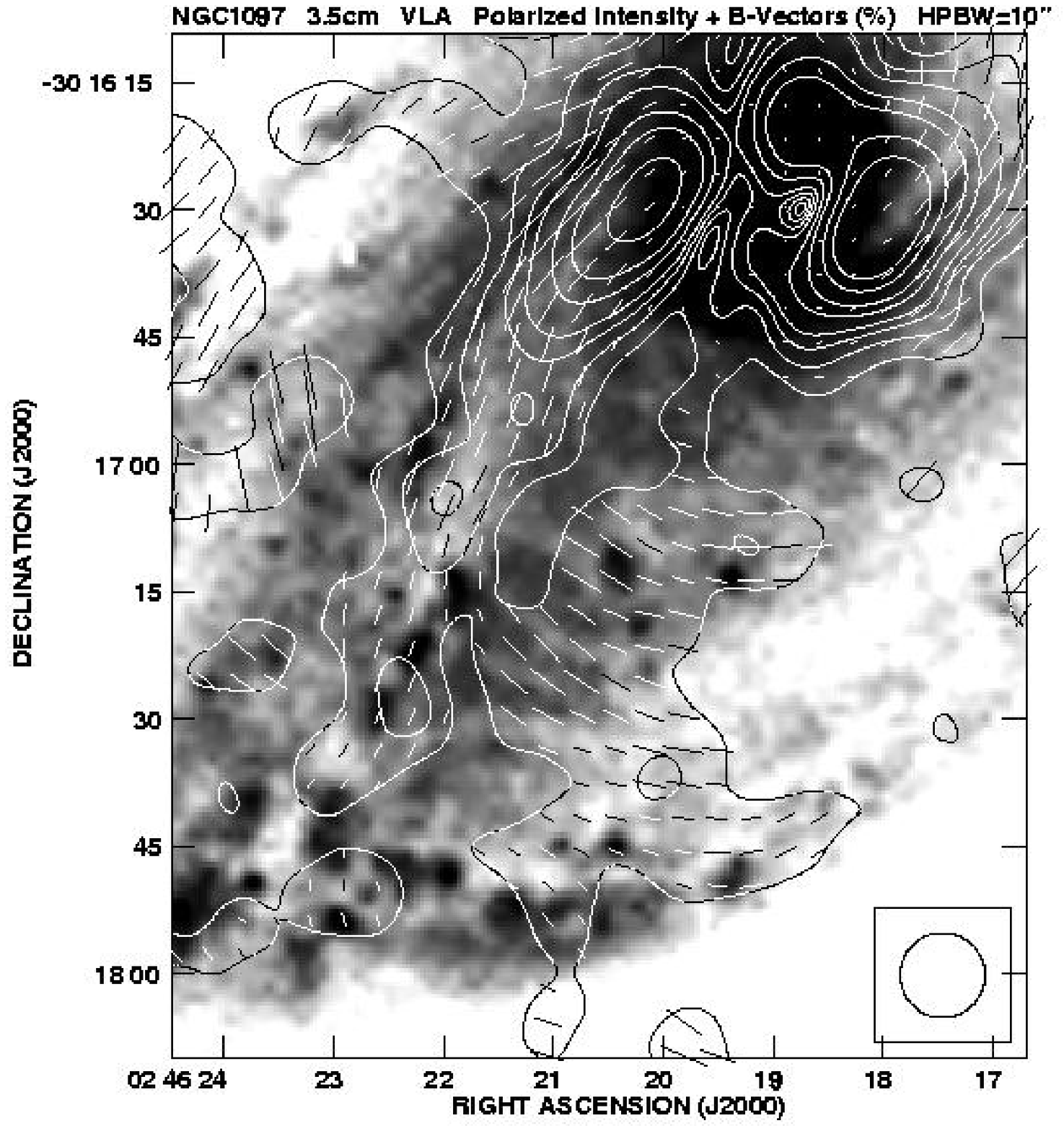}
\hfill
\includegraphics[bb = 62 135 543 645,width=0.465\textwidth,clip=]{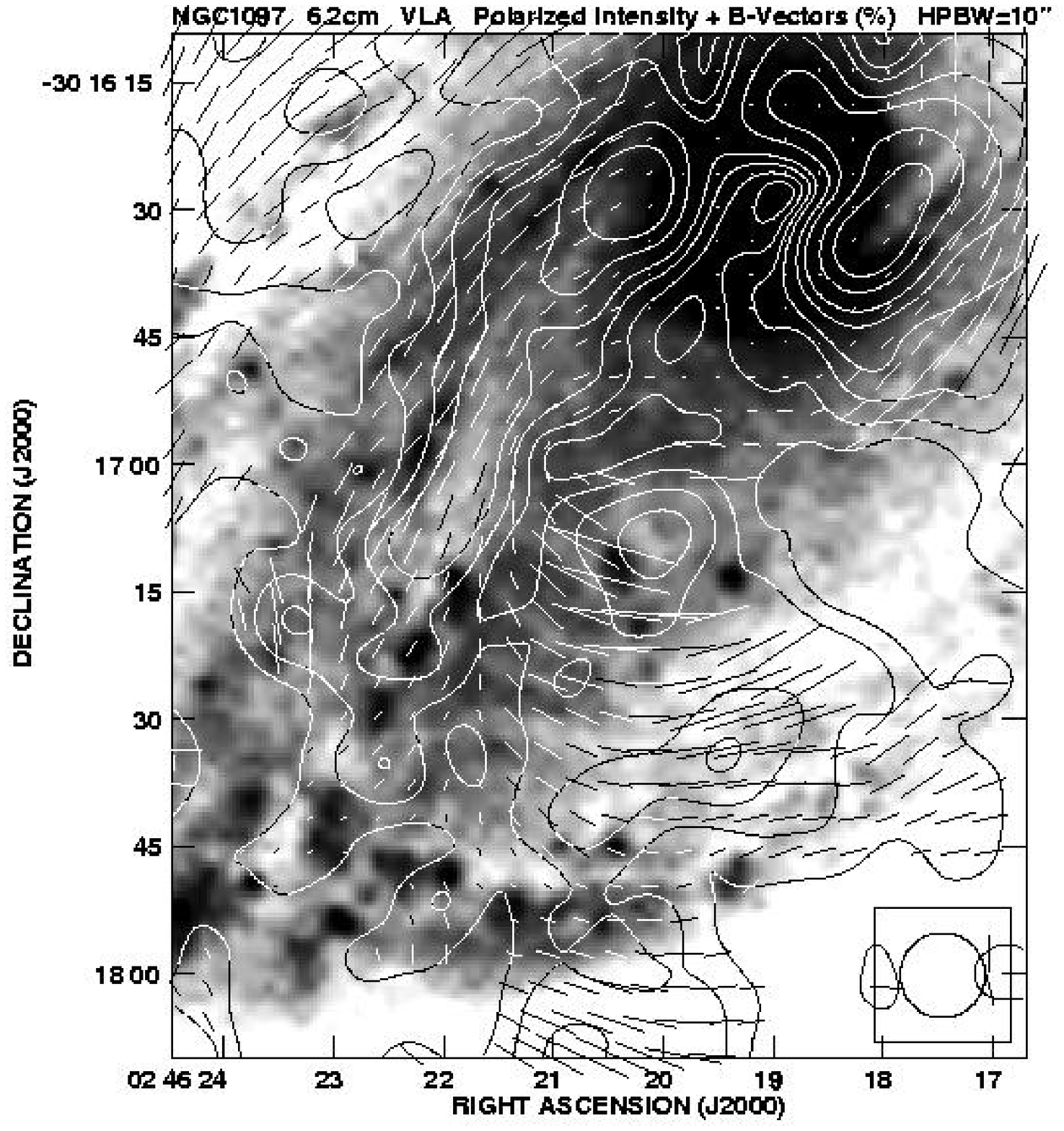}
\caption{{\it Top row:\/} Total intensity contours and observed
   $B$-vectors ($E+90^\circ$) of the central and southern parts of
   NGC~1097 at $\lambda3.5$~cm (left) and $\lambda6.2$~cm (right) at
   10\arcsec\ resolution.
The contour intervals are as in Fig.~\ref{n1097s6}, with the lowest
 contour level at 80~$\mu$Jy/beam area.
The vector length is proportional to
   polarized intensity, 3\arcsec\  length corresponds to 30~$\mu$Jy/beam area.
   The vector orientations are not corrected for Faraday rotation.
{\it Bottom row:\/} Polarized intensity contours and observed $B$-vectors
   at $\lambda3.5$~cm (left) and $\lambda6.2$~cm (right)
   at 10\arcsec\ resolution. The contour intervals are as in
   Fig.~\ref{n1097s6},
   with the lowest contour level at 20 $\mu$Jy/beam area.
   The vector length is proportional to fractional polarization,
   3\arcsec\ length corresponds to 20\%.
   The vector orientations are not corrected for Faraday rotation.}
\label{n1097s10}
\end{figure*}

%---------------------------------------------

{\bf NGC~1365}, at 18.6~Mpc distance ($1\arcsec\approx93\,\p$),
is probably the best studied barred galaxy (see the review by
Lindblad\ \cite{L99}). The \HI\ emission and
velocity field have been studied in detail by Ondrechen \& van der
Hulst (\cite{OH89}) and J\"ors\"ater \& van Moorsel (\cite{JM95}).
The galaxy plane is inclined by about $40^\circ$ to the line of sight
and its line of nodes has a position angle of about $40^\circ$ with
respect to the
north-south direction. The north-western side is closer to the observer.
The bar has a length of about 22~kpc and a position angle of about $92^\circ$.

NGC~1365 has an active Seyfert nucleus like NGC~1097, but the
circumnuclear ring in radio continuum is incomplete (Sandqvist et al.\
\cite{SJ95}). Radio continuum emission from the bar has first been
detected
by Beck et al.\ (\cite{BS02}), it coincides with the dust lanes, as in
NGC~1097. The map of NGC~1365 at $\lambda3.5$~cm with 30\arcsec\,
resolution (Beck et al.\ \cite{BS02}, their Fig.~7) shows a deflection
of the regular magnetic field in its eastern bar, though smoother than in
NGC~1097.  Data at higher resolution and corrected for Faraday
rotation are presented in this paper (Sect.~\ref{sectRM}).

%------------------
\section{Observations}
%------------------
\subsection{Radio observations}
\label{sectObs}

%-------------------------
\begin{table*}           % Table 1.
\caption{\label{Table1}Observational parameters of VLA observations}
\centering
\begin{tabular}{lcc}
\hline\hline
\  & NGC 1097 & NGC 1365 \\
\hline
Frequency (GHz)          & 4.835 \& 4.885  & 4.835 \& 4.885 \\
Array                    & DnC             & DnC \\
Pointing (J2000)         & $02^{\rm h}46^{\rm
m}19\fs77$--$30\degr16'28\farcs55$ & $03^{\rm h}33^{\rm
m}35\fs50$--$36\degr07'59\farcs88$  \\
Observing date
%AS :
\hfill    & 1996 May 20     & 1996 May 20 and 1999 Feb
19 \\
Net observing time (min) & 32              & 20 and 280 \\
Shortest baseline        & 506$\lambda$    & 490$\lambda$ and
404$\lambda$ \\
\hline

Frequency (GHz)          & 4.835 \& 4.885  & \\
Array                    & CnB             & \\
Pointing (J2000)         & $02^{\rm h}46^{\rm
m}19\fs87$--$30\degr16'27\farcs30$  & \\
Observing date
%AS :
\hfill    & 1998 Nov 13 \& 15  & \\
Net observing time (min) & 560 & \\
Shortest baseline        & 765$\lambda$    & \\
\hline

Frequency (GHz)          & 8.435 \& 8.485  & 8.435 \& 8.485 \\
Array                    & DnC             & DnC \\
Pointing (J2000)         & $02^{\rm h}46^{\rm
m}19\fs81$--$30\degr16'28\farcs05$ & $03^{\rm h}33^{\rm
m}35\fs50$--$36\degr07'59\farcs88$ \\
Observing date
%AS :
\hfill    & 1996 May 20 and 1997 Oct 07 \& 09  & 1996
May 20 and 1997 Oct 07 \& 09 \\
Net observing time (min) & 38 and 312        & 24 and 311 \\
Shortest baseline        & 794$\lambda$ and 704$\lambda$    &
790$\lambda$ and 704$\lambda$ \\
\hline

Frequency (GHz)          & 8.435 \& 8.485  & \\
Array                    & CnB             & \\
Pointing (J2000)         & $02^{\rm h}46^{\rm
m}19\fs87$--$30\degr16'27\farcs30$ & \\
Observing date
%AS :
\hfill    & 1998 Nov 11 \& 14  & \\
Net observing time (min) & 560             & \\
Shortest baseline        & 1330$\lambda$   & \\
\hline
\end{tabular}
\end{table*}
%-----------------------------------

%-----------------------------------
\begin{table*}           % Table 2.
\caption{\label{Table2}Weights and r.m.s.\ noise values of total
intensity $\sigma_{I}$ and of polarized intensity $\sigma_{Q,U}$ (for
Stokes parameters $Q$ and $U$) (in $\mu$Jy/beam) in the final maps}
\centering
\begin{tabular}{lccccccccc}
\hline\hline
                 &\multicolumn{5}{c}{NGC~1097}&
&\multicolumn{3}{c}{NGC~1365} \\
                 \cline{2-6}                     \cline{8-10}
{\bf Beam size}\rule{0pt}{11pt} & 2\arcsec & 4\arcsec & 6\arcsec &
10\arcsec & 15\arcsec && 9\arcsec & 15\arcsec & 25\arcsec \\
\hline
         &\multicolumn{9}{c}{\rule{0pt}{11pt}{$\lambda${\bf 3.5~cm}}} \\
Weighting &  uniform & robust & natural & natural & natural && robust &
natural & natural \\
$\sigma_{I}$  &  9 & 7 & 8 & 11 & 17 && 12 & 12 & 20 \\
$\sigma_{Q,U}$ &  9 & 7 & 6 &  8 & 11 && 12 & 12 & 15 \\
\hline
         &\multicolumn{9}{c}{\rule{0pt}{11pt}{$\lambda${\bf 6.2~cm}}} \\
Weighting & -- & uniform & natural & natural & natural && -- & robust &
natural \\
$\sigma_{I}$  & -- & 12 & 8 & 15 & 20 && -- & 14 & 14 \\
$\sigma_{Q,U}$ & -- & 12 & 8 &  9 & 13 && -- & 14 & 14 \\
\hline
\end{tabular}
\end{table*}
%----------------------------------

NGC~1097 and NGC~1365 were observed with the Very Large Array
(VLA) operated by the NRAO
\footnote{The NRAO is a facility of the
National Science Foundation operated under cooperative agreement by
Associated Universities, Inc.}
in its DnC and CnB arrays at
4.86\GHz ($\lambda6.2$~cm) and 8.46\GHz ($\lambda3.5$~cm). At both
frequencies two IFs, separated by 50\MHz and with a bandwidth of
50\MHz each, were recorded.  3C48 and 3C138 were used as primary flux
calibrators, 3C138 was also used for polarization angle calibration;
0240-231 was our phase calibrator. Details of the observations are
given in Table~\ref{Table1}.

Data processing was done with the standard procedures of {\sc aips}.
Visibility data in the same frequency band, obtained from the different
observation periods, from the different arrays and from both IF
channels, were combined in the $uv$ plane. Maps in Stokes parameters
$I$, $Q$ and $U$ were obtained by {\sc imagr} from the combined data at each
frequency band. Different weightings were applied to obtain maps with
different resolutions (Table~\ref{Table2}). Uniform weighting
(which gives the same weights to each cell in the $uv$ plane)
reveals the best angular resolution, but larger r.m.s.\ noise than
`natural' weighting
(which gives the same weights to all antenna pairs and
hence emphasizes the inner $uv$ plane)
or `robust' weighting (an intermediate case).

At both wavelengths, several values of the `zero-spacing' flux and its
weight were tested in the {\sc imagr} cleaning process to minimize the
negative-bowl effect in the total-intensity images caused by missing
short-baseline visibility data in the $uv$ plane.

%---------------------------------------------

%FIGS NGC1097 15\arcsec\
\begin{figure*}[htbp]
\includegraphics[bb = 62 113 528 666,width=0.495\textwidth,clip=]{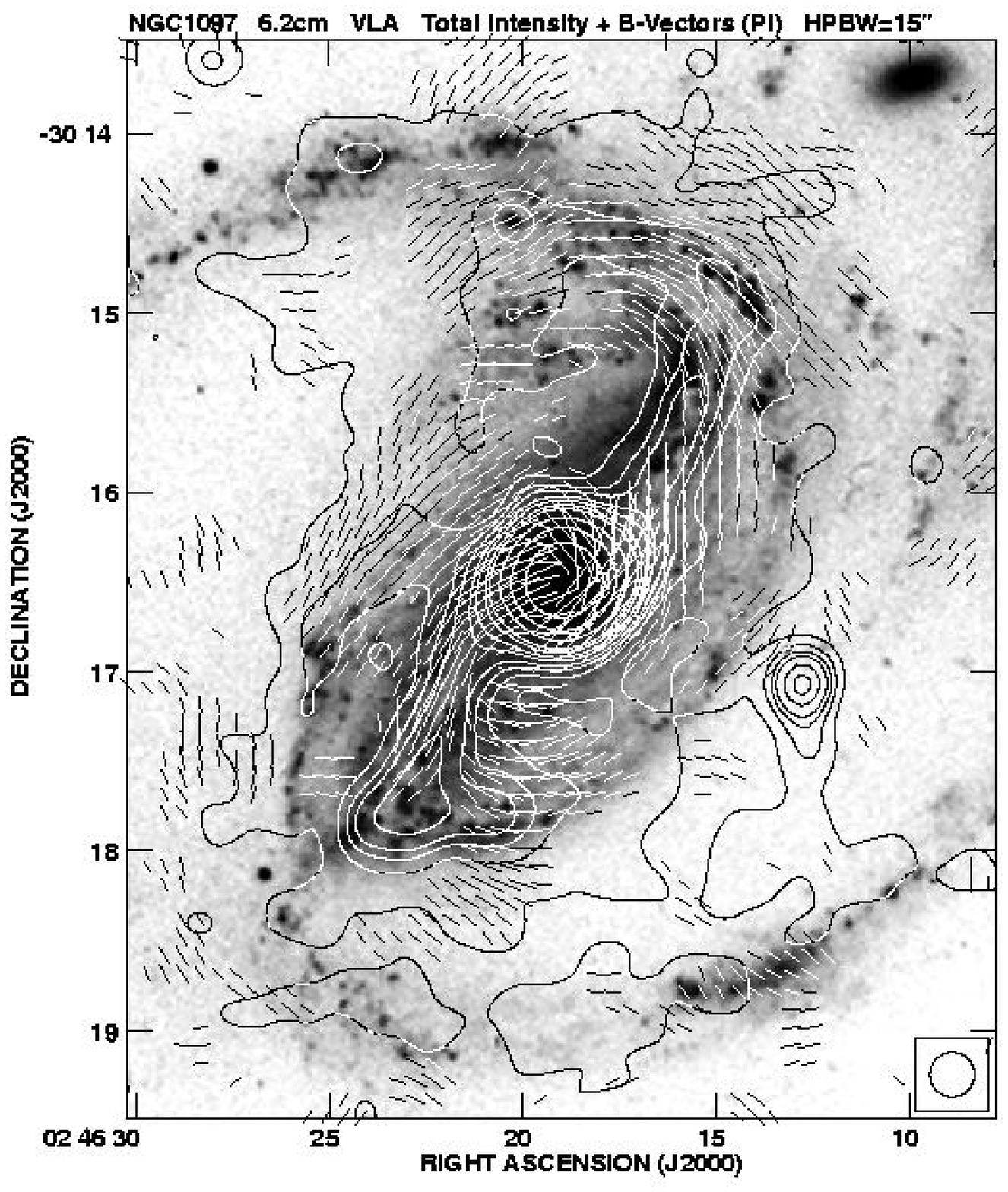}
\hfill
\includegraphics[bb = 62 113 528 666,width=0.495\textwidth,clip=]{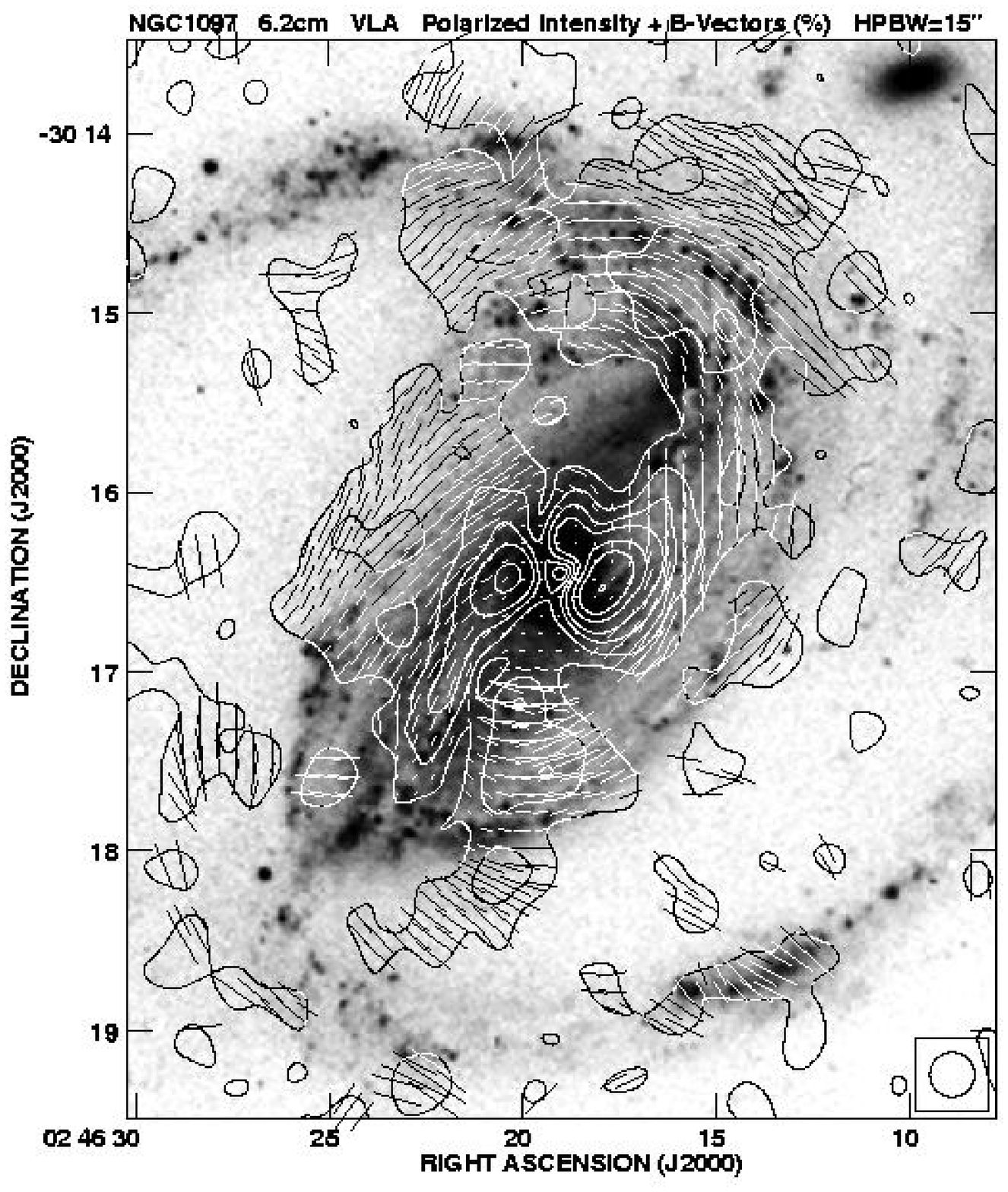}
\caption{{\it Left:\/} Total intensity contours and observed $B$-vectors
($E+90^\circ$) of NGC~1097 at $\lambda6.2$~cm at 15\arcsec\ resolution.
The contour intervals are as in Fig.~\ref{n1097s6}, with the lowest contour
   level at 150~$\mu$Jy/beam area.
The vector length is proportional to polarized
   intensity, 3\arcsec\ length corresponds to
   30~$\mu$Jy/beam area. The vector orientations are not corrected for Faraday
   rotation. The companion galaxy NGC~1097A is visible in the top right
   corner.
{\it Right:\/} Polarized intensity contours and
   observed $B$-vectors at 15\arcsec\ resolution. The vector length is
   proportional to fractional polarization, 3\arcsec\ length
   corresponds to 10\%. The contour
   intervals are as in Fig.~\ref{n1097s6}, with the basic contour level at
   50~$\mu$Jy/beam area. The vector orientations are not corrected for Faraday
   rotation.}
\label{n1097s15}
\end{figure*}

%--------------------------------------

The largest structures visible to the VLA D-array are about 3\arcmin\
at $\lambda3.5$~cm and 5\arcmin\ at $\lambda6.2$~cm, less than the
full extent of the galaxies as observed at $\lambda20$~cm (Beck et
al.\ \cite{BS02}). No single-dish maps are available to add the missing
large-scale emission component in total intensity. Hence our maps can only be
used for spectral index studies of the bright ridges and central regions
(Figs.~\ref{n1097spi10} and \ref{n1097spi4}).
We do not expect that any large-scale structure is missing in the maps of
the Stokes parameters $Q$ and $U$ (and, hence, in the polarized intensity maps)
because the magnetic field orientation and hence the polarization angle
is not constant across the galaxies.

%--------------------------------------

%FIGS NGC1365 15\arcsec\
\begin{figure*}[htbp]
\includegraphics[bb = 62 171 528 609,width=0.475\textwidth,clip=]{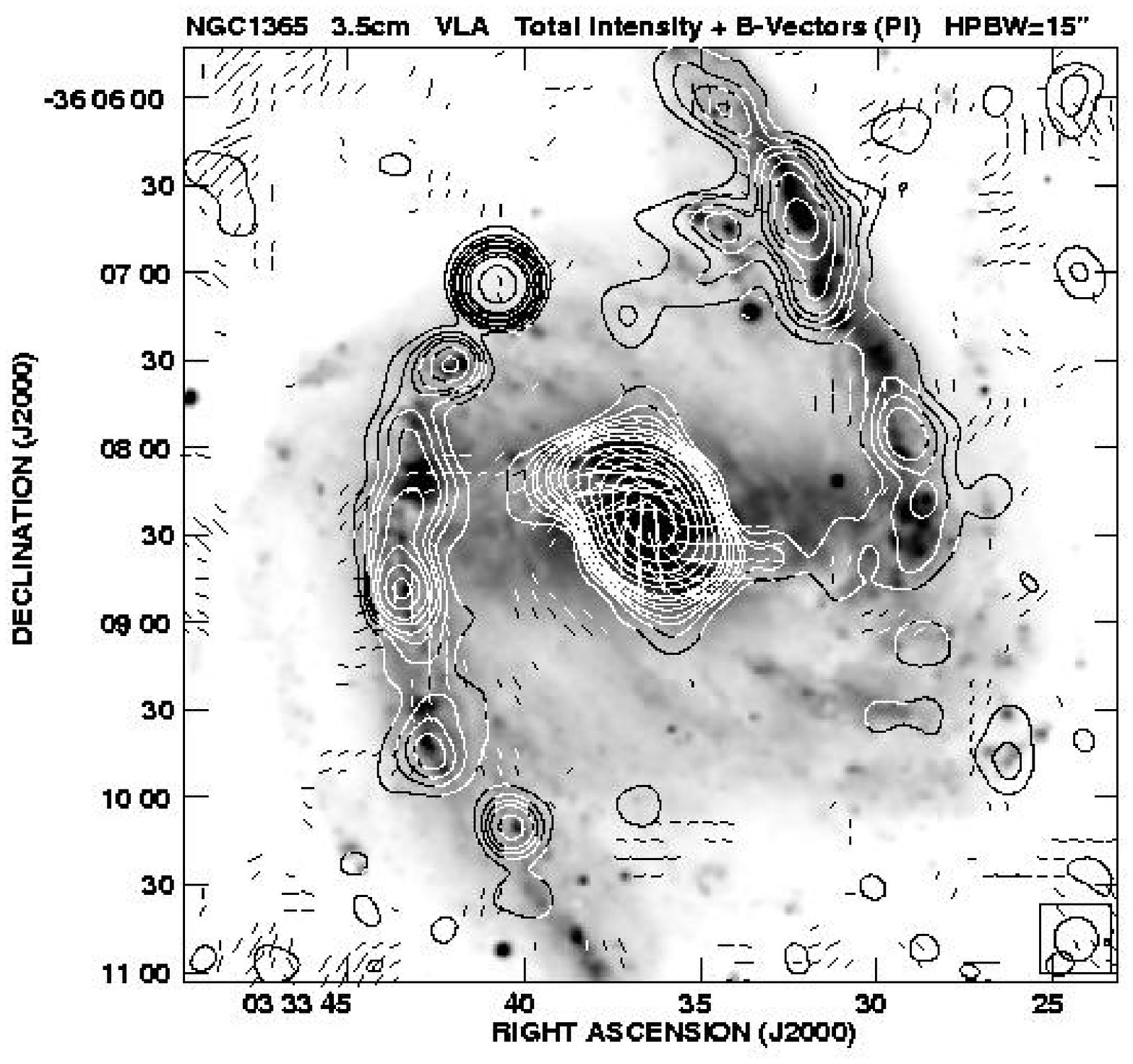}
\hfill
\includegraphics[bb = 62 171 528 609,width=0.475\textwidth,clip=]{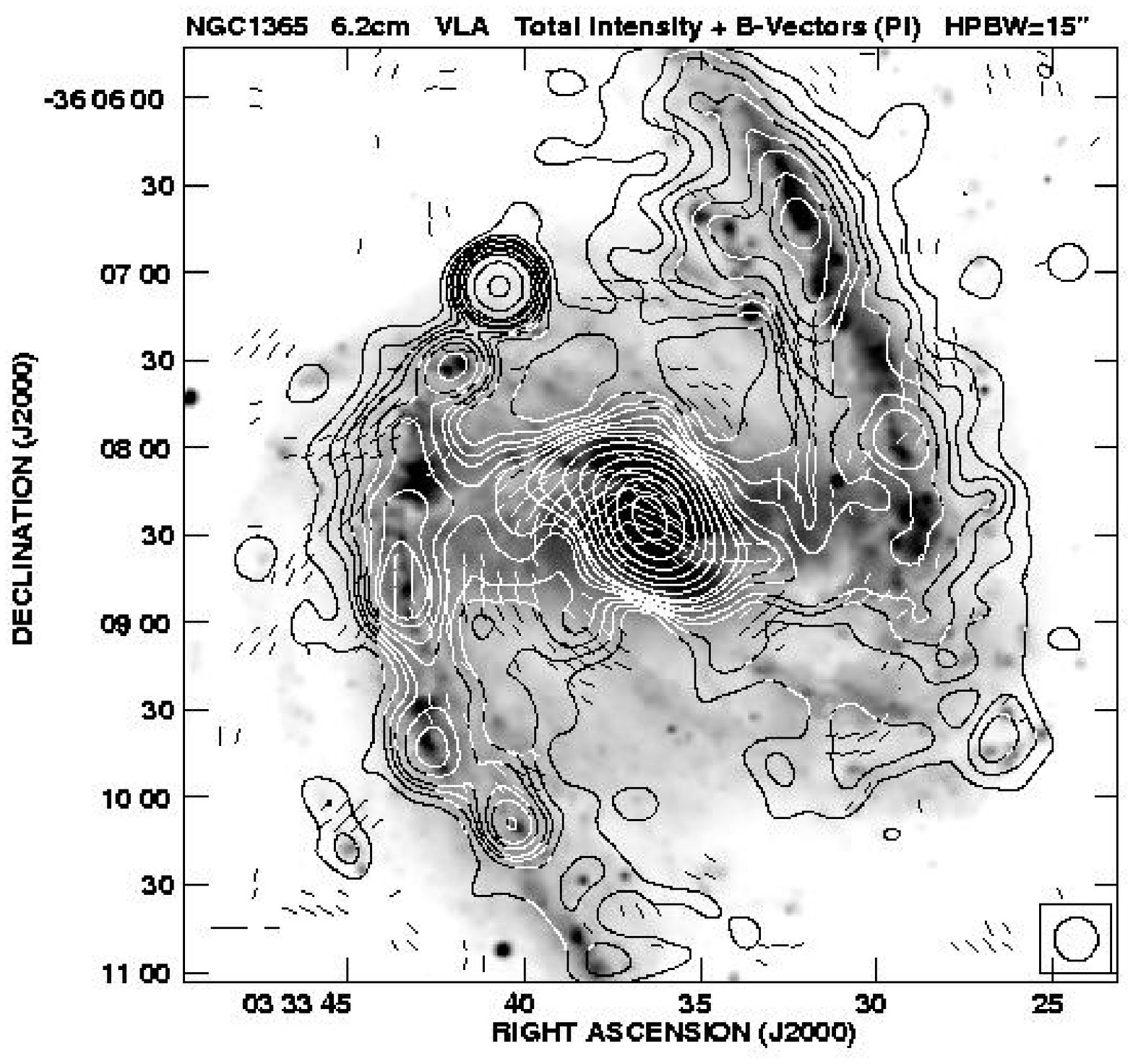}

\vspace{0.2cm}

\includegraphics[bb = 62 171 528 609,width=0.475\textwidth,clip=]{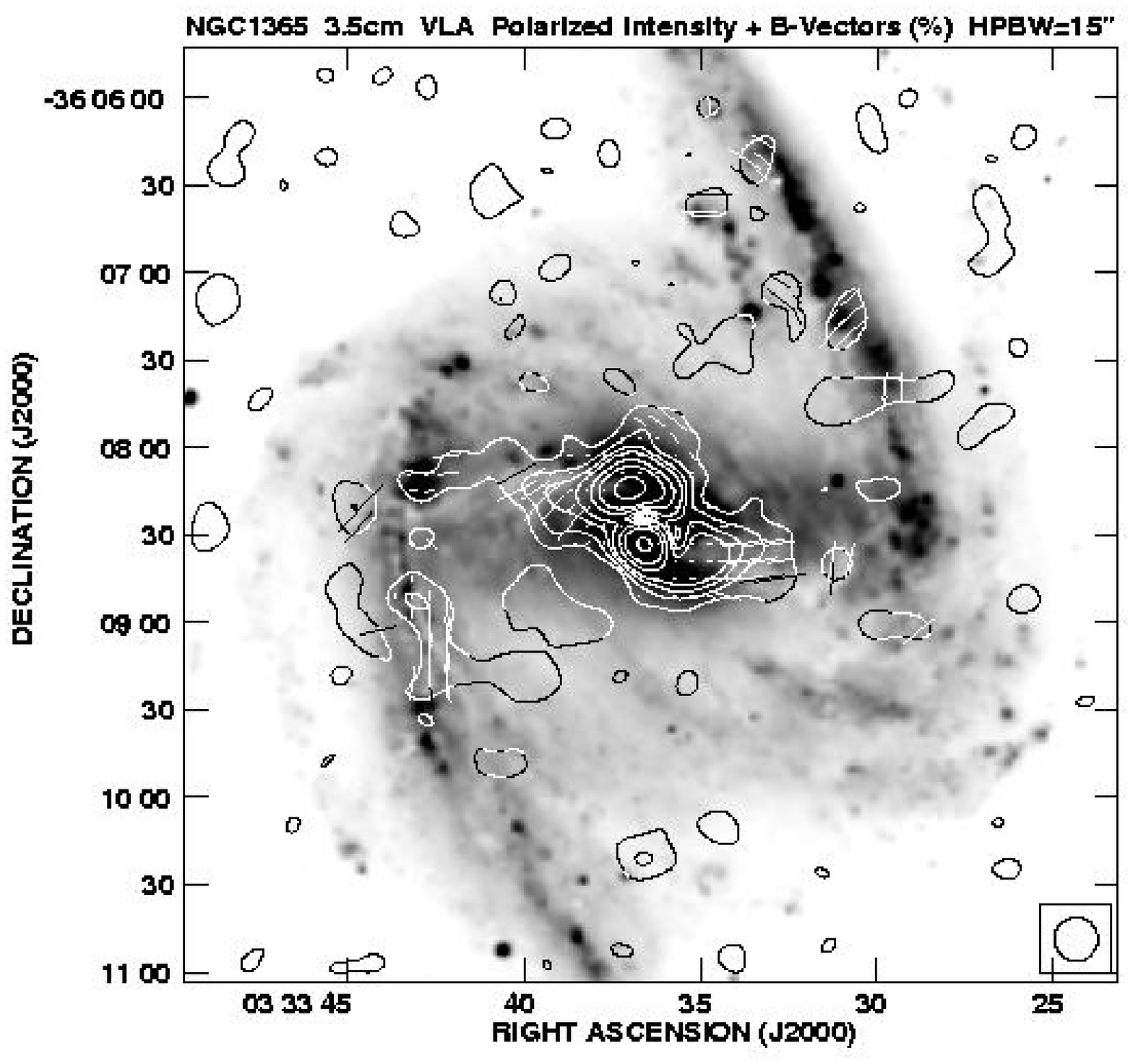}
\hfill
\includegraphics[bb = 62 171 528 609,width=0.475\textwidth,clip=]{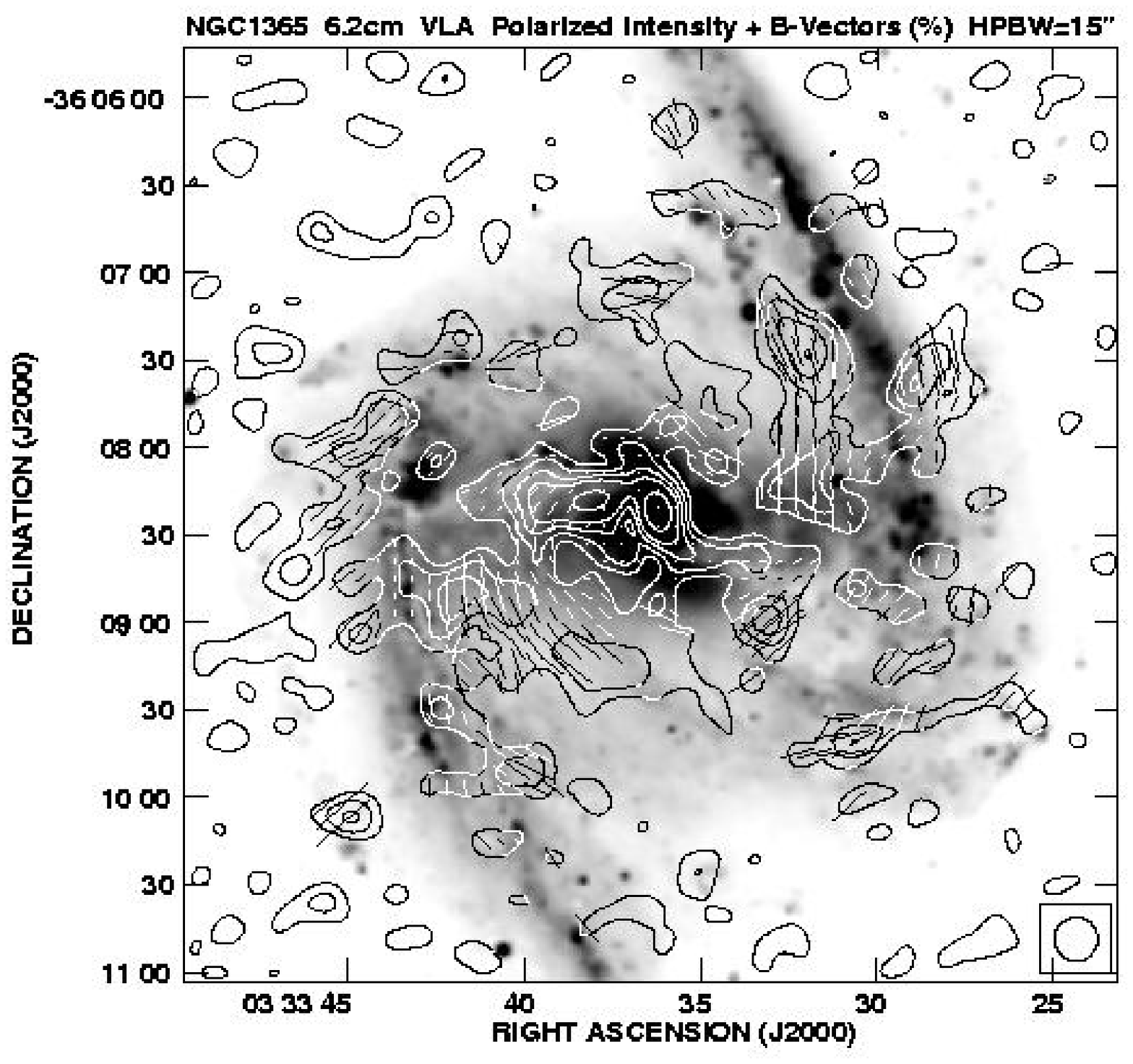}
\caption{%
{\it Top row:\/} Total intensity contours (corrected for
primary beam attenuation) and observed $B$-vectors ($E+90^\circ$)
of NGC~1365 at $\lambda3.5$~cm (left) and $\lambda6.2$~cm (right) at
   15\arcsec\ resolution.
The contour intervals are $1, 2, 3, 4, 6, 8, 12, 16, 32, 64, 128, 256, 512$
times 60~$\mu$Jy/beam area.
   The vector length is proportional to
   polarized intensity, 3\arcsec\ length corresponds to 30~$\mu$Jy/beam area.
   The vector orientations are not corrected
   for Faraday rotation. The background optical {\it ESO\/} image was
   kindly provided by Per Olof Lindblad.
 {\it Bottom row:\/} Polarized intensity contours and observed $B$-vectors
   at $\lambda3.5$~cm (left) and $\lambda6.2$~cm (right) at 15\arcsec\ resolution.
The contour intervals are $1, 2, 3, 4, 6, 8, 12$ times 25~$\mu$Jy/beam area.
   The vector length is proportional to fractional
   polarization, 3\arcsec\ length corresponds to 10\%.
The vector orientations are not corrected for Faraday rotation.}
\label{n1365s15}
\end{figure*}

%%--------------------------------------

Maps of the linearly polarized intensity (\PI) and the polarization
angle (PA) were obtained from the $Q$ and $U$ maps. The positive bias in
\PI\ due to noise was subtracted by applying the {\sc polco} correction.

The final maps of NGC~1097 were smoothed to circular FWHM Gaussian beams of
2\arcsec\ (central region, Fig.~\ref{n1097center}), 4\arcsec\
(Figs.~\ref{n1097s4} and \ref{n1097center}), 6\arcsec, 10\arcsec\ and
15\arcsec\ (Figs.~\ref{n1097s6}--\ref{n1097s15}). Polarization angles at 10\arcsec\
resolution were used to compute the map of Faraday rotation (Fig.~\ref{rm}).

%%--------------------------------------

%FIGS NGC1365 25\arcsec\
\begin{figure*}[htbp]
\includegraphics[bb = 62 171 528 609,width=0.495\textwidth,clip=]{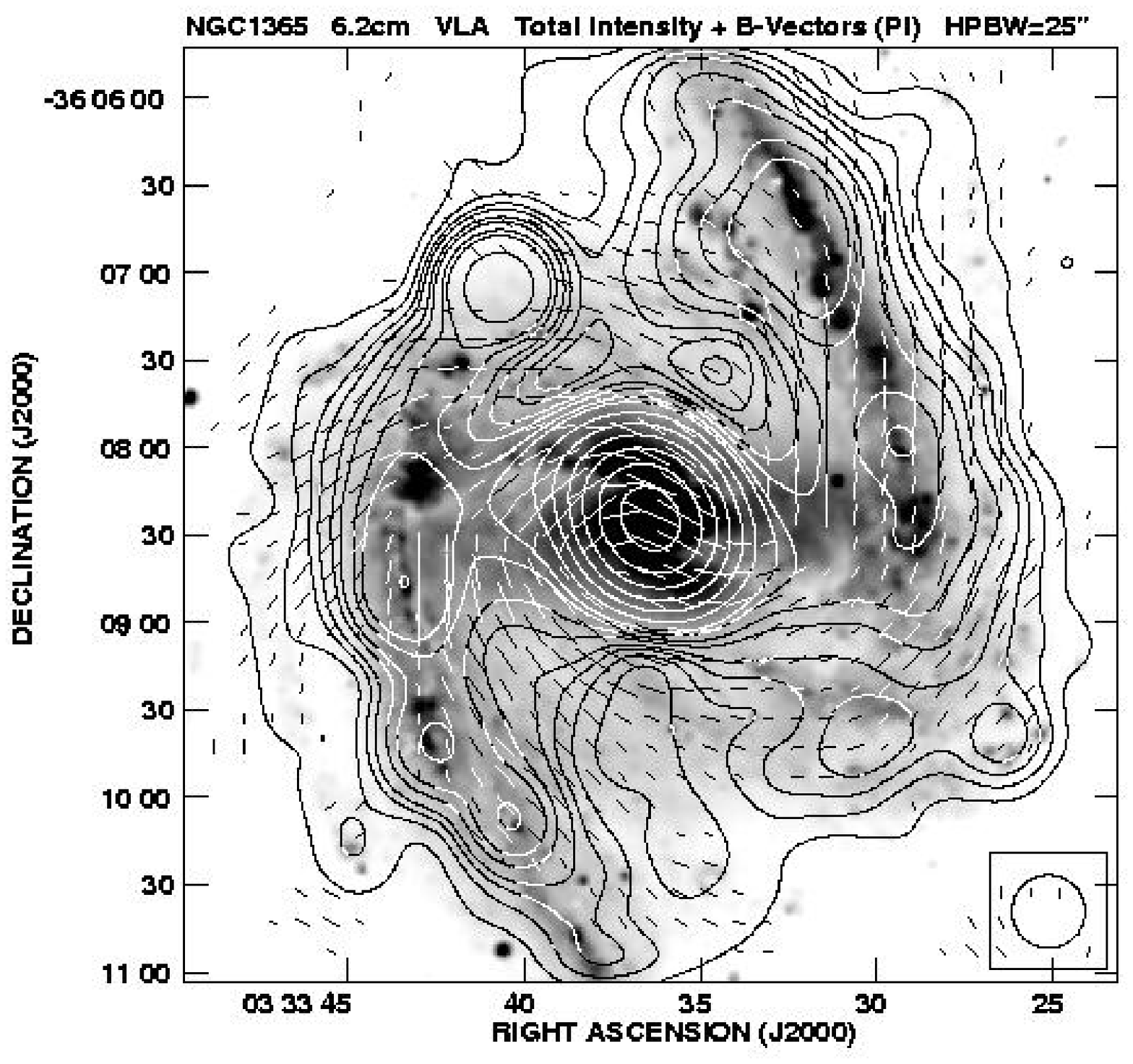}
\hfill
\includegraphics[bb = 62 171 528 609,width=0.495\textwidth,clip=]{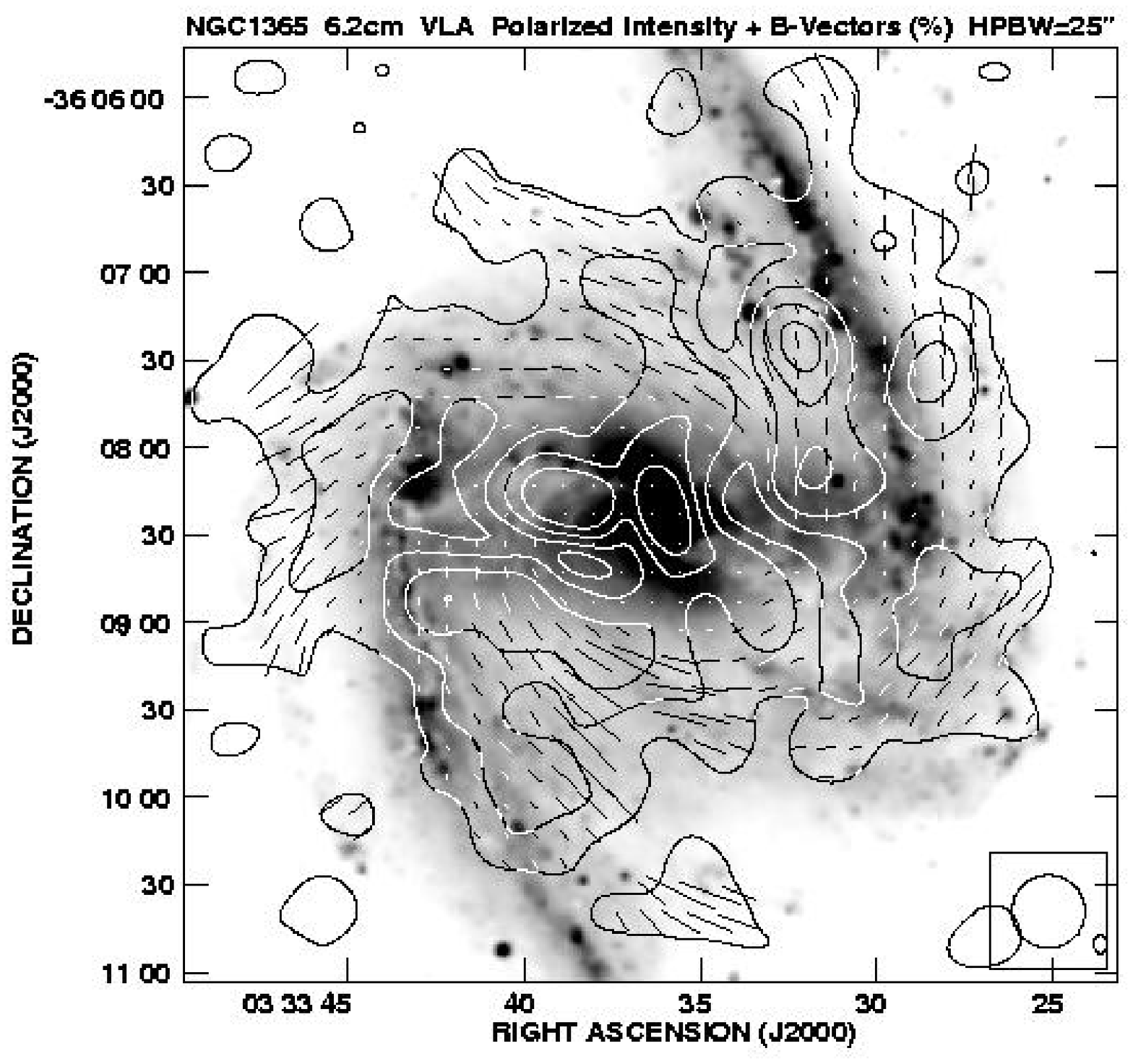}
\caption{{\it Left:\/} Total intensity contours (corrected for primary
   beam attenuation) and observed $B$-vectors ($E+90^\circ$) of NGC~1365
   at $\lambda6.2$~cm at 25\arcsec\ resolution.
   The contour intervals are as in
   Fig.~\ref{n1365s15}, with the lowest contour level at 80~$\mu$Jy/beam area.
   The vector length is
   proportional to polarized intensity, 3\arcsec\ length corresponds to
   30~$\mu$Jy/beam area.
   The vector orientations are not corrected for Faraday rotation.
{\it Right:\/}
   Polarized intensity contours and observed $B$-vectors at 25\arcsec\ resolution.
   The contour intervals are as in
   Fig.~\ref{n1365s15}, with the basic contour level at 40~$\mu$Jy/beam area.
   The vector length is proportional to fractional
   polarization, 3\arcsec\ length corresponds to 10\%.
   The vector orientations are not corrected for Faraday rotation.
}
\label{n1365s25}
\end{figure*}

%-------------------------------------

For NGC~1365, no data with the CnB array were obtained, so that the
highest achievable resolution (uniform weighting) is 7\arcsec\ at
$\lambda3.5$~cm and 13\arcsec\ at $\lambda6.2$~cm. However, the maps
at these resolutions have low signal-to-noise ratios and show little
extended emission.  In this paper, we present only maps obtained with
robust and natural weighting (see Table~2). Diffuse emission of
NGC~1365 is best visible in the maps at 15\arcsec\ resolution
(Fig.~\ref{n1365s15}) and especially in the $\lambda6.2$~cm maps at
25\arcsec\ resolution (Fig.~\ref{n1365s25}).  Maps of the central
region at 9\arcsec\ resolution are shown in Fig.~\ref{n1365center}.
Polarization angles at 25\arcsec\ resolution were used to compute the map of
Faraday rotation (Fig.~\ref{rm}).

The total intensity maps of NGC~1365 were corrected for primary beam
attenuation.
This correction was not necessary for the total intensity maps of
NGC~1097 and the polarization
maps of both galaxies because the extent of the
visible emission is smaller than the diameter of the primary beam
(5\farcm4\ at $\lambda3.5$~cm and 9\arcmin\ at $\lambda6.2$~cm).
However, the spectral index is sensitive to small systematic effects,
and thus the total intensity of NGC~1097 used for the maps
of spectral index (Figs.~\ref{n1097spi10} and \ref{n1097spi4})
had to be corrected for primary beam attenuation.

The values for resolution and r.m.s.\ noise of the maps shown in this
paper are given in Table~\ref{Table2}.

%------------------------------------------
\subsection{X-ray observations of NGC~1097}
\label{sectX}

NGC~1097 was observed on Dec 27, 1992, for 2.7 hours with the Position Sensitive
Proportional Counter (PSPC, Pfeffermann et al.\ \cite{PB87}) on board
the {\it ROSAT\/} X-ray satellite (observation request 600449, PI: H.~Arp). We
retrieved the data from the ROSAT Data Archive and performed a
standard spatial and spectral analysis with the Extended X-ray
Scientific Analysis System (EXSAS, Zimmermann et al.\ \cite{ZB98}).

To show the extended soft-band X-ray emission, the data was spatially
binned, exposure corrected and smoothed in energy sub-bands (cf.\ Ehle
et al.\ \cite{EP98}). The smoothing was done with a Gaussian filter with
the FWHM corresponding to the average resolution of the point spread
function at the PSPC centre in the individual energy sub-bands. The
two lowest sub-bands (0.11--0.19~keV and 0.20--0.41~keV), smoothed
to 52\arcsec\ and 38\arcsec\ resolutions, respectively, were
added to create a standard soft-band (0.1--0.4~keV) map of the X-ray emission of
NGC~1097 (Fig.~\ref{n1097x}).

The spatial distribution and spectral characteristics
of the soft X-ray emission are discussed in Sect.~\ref{sectHalo}.

%-------------------------------------
\section{The radio maps}
%-------------------------------------

\subsection{Discovery of a radio galaxy in the field of NGC~1097}
\label{sectRGal}

%% --------------------------------------------
% FIG Radio Galaxy
\begin{figure}[htbp]
\centerline{\includegraphics[bb = 126 64 464 681,width=0.35\textwidth,clip=]{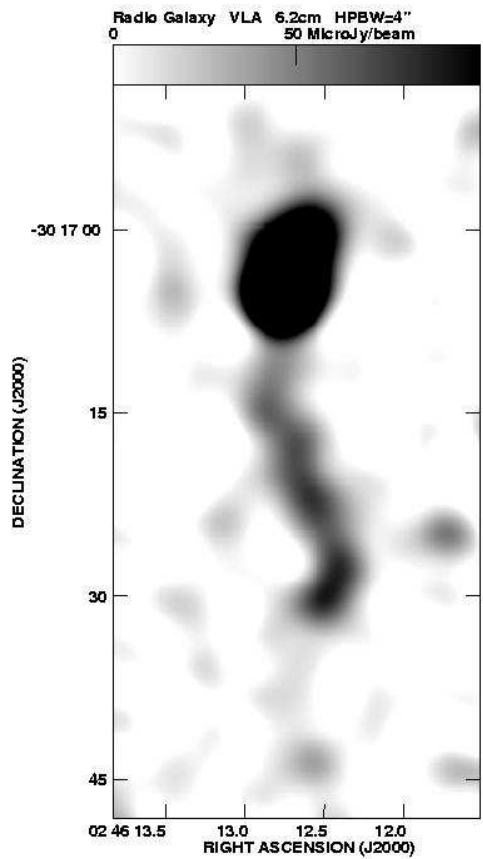}}
\caption{Radio galaxy west of NGC~1097 (total intensity), observed
at $\lambda6.2$~cm with 4\arcsec\ resolution.}
\label{radiogal}
\end{figure}
%% ----------------------------------------------

A radio galaxy is visible about 90\arcsec\ southwest
of the nucleus of NGC~1097 (Figs.~\ref{n1097s6} and ~\ref{radiogal}).
It is not included in the NED and NVSS catalogues. The position of the
nucleus (which is still unresolved at 2\arcsec\ resolution) is
RA, DEC(J2000) = 02~46~12.75, $-30$~17~05.3, its flux density is
$753\pm10\,\mu$Jy at
$\lambda3.5$~cm, and its spectrum is flat (with a spectral index
$\alpha\simeq+0.1$)
which is typical of an active galactic nucleus.
The northern component is located at
RA, DEC(J2000) = 02~46~12.7, $-30$~17~01, its flux density is $283\pm36\,\mu$Jy
at $\lambda3.5$~cm, and its
spectum is steep ($\alpha\simeq-0.5$). The southern jet shows
a spectral steepening from $\alpha\simeq+0.1$ to $\alpha\simeq-1.5$
with increasing distance from the nucleus.
No polarized emission was detected from any component.

%%--------------------------------------

%FIG NGC1097 spectral index 10\arcsec\
\begin{figure}[htbp]
\centerline{\includegraphics[bb = 62 106 528 666,width=0.495\textwidth,clip=]{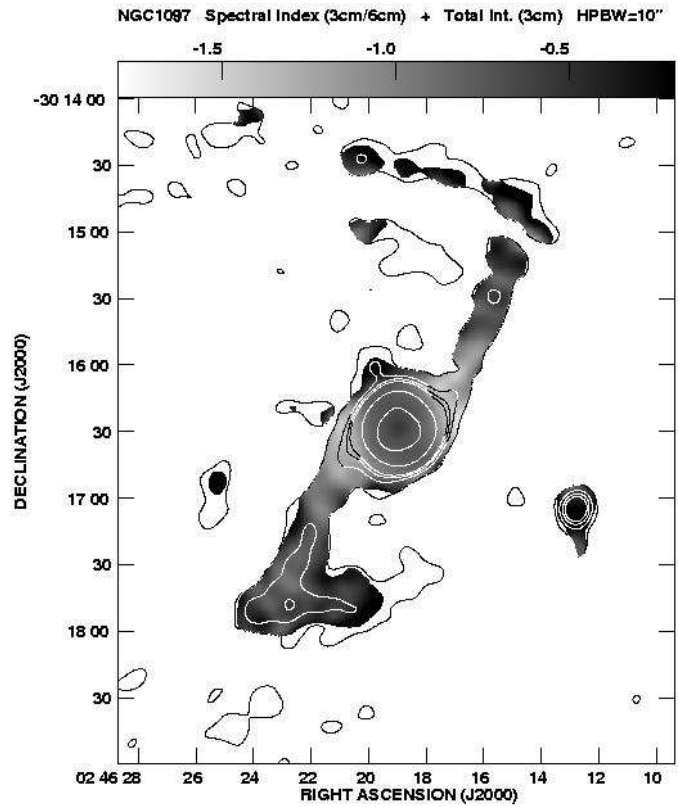}}
\caption{%
Spectral index
of NGC~1097 between $\lambda3.5$~cm and $\lambda6.2$~cm at 10\arcsec\
resolution, shown where the signal at each wavelength exceeds $15\sigma_I$
(to avoid systematic effects due to the missing large-scale structures).
Contours represent $\lambda3.5$~cm total intensity at
$1, 2, 3, 4, 16, 64$ times 150~$\mu$Jy/beam area and the same resolution.}
\label{n1097spi10}
\end{figure}

%%--------------------------------------

%% ----------------------------------------------
\subsection{The radio ridges in the bar regions}
\label{sectRidge}

\begin{figure}[htbp]
\centerline{\includegraphics[bb = 62 128 528 645,width=0.495\textwidth,clip=]{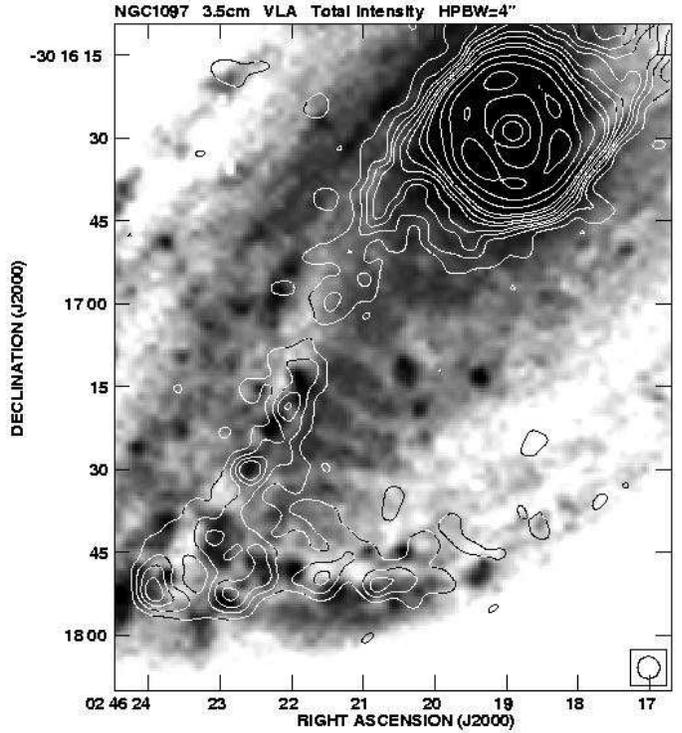}}
\caption{Total intensity contours of the central and southern parts
   of NGC~1097 at $\lambda3.5$~cm at 4\arcsec\ resolution. The contour
   intervals are as in Fig.~\ref{n1097s6}, with the basic contour level
   at 20~$\mu$Jy/beam area (the $3\sigma$ level).}
\label{n1097s4}
\end{figure}

The total and polarized radio continuum emission of NGC~1097 and NGC~1365 is
strongest near the galactic centres (see Sects.~\ref{sectCenter1097} and
\ref{sectCenter1365}) and in \emph{radio ridges\/} that approximately coincide
with the prominent dust lanes seen in the optical images
(Figs.~\ref{n1097s6}--\ref{n1097s4}). The lengths of the ridges are close
to the bar lengths. As gas and stars
in the bar region rotate faster than the bar pattern in the clockwise direction
(Ondrechen et al.\ \cite{OHH89}), the ridges are located on the
`downstream' side of the bar which is {\it leading\/} with respect to
the sense of rotation, and the regions `upstream' of the shock front are
located on the {\it following\/} side of the bar.
The ridges enter the central region
tangentially to the circumference at the two points where
the absolute maxima of polarized intensity occur, as can be
seen in the lower panels of Figs.~\ref{n1097s6} and \ref{n1365s15}.
Significant polarized emission \PI\ is also detected outside of the ridges
in the upstream regions (Figs.~\ref{n1097s10} and \ref{n1365s15})
and from around the whole bar region, forming a smooth envelope
(Figs.~\ref{n1097s15} and \ref{n1365s25}, see Sect.~\ref{sectMFArms}).

The general properties of the radio ridges, summarized in Table~\ref{Table3}, have
been determined from the maps at $\lambda6.2$~cm, with 6\arcsec\
($\approx500\p$) resolution for NGC~1097 and
with 15\arcsec\ ($\approx1.4\kpc$) resolution for NGC~1365.

The \emph{thermal fractions\/} of the radio emission $f_\mathrm{th}$ in the ridges
were estimated from the observed spectral index (Fig.~\ref{n1097spi10}),
assuming a constant synchrotron spectral index of $\alpha_\mathrm{s}=-1.0$.
A synchrotron spectral index of $\alpha_\mathrm{s}=-1$ is expected
from particle acceleration in strong shocks ($\alpha_0=-0.5$ intrinsically,
steepened by $\Delta \alpha=-0.5$ due to synchrotron and/or inverse
Compton losses).
We attribute the observed flattening of the spectrum with increasing distance
from the centre of NGC~1097 (Table~\ref{Table3}) to a larger thermal
fraction in the outer parts of the ridge.
This is in line with the H$\alpha$ intensity (Quillen et al.\
\cite{QF95}) which is very small in the inner bar and stronger in the
middle bar of NGC~1097.
In NGC~1365, no similar effect was detected.
However, the separation of thermal and nonthermal emissions
is uncertain because of the possible error in spectral index due to
the uncertainty in $\alpha_\mathrm{s}$, and also due to missing
large-scale emission in the total intensity maps.
Hence the values for $f_\mathrm{th}$ in Table~\ref{Table3} should be
regarded as crude estimates.
For a more detailed study, high-resolution observations at further
frequencies or sensitive, extinction-corrected H$\alpha$ data are required.

\begin{figure}[htbp]
\centerline{\includegraphics[bb = 62 106 528 666,width=0.495\textwidth,clip=]{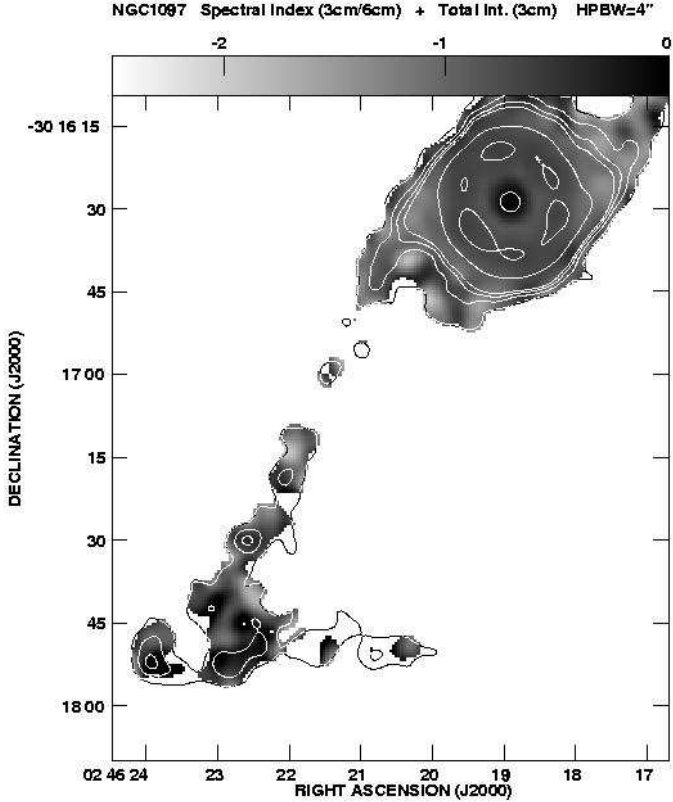}}
\caption{
Spectral index
   of the central region and southern ridge of NGC~1097
 between $\lambda3.5$~cm and $\lambda6.2$~cm at 4\arcsec\
   resolution, shown where the signal
exceeds $5\sigma_I$ at each wavelength. Contours show the
$\lambda3.5$~cm total intensity at
$1, 2, 3, 4, 16, 64$ times 40$\mu$Jy/beam area and the same
resolution.}
\label{n1097spi4}
\end{figure}

At the highest available resolution (4\arcsec, corresponding to
about 330~pc) the southern total-intensity radio ridge of NGC~1097 is
resolved into several features (Fig.~\ref{n1097s4}) which are not strictly
aligned with the ridge axis, neither do they coincide with the dust clouds
visible in the optical image. The strongest peaks are located
upstream of the dust lanes, and some of them coincide with
star formation regions visible in the optical images. According to the radio spectral
index map (Fig.~\ref{n1097spi4}) the emission from the peaks is a mixture of
thermal and nonthermal components. The observed spectral index $\alpha$ reveals
strong variations along the bar, with values ranging between $-0.1$
(expected for purely thermal emission) and less than $-1$ (typical of purely
nonthermal emission by cosmic-ray electrons suffering from synchrotron energy
loss in a strong magnetic field).

The location of the star formation complexes upstream of the
radio ridges seems to be in conflict with the picture of the onset of
star formation in the shock, as observed in other barred galaxies
(Sheth et al.\ \cite{SV02}). Our observations indicate that some compression
of gas and magnetic fields occurs ahead of the shock front,
perhaps in the narrow dust filaments mentioned in Sect.~\ref{sectMF}.

The upstream and downstream regions in \PI\ are separated by valleys of
low polarized emission, the \emph{depolarization valleys\/}, which are discussed
in Sect.~\ref{sectMF}.
The polarized intensities are similar on both sides of the
depolarization valleys, whereas the total intensity is much larger on the
ridges. Thermal emission cannot account for the enhanced total
emission in the ridges (see Table~\ref{Table3}).
This indicates that mainly the turbulent component of
the magnetic field is compressed in the shock, while the regular field
remains almost unaffected. This surprising result will be discussed in
Sects.~\ref{sectCompr} and \ref{sectContrast}.

In NGC~1097, the \emph{average degree of polarization\/} $p$ of the synchrotron
emission $p=\PI/I_\mathrm{s}$ at $\lambda6.2$~cm decreases from
about 30\% in the middle southern ridge to about 5\% at the end of the
bar (Table~\ref{Table3}), while it is almost constant
(approximately $20$\%) along the northern ridge
(here $I_\mathrm{s}$ is the synchrotron intensity).
In NGC~1365, $p$ is low
(5--15\%) along both ridges. The degrees of polarization are
generally similar at $\lambda3$~cm and $\lambda6$~cm, indicating that
Faraday depolarization is small (see the right-hand panel of Fig.~\ref{rm}),
except in the inner ridges and in the central regions.
In the region upstream of the ridge $p$ is 30--40\% (in both galaxies),
larger than in the ridge.

The observed \emph{half-maximum full width\/} $w$ (in both total and polarized
intensity -- the difference is insignificant)
of the northern and southern ridges of NGC~1097 increases from
about $600\p$ near the circumnuclear ring to about $900\p$ near the
end of the bar. Assuming a Gaussian ridge profile,
the observed width $w$ was corrected for smearing by the telescope
beam (with a Gaussian half-power width $\Theta$)
to obtain the intrinsic width $w_0$ via $w_0^2=w^2-\Theta^2$.
The intrinsic width varies from about $400\p$ to about
$700\p$ (Table~\ref{Table3}).
The radio ridges of NGC~1365 are broader and shorter than in NGC~1097,
with an intrinsic width of $\simeq1000\p$ and a length of about
1\farcm5 (8~kpc).

%%---------------------------------------
%FIG NGC1097 4\arcsec\
%%-------------------------------------

%FIG NGC1097 spectral index 4\arcsec\
%%------------------------------------

%---------------------------------------

\begin{table*}           % Table 3.
\caption{\label{Table3}Properties of the radio ridges in the bars of
NGC~1097 and NGC~1365 from the $\lambda6.2$~cm data.
$I$ is the total intensity, \PI\ the polarized intensity.
$\epsilon_I$ and $\epsilon\reg$ are the contrasts
in $I$ and \PI\ between the ridges and the surrounding regions.
$\alpha$ is the spectral index between $\lambda6.2$~cm and $\lambda3.5$~cm,
$f_\mathrm{th}$ is the thermal fraction of $I$, $I_\mathrm{s}=I(1-f_\mathrm{th})$
is the synchrotron intensity, and $p=P/I_\mathrm{s}$ is the
fractional polarization of the synchrotron emission.
$B\tot$ and $B\reg$ are the equipartition strengths of the total and
regular + anisotropic random magnetic fields derived from $I_\mathrm{s}$ and \PI\
(see Sect.~\ref{sectMFbar}).
The distance along the NGC~1097 ridge is measured from the galaxy's centre.
For NGC~1365, average values along the ridges are given.
}
\centering
\begin{tabular}{@{}lccccccccc@{}}
\hline\hline
      & Width in & Contrast & Width in & Contr. & Spectral & Thermal & Fractional & $B\tot$ & $B\reg$\\
      & $I$ [pc] & $\epsilon_I$ & \PI\ [pc] & $\epsilon\reg$ & index $\alpha$ & fract.\ $f_\mathrm{th}$ & polarization\ $p$ & [$\mu$G] & [$\mu$G]\\
\hline
{\bf NGC~1097}\\
Inner ridge (2~kpc south) & 400 &  10 & 400 & 7   & $-1.0$ & $<0.05$ & 0.10 & 30 & 9\\
Middle ridge (4~kpc south)& 500 &  7  & 700 & 1.5 & $-0.9$ & 0.05 & 0.28 & 20 & 12\\
Outer ridge (6~kpc south) & 700 &  5  & 800 & 0.9 & $-0.7$ & 0.25 & 0.15 & 22 & 9\\
\hline
{\bf NGC~1365}\\
Eastern ridge             & 900 &  10 & 900 & 0.5 & $-0.75$ & 0.2 & 0.13 & 24 & 10\\
Western ridge             &1000 &  5  & 900 & 3   & $-0.75$ & 0.2 & 0.04 & 30 & 7\\
\hline
\end{tabular}
\end{table*}
%%---------------------------------------

The \emph{total intensity contrast\/} in NGC~1097 between the ridges and the
surrounding
regions, at $\pm(7$\arcsec\--12\arcsec) distance from the ridge axis,
is about $10$ in the inner ridge and decreases to about $5$ towards the bar's
end (Table~\ref{Table3}). These estimates apply to
both the northern and the southern ridges, although the radio
emission in the northern ridge is fainter. In NGC~1365, the contrast is
about $10$ in the eastern and $5$ in the western ridge, while the total radio
intensity is higher in the west (Table~\ref{Table3}).

The \emph{contrast in polarized intensity\/} between the ridges and their
surroundings is quite different from that in total intensity
(Table~\ref{Table3}). Strong polarized emission is observed upstream of
the radio ridge and dust lane in one half of each galaxy (especially to the
west of the southern ridge in NGC~1097 and south of the eastern ridge in
NGC~1365), separated from the polarization
ridge by a narrow `depolarization valley'
(see Sect.~\ref{sectMF}). Near the outer end of the ridges,  the
polarized intensity upstream is even stronger than in the ridge,
so that the contrast is smaller than unity. In the northern
and inner southern ridges of NGC~1097, the polarized intensity is
about 7 times higher than in the upstream regions. The difference
from the total intensity ridges suggests that turbulent and regular
magnetic fields respond differently to compression and shear in the dust
lane region (see Sect.~\ref{MFCS}).

The ridge of total intensity observed in NGC~1097 is systematically
\emph{shifted\/} from that in polarized intensity (Fig.~\ref{fig:vr}).
As determined from several one-dimensional cuts
of the southern ridge at 10\arcsec\ resolution, the
peak in total intensity is about 5\arcsec\ (400\p) upstream -- i.e.\ to the
south-west -- compared to the
peak in polarized intensity. Note that the position of the peak of an
emission structure can be determined
with a better accuracy than the beam width if the
signal-to-noise ratio is high (Harnett et al.\ \cite{HE04}).
Figure~\ref{n1097s10} shows that the polarized intensity follows the
optical dust lanes very closely.

In the northern ridge of NGC~1097, a similar shift between the ridges
of total and polarized emission is visible, though less clearly due to the
weaker polarized emission.
In NGC~1365, an offset has not been detected.  However, the resolution of our data
is lower than for NGC~1097, so that a shift of $\le10$\arcsec\ cannot
be excluded.

%-----------------------------------------
\subsection{The magnetic field structure in the bar regions}
\label{sectMF}

The $B$-vectors of polarized emission in the bar region change their
orientation rapidly
upstream of the dust lanes in both galaxies
(Figs.~\ref{n1097s6}--\ref{n1365s15}).
This leads to depolarization within the telescope
beam, observed as a `depolarization valley' parallel to the emitting
ridge.  This was first observed in NGC~1097 by Beck et al.\
(\cite{BE99}).
Our new data show that the valley persists when observed with
higher angular resolution. Its structure is similar at both
wavelengths so that it cannot be produced by Faraday depolarization.
The average distance between the depolarization valley and the ridge
in polarized intensity is 10\arcsec\ ($\approx800\p$) in NGC~1097 and
11\arcsec\ ($\approx1.0$~kpc) in NGC~1365.

The observed width of the depolarization valley, defined as the distance
between points where the polarized intensity drops to half the value
outside the valley, is about 4\arcsec\ ($\approx300\p$) in NGC~1097
(Fig.~\ref{n1097s6}) and about 6\arcsec\ ($\approx500\p$) in NGC~1365.
This is the scale at which the regular magnetic field is deflected.
Note that a value of the width smaller than the resolution is reliable because
the deflection valley appears only in the map of polarized intensity
(which is not a directly observed quantity). In the maps of the
observed quantities, the Stokes parameters $Q$ and $U$, the valley is represented by
a smooth variation with a sign reversal at the location of the
depolarization valley. The width of the valley depends on the
gradients in $Q$ and $U$. Beam smearing decreases
these gradients, so that the intrinsic width is smaller than
that given above.

According to our $\lambda6.2$~cm data of NGC~1097 with high
signal-to-noise ratio (right-hand panel of Fig.~\ref{n1097s10}), the field
orientation turns smoothly within a distance of about 20\arcsec\ (1.6~kpc)
upstream of the ridge.
Both depolarization and field bending are discussed in Sect.~\ref{sectRMShear}.
In NGC~1365 the turning of the field lines
is smoother and the depolarization valley is broader than in NGC~1097.

%%-------------------------------------------
%FIGS Deprojection
\begin{figure*}[htbp]
\includegraphics[bb = 112 149 528 637,width=0.495\textwidth,clip=]{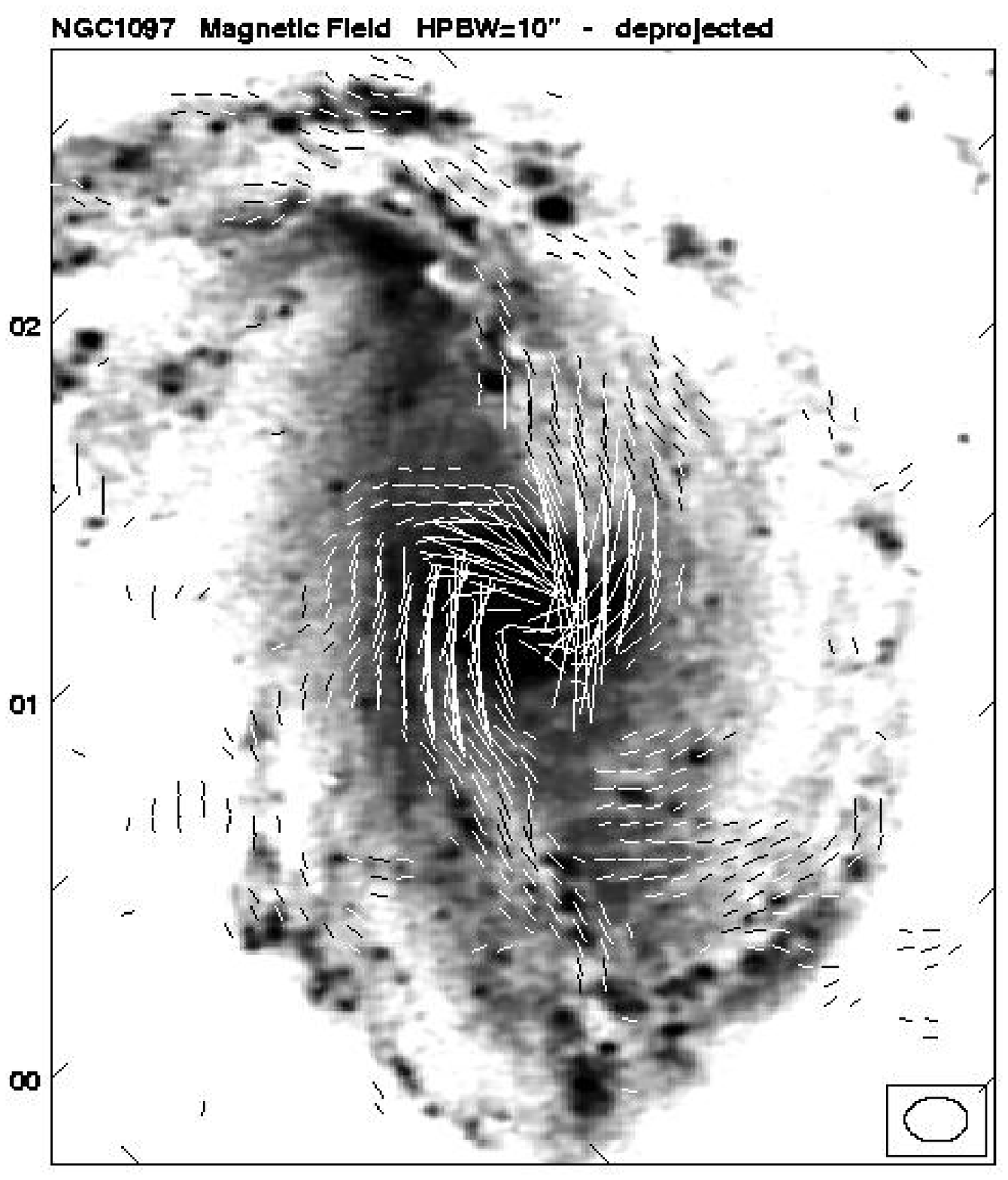}
\hfill
\includegraphics[bb = 84 193 521 501,width=0.495\textwidth,clip=]{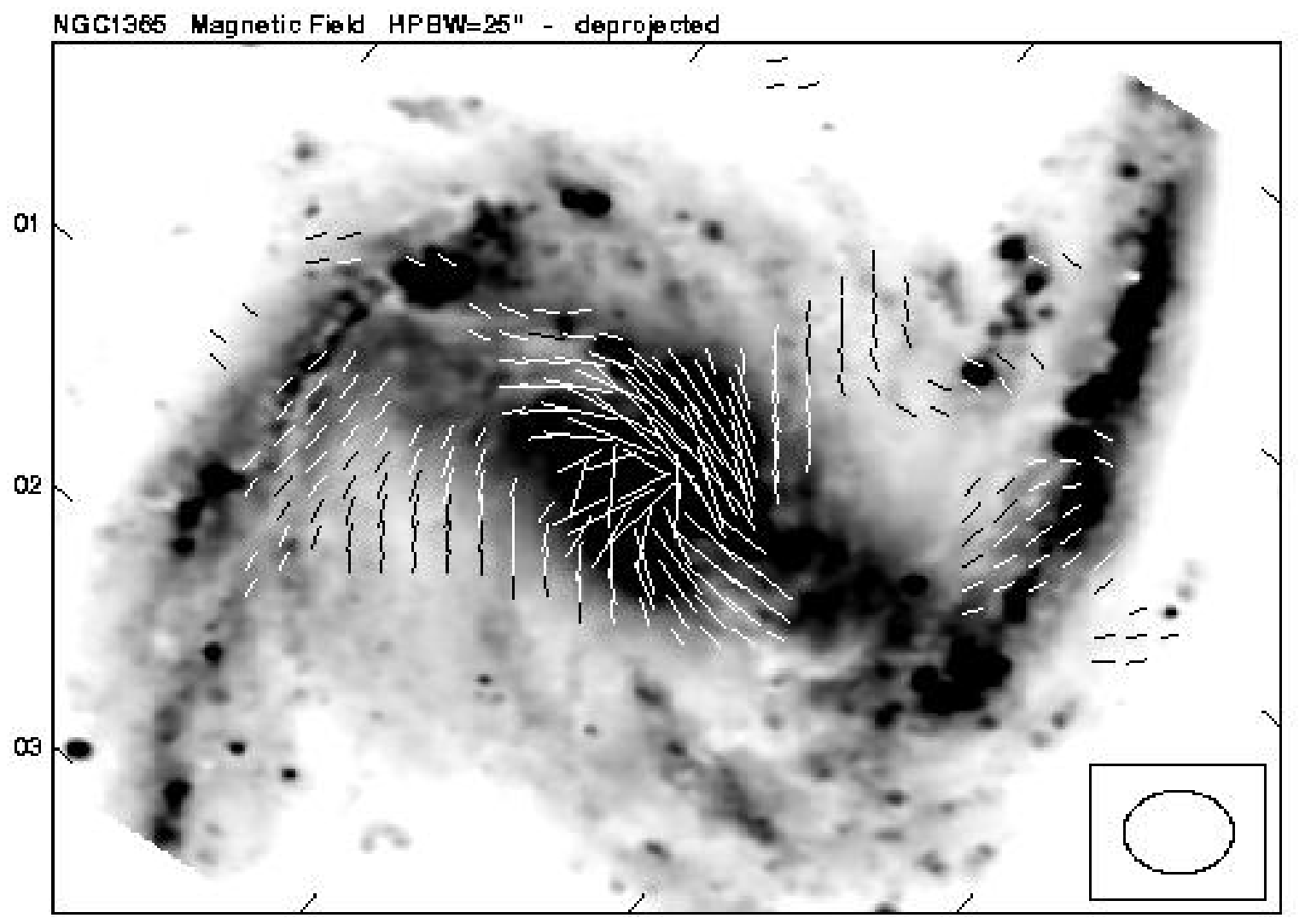}
\caption{Intrinsic orientation of the magnetic field
$B\reg$ (regular and/or anisotropic random),
obtained from polarized synchrotron emission
in NGC~1097 (left) and NGC~1365 (right), corrected for Faraday rotation and
deprojected into the galaxies' planes, with the major axis oriented
north-south. The vector length is proportional to polarized
intensity at $\lambda3.5$~cm.  The resolutions are 10\arcsec\ and
25\arcsec, respectively; the deprojected beam is shown in the lower right corner.
North is at top left and top right of the NGC~1097 and NGC~1365 images, respectively.
Distances given in axis labels are in arcmin.
}
\label{deproj}
\end{figure*}
%%------------------------------------------

Figure~\ref{deproj} shows the intrinsic magnetic field orientations in
both galaxies, derived from the $B$-vectors at $\lambda3.5$~cm and
$\lambda6.2$~cm, corrected for Faraday rotation (see Sect.~\ref{sectRM}) and
deprojected into the galaxies' planes.

The intrinsic magnetic field in the southern ridge of NGC~1097
follows the dust lane orientation (left-hand panel of Fig.~\ref{deproj}).
The orientation of the field and that of the dust lane generally agree in the
northern bar of NGC~1097 as well, but both are
less pronounced than in the south. As the northern half appears
distorted in all spectral ranges, we suspect gravitational distortion
by the companion galaxy NGC~1097A. In NGC~1365 the field in the
eastern ridge is along the dust lane, but the resolution is
insufficient to trace details.

In the southern bar of NGC~1097 (at 35\arcsec--50\arcsec\ or
3--4~kpc distance from the centre) the intrinsic pitch angle of the
magnetic field obtained from the polarized emission
(i.e., the angle between the field orientation and circumference) jumps from
about 15\degr\ (almost azimuthal) in the upstream region (west of the
southern ridge) to about 75\degr\ (almost radial) in the southern
ridge (see the left-hand panel of Fig.~\ref{deproj}),
which yields a deflection angle of about 60\degr. The
deflection angle decreases to about 40\degr\ at around 60\arcsec\
radius; this decrease is accompanied by a reduction in the
contrast in magnetic field strength (see Table~\ref{Table3}).
In NGC~1365 (see the right-hand panel of Fig.~\ref{deproj}),
the deflection angle is about 70\degr\ in the inner eastern bar.

The orientations of the field lines do not everywhere follow those of
the gas streamlines (in the corotating frame) of the hydrodynamic models of
Athanassoula (\cite{A92b}) and the models for NGC~1365 by Lindblad et
al.\ (\cite{LL96}).
It is instructive to compare the right-hand panel of Fig.~\ref{deproj}
and Fig.~26b of Lindblad et al.\ (\cite{LL96}), where it appears that the
alignment is reasonably tight in the northern part of the galaxy, but
not in the southern part.
This is discussed in detail in Sect.~\ref{sectModel2}.

The intrinsic field orientation upstream of the southern bar of NGC~1097
(left-hand panel in Fig.~\ref{deproj})
follows the small, feather-like dust filaments located upstream of the
main dust lanes as seen in the optical image (Fig.~\ref{n1097s10},
around RA, DEC(J2000) = 02~46~21, $-30$~16~25).
The reason for this alignment and the origin of the filaments are unknown.

The structure of the incoherent and coherent magnetic fields is further
discussed in Sects.~\ref{sectAniso} and \ref{sectRMShear}.

%-----------------------------------------
\subsection{Faraday rotation and depolarization in the bar region}
\label{sectRM}

%-----------------------------------------
% FIGS RM and DP
\begin{figure*}[htbp]
\includegraphics[bb = 62 106 528 673,width=0.475\textwidth,clip=]{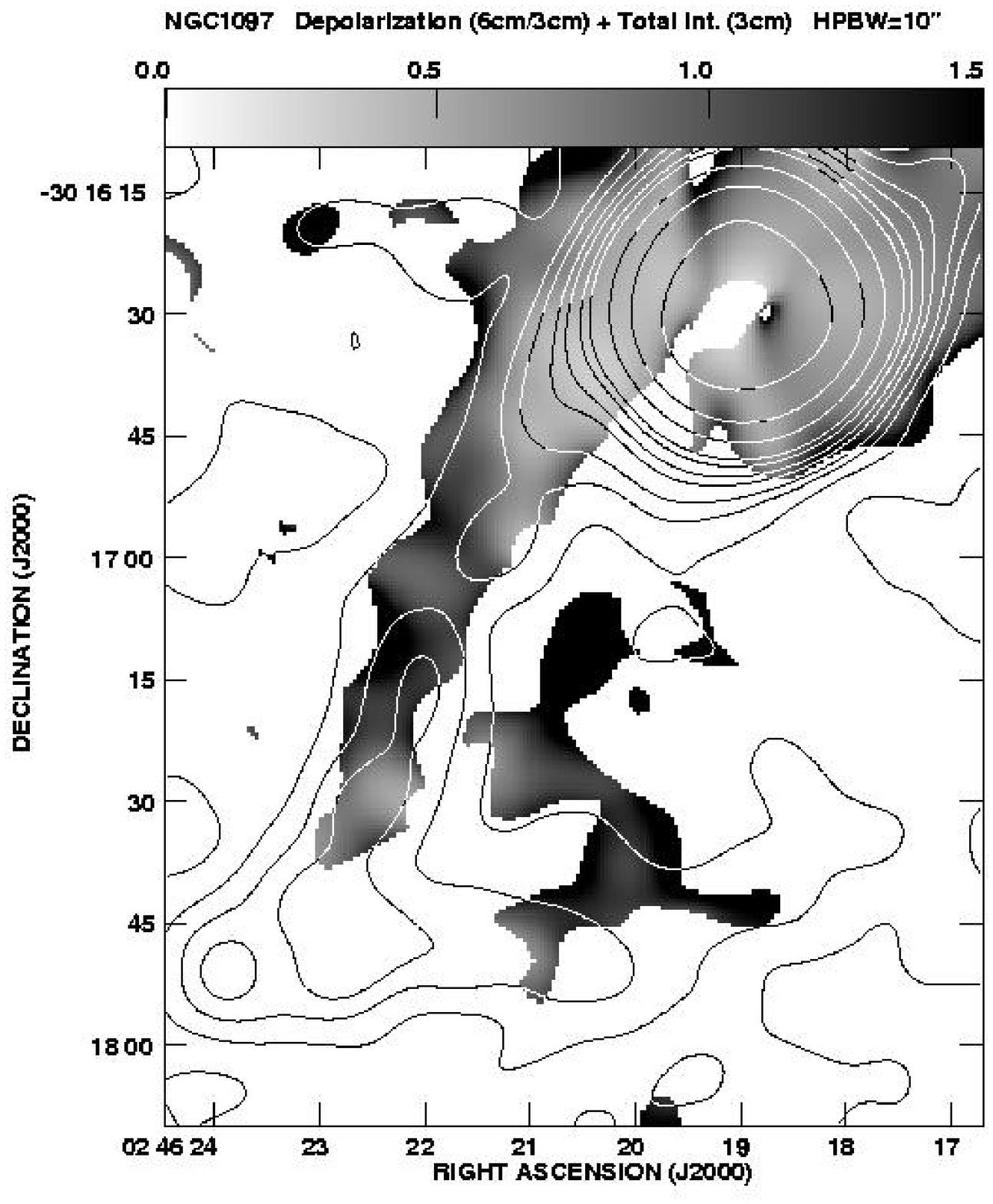}
     \hfill
\includegraphics[bb = 62 106 528 673,width=0.475\textwidth,clip=]{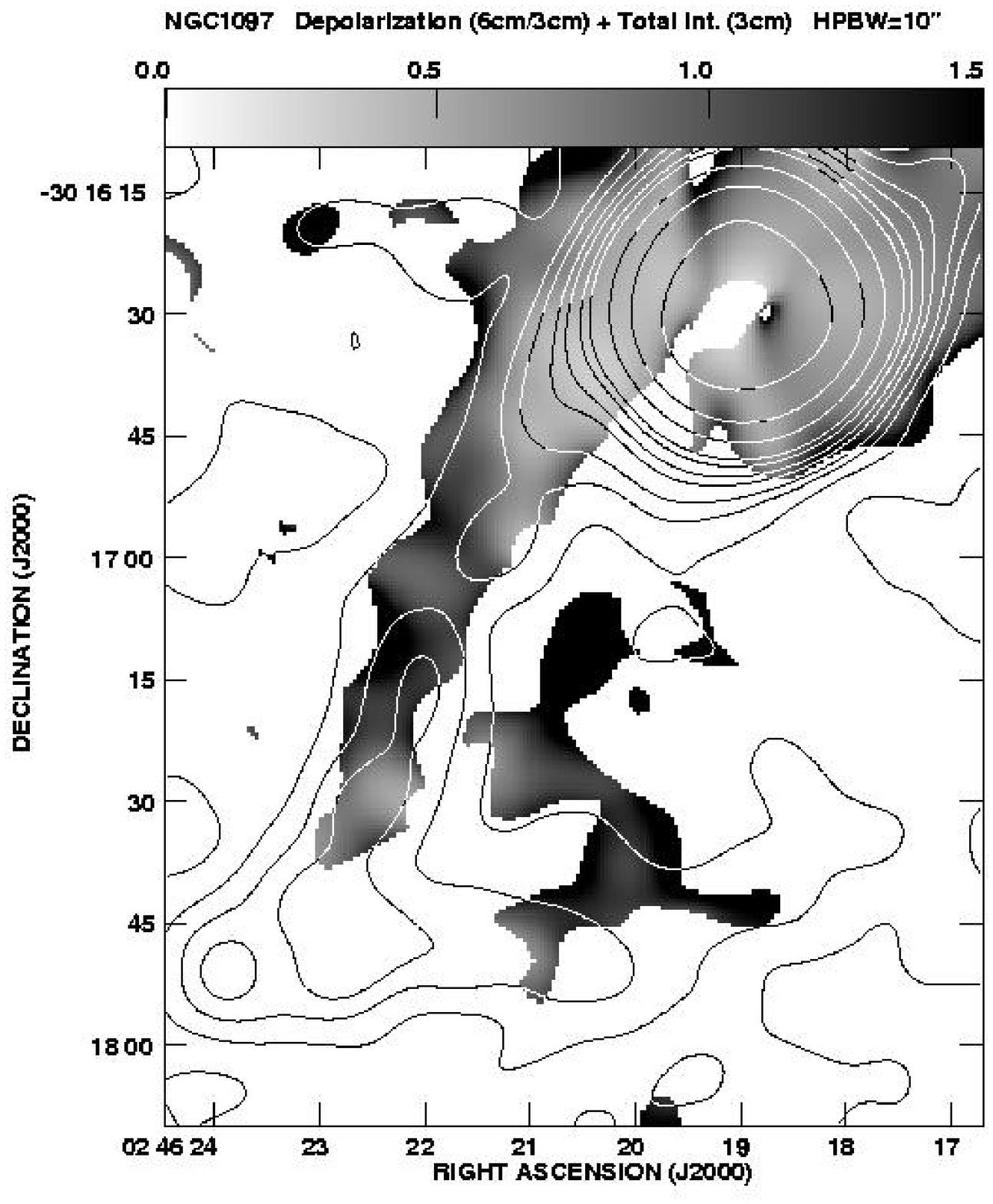}

\vspace{0.2cm}

\includegraphics[bb = 62 149 536 630,width=0.475\textwidth,clip=]{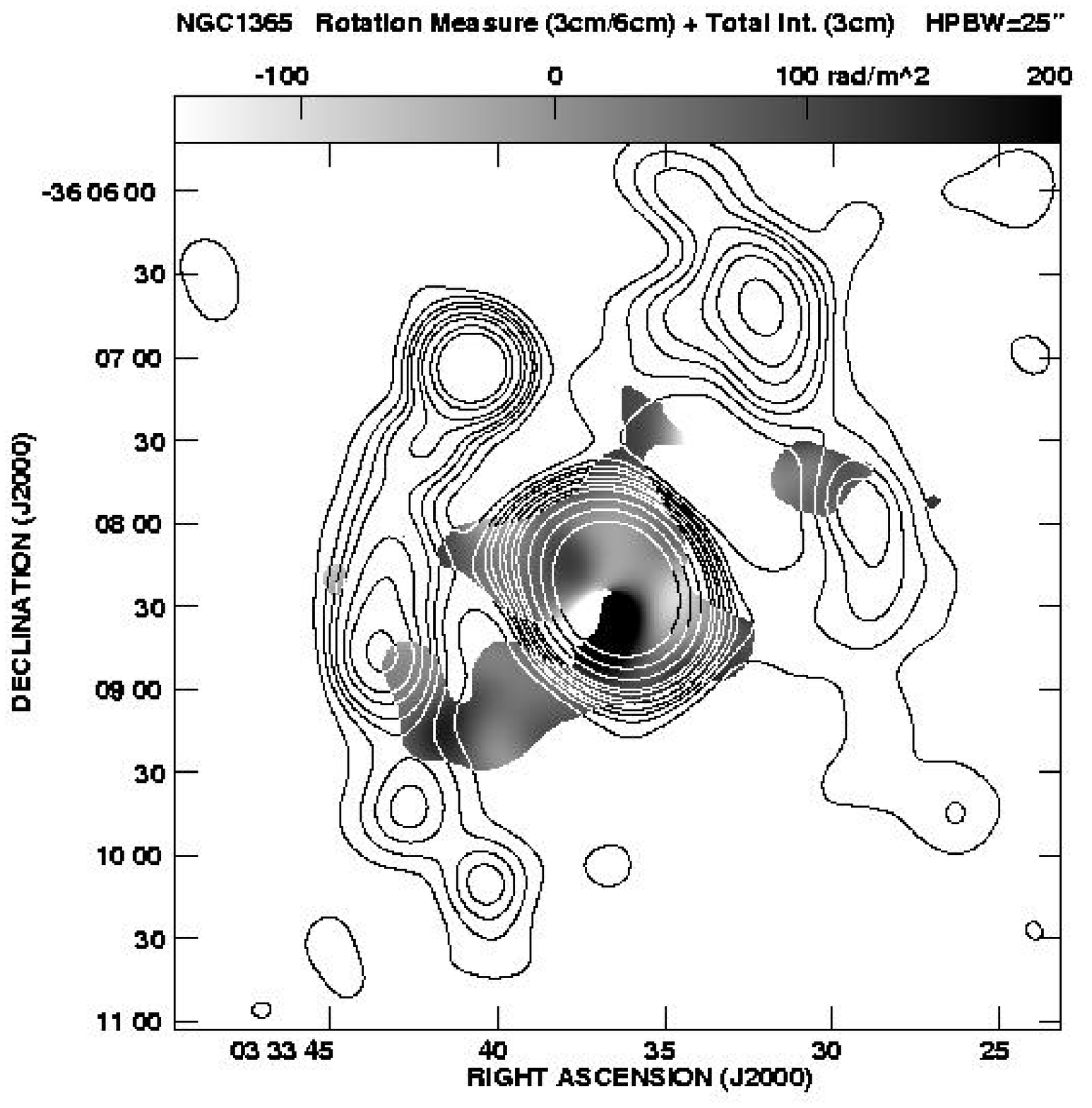}
\hfill
\includegraphics[bb = 62 149 536 630,width=0.475\textwidth,clip=]{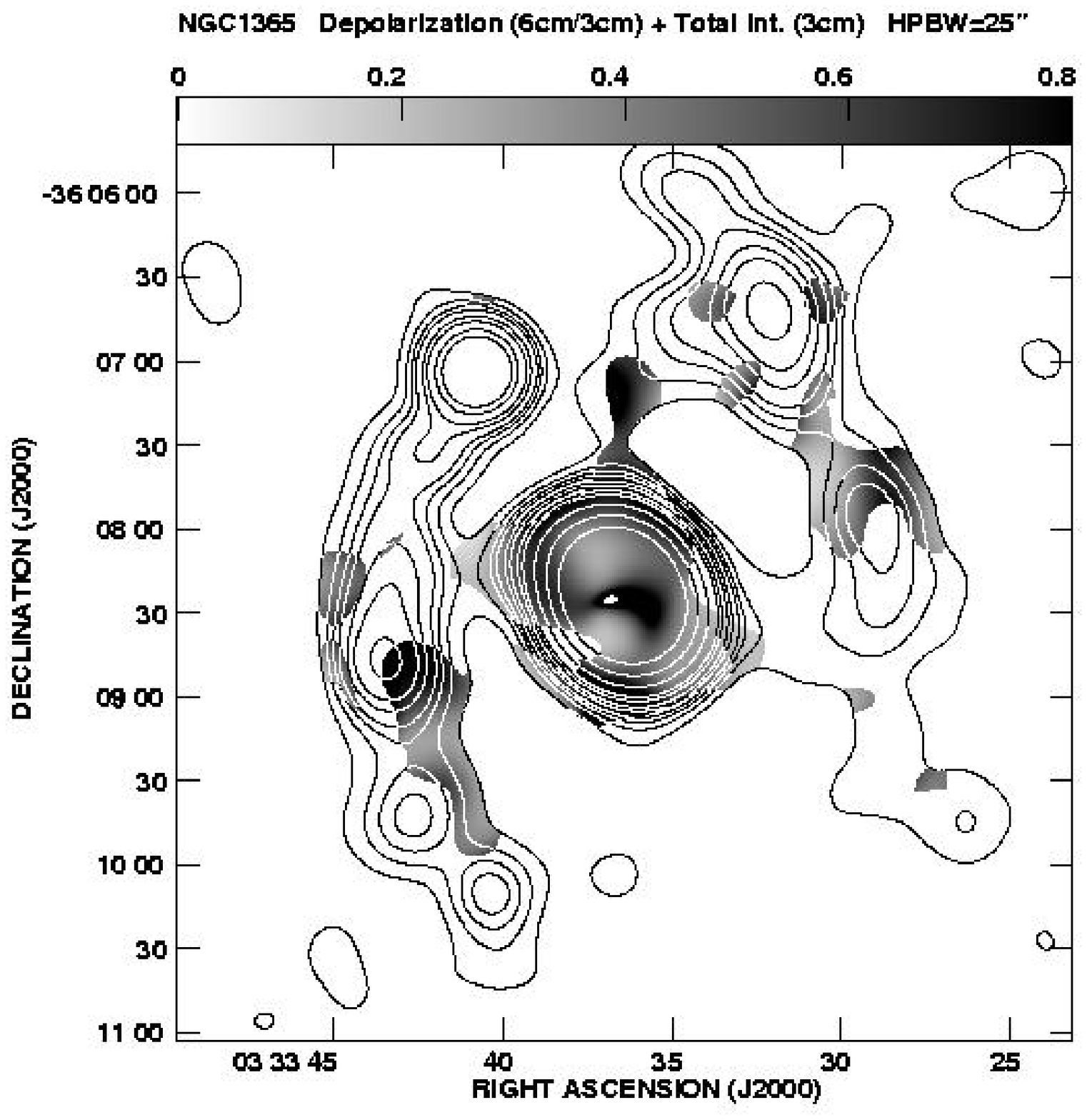}
   \caption{{\it Top row:\/} Faraday rotation measure (left) and Faraday
     depolarization (right) of NGC~1097 in grey scales between
     $\lambda3.5$~cm and $\lambda6.2$~cm at 10\arcsec\
     resolution, overlaid onto contours of total intensity at
     $\lambda3.5$~cm.  {\it Bottom row:\/} Faraday rotation measure
     (left) and Faraday depolarization (right) of NGC~1365 in grey
     scales at 25\arcsec\ resolution, overlaid onto contours of total
     intensity at $\lambda3.5$~cm.}
   \label{rm}
\end{figure*}
%% ----------------------------------------

% FIG NGC1097 AZIMUTHAL RM VARIATION
\begin{figure*}[htbp]
\includegraphics[width=0.495\textwidth]{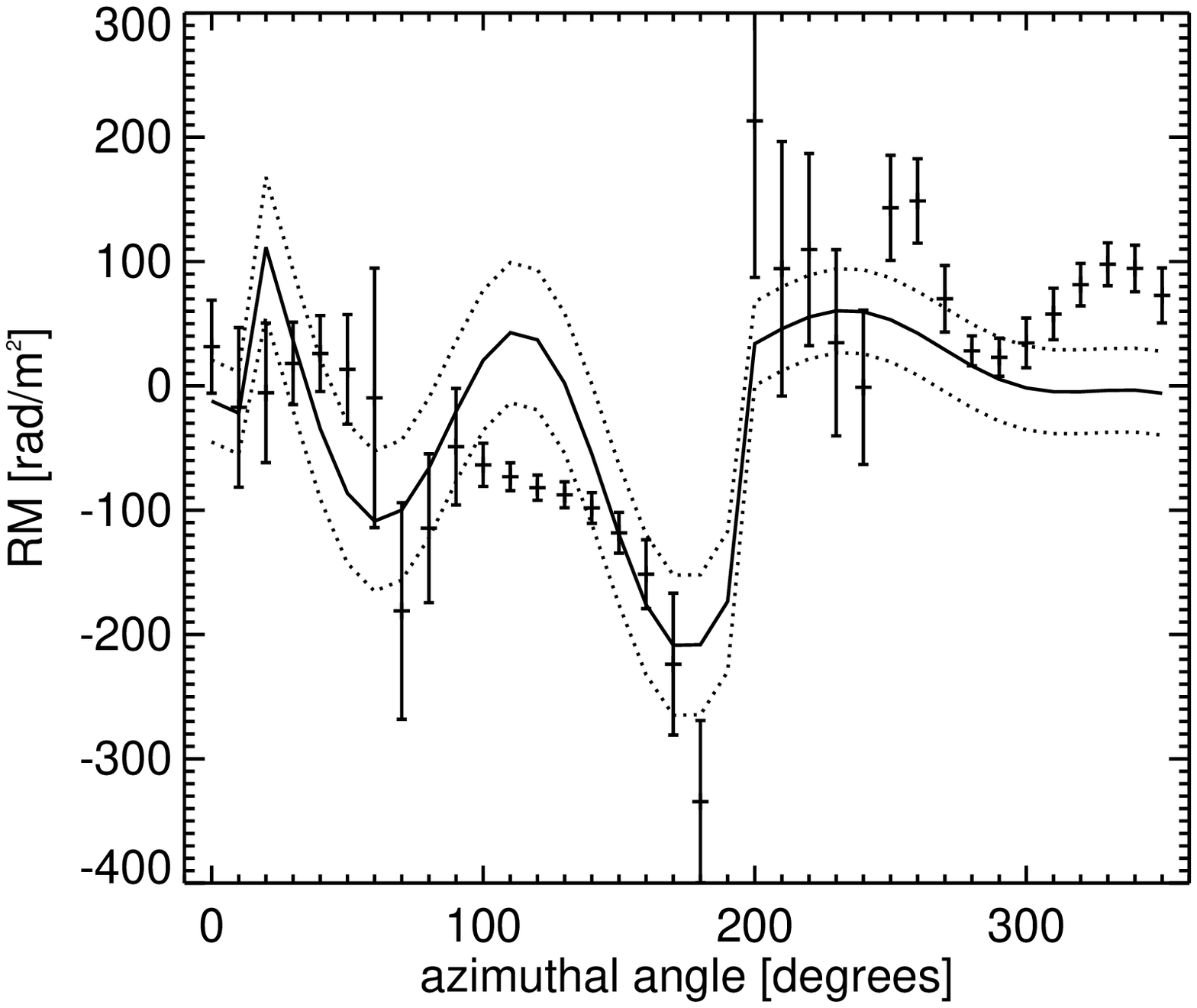}
\hfill
\includegraphics[width=0.495\textwidth]{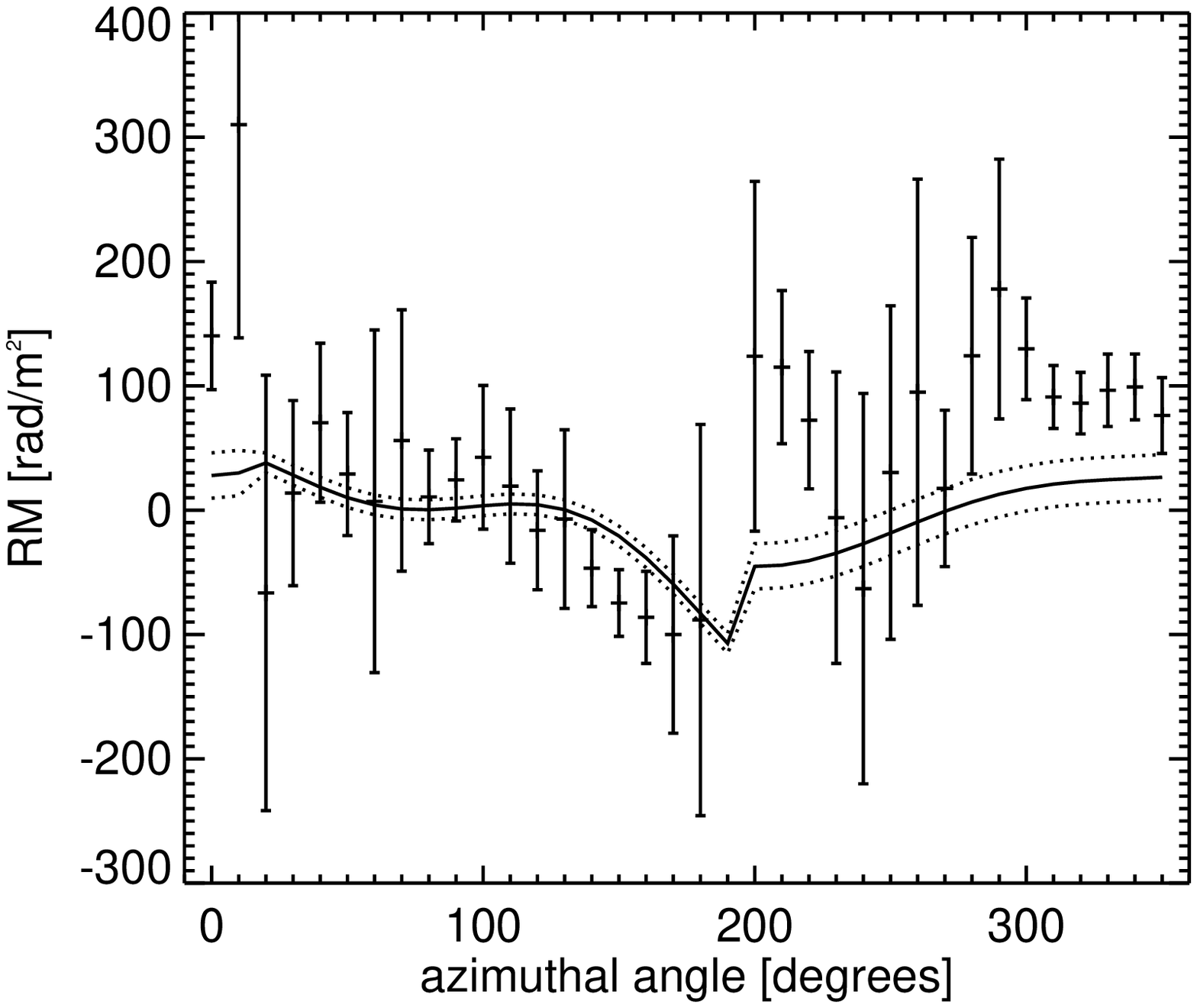}
\caption{The azimuthal variation of \RM\ in NGC~1097, obtained between $\lambda3.5$~cm
and $\lambda6.2$~cm at 10\arcsec\ resolution in the rings
$1.25<r<2.5\kpc$ (15\arcsec--30\arcsec, left) and $2.5<r<3.75\kpc$
(30\arcsec--45\arcsec, right) where $r$ is the radius from the
centre of NGC~1097. The error bars show the standard deviation
of \RM\ in each sector. Solid lines are \RM\ values obtained from
the fits to polarization angles described in Sect.~\ref{sectGlobal}, with the
dotted lines showing the range of \RM. The azimuthal angle is measured in the
galaxy's plane (position angle of the major axis of $-45\degr$,
inclination of $45\degr$) counterclockwise from the northern major
axis. The southern ridge line of polarized intensity is located at
$130\degr$ azimuthal angle (left) and at $160\degr$ (right).}
\label{n1097rm}
\end{figure*}
%% -----------------------------------------

The polarization angles at $\lambda3.5$~cm and $\lambda6.2$~cm
of NGC~1097 and NGC~1365 with 10\arcsec\ and 25\arcsec\ resolution,
respectively, were used to derive
maps of the Faraday rotation measure \RM\ (measured in
$\!\radm$) shown in Fig.~\ref{rm}.
The ambiguity in \RM\ due to the $\pm\pi$ uncertainty
in polarization angle is $\pm1232\radm$. The Faraday rotation measure
\RM\ is proportional to the
product of thermal electron density $\ne$ (in $\!\cmc$) and the
line-of-sight component $B_{\parallel}$ of the regular magnetic
field (in $\!\mkG$), integrated over the path length $L$ (in pc):
$\RM=0.81\int \ne B_{\parallel}\,dL$.

\RM\ in NGC~1097 (top left panel of Fig.~\ref{rm}) is generally
positive on the western and negative on the eastern side of the inner
southern ridge, with a jump at the position of the depolarization valley,
where the orientation of the field $B_{\perp}$ in the sky plane also
turns rapidly (see Sect.~\ref{sectMF}). This discontinuity in the magnetic
field is weaker in the middle ridge and vanishes in the outer ridge. However,
the errors in \RM\ increase with increasing distance from the centre due to
the decreasing polarized intensities.

Figure~\ref{n1097rm} shows the azimuthal variation of \RM\ in two
rings in the plane of NGC~1097. In the inner ring (left-hand panel), west
of the southern ridge (at $190\degr$ azimuthal angle), \RM\ {\it
reverses its sign\/} and jumps by about $500\radm$, which corresponds to
a change of the polarization angle by about $110\degr$ at $\lambda6.2$~cm
and by about $35\degr$ at $\lambda3.5$~cm. At larger radii
(right-hand panel of Fig.~\ref{n1097rm}), the \RM\ jump decreases to about
$200\radm$. In the outer bar (45\arcsec--75\arcsec\ radius, not shown), \RM\
does not jump, but changes smoothly by about $300\radm$ within a range
of $50\degr$ in azimuthal angle.

Negative \RM\ in the southern bar means that the line-of-sight
component of the magnetic field points away from us. Knowing that the
position angle of the major axis of the projected plane of NGC~1097 is
$-45^\circ$ with the south-western side being nearer to us,
the radial component of the magnetic field in the ridge points
{\it inwards\/} (i.e. towards the galaxy's centre).

In the northern ridge of NGC~1097, signal-to-noise ratios are low,
so that \RM\ can be determined only with low accuracy. \RM\ values of
$+100\radm$ around $330^\circ$ azimuth (Fig.~\ref{n1097rm}) show
that the regular magnetic field along the ridge also points towards
the centre and has a similar strength to that in the southern ridge.

\RM\ in NGC~1365 (bottom left panel of Fig.~\ref{rm}) is measured only
in the central region (Sect.~\ref{sectCenter1365})
and a few patches between the spiral arms.
The polarized emission in NGC~1365 is generally weaker than
that in NGC~1097.

Faraday depolarization $\DP$ (defined as the ratio of the degrees of
polarization at $\lambda6.2$~cm and $\lambda3.5$~cm) is shown in
the right-hand column of Fig.~\ref{rm}. The amount of depolarization
is small ($\DP\simeq1$) in
the outer ridge and upstream regions of NGC~1097, but increases
towards the centre ($\DP\simeq0.5$) where the turbulent field strength
is presumably larger.  Depolarization in NGC~1365 is stronger
($\DP=0.3$--0.6 in the
ridges and in the spiral arms) than in NGC~1097, possibly due to
larger thermal plasma density, but the data are of limited quality as
polarized intensity is low at $\lambda3.5$~cm.

%-----------------------------------------------
\subsection{Field structure in the outer regions and spiral arms of NGC~1097}
\label{sectArms}

The optical image of NGC~1097 (Fig.~\ref{n1097s6}) shows an elliptical
structure of about $3\farcm5 \times 2\arcmin$ in size, connecting the
ends of the bar.
The ellipse is also the outer edge of the disc-like diffuse optical
emission. This ellipse can reflect
one family of orbits in a typical bar potential
(see, e.g., Fig.~12 in Athanassoula\ \cite{A92a}).
Highly polarized radio emission
emerges from along most of this feature (Fig.~\ref{n1097s15}),
with $B$-vectors mostly tangent to it.
This indicates that the regular magnetic field is aligned with the gas
flow in this region.

The spiral arms are clearly visible in radio continuum, at least the
parts which are near to the galaxy's centre.
The orientation of the $B$-vectors at
$\lambda6.2$~cm is along the inner northern spiral arm, but almost
perpendicular to the outer arm in the south-west
(Fig.~\ref{n1097s15}). Strong Faraday
rotation seems improbable at such large radii, hence the field must be
distorted, presumably together with the gas flow.

%---------------------------------------------
\subsection{The magnetic field strength in the bar regions}
\label{sectMFbar}

We computed the strengths of the total $B_I$ and
regular + anisotropic random fields $B\reg$ in the radio ridges
of NGC~1097 and NGC~1365 from the total synchrotron intensity
$I_\mathrm{s}$ and the degree of polarization $p$ of the
synchrotron emission, using the maps at $\lambda6.2$~cm at 10\arcsec\
and 15\arcsec\ resolution, respectively.
(Although at $\lambda6.2$~cm Faraday depolarization may be significant
in some regions, we preferred to use this wavelength because of the higher
signal-to-noise ratios than at $\lambda3.5$~cm.)
$I_\mathrm{s}$ followed from the total intensity by subtracting
the thermal fraction estimated from the observed spectral
index $\alpha$ between $\lambda3.5$~cm and $\lambda6.2$~cm
(Table~\ref{Table3}).

We further assumed equipartition between the energy of the total magnetic
field and that of the cosmic rays (protons + electrons),
with a number density ratio $K$ of cosmic ray protons to
electrons of 100 in the relevant (GeV) energy range, a path length
through the synchrotron-emitting medium of $L=500\p$
(similar to the intrinsic width of the ridges of NGC~1097,
see Sect.~\ref{sectRidge}),
and a synchrotron spectral index of $\alpha_\mathrm{s}=-1$.
We applied the revised equipartition formula of Beck \& Krause (\cite{BK05}).
(For field strengths beyond $10\mkG$ and for $\alpha_\mathrm{s}\le-1$
the revised formula gives smaller field strengths than the classical estimate.)
The results for $B\tot$ and $B\reg$ are given in Table~\ref{Table3}.
These estimates scale with $(K / L)^{1/(3-\alpha_\mathrm{s})}$,
so that even an uncertainty in $K$ or $L$ of 40\% would
cause an error in $B$ of only 10\% (for $\alpha_\mathrm{s}=-1$).

The ridges in total and polarized intensity are produced by an
increase in $B\tot$ and $B\reg$ due to compression and shear.
The strong anisotropy in the random magnetic field
caused by compression is the main contributor to the increased
intensity (Sects.~\ref{sectCompr} and \ref{sectAniso}).

The observed \RM\ can be used to estimate the strength of the
component $\mean{B}_{\parallel}$ of the regular field $\mean{\vec{B}}$
along the line of sight
if the thermal electron density and the path length are known.
In principle the thermal electron density of the Faraday rotating layer can be
derived from an estimate of the emission measure of the thermal radio emission.
The resulting estimates for $\mean{B}_{\parallel}$ depend on the uncertain
quantities \EM, $L$ and the volume filling factor $\eta$ of thermal electrons
to the power $-1/2$, and are thus more sensitive to
the assumed parameters than in the equipartition estimate. Bearing this in mind,
for an estimated \EM\ of $100\cm^{-6}\p$ in the southern ridge of NGC~1097
(where the thermal fraction $f_\mathrm{th}$ is about 0.05, see Table~\ref{Table3}),
using Eq.~(1) in Ehle \& Beck (\cite{EB93}),
assuming that $20\%$ of this emission measure is due to diffuse gas
($\eta=0.2$, as in normal spiral galaxies, e.g. Greenawalt et al.\
\cite{Greenawalt98}), and taking a path length $L=500\p$,
we obtain $\mean{n}_\mathrm{e}=(\EM \eta / L)^{1/2} \simeq 0.2\cmc$. For an
\RM\ of $100\radm$ observed in the southern ridge, this gives a
strength of the regular magnetic field component along the line of sight of
$\mean{B}_{\parallel}=\RM / (0.81\mean{n}_\mathrm{e} L) \simeq 1\mkG$.

To compute the full regular magnetic field strength $\mean{B}$, the
angle $\phi_1$ between the field and the sky plane
has to be known. If the regular field is directed
along the ridge, the position angle of the field in the sky plane with respect to the
galaxy's major axis is $\phi_2 \simeq 20^{\circ}$. However, the left-hand panel of
Fig.~\ref{fig:global} indicates that the regular field may have a position angle
of about $45^{\circ}$ in the galaxy plane, or $\phi_2 \simeq 35^{\circ}$ in the sky plane.
The angle $\phi_1$ between the regular field and the sky plane follows
from $\tan \phi_1 = \tan \phi_2 \sin i$ where $i=45\degr$ is the galaxy's inclination,
and the regular field strength follows from
$\mean{B}=\mean{B}_{\parallel} / \sin \phi_1$. For the range of probable
$\phi_1$ values, we derive $\mean{B} \approx 2-4\mkG$.
This is significantly lower than the
value obtained from polarized intensity, $B\reg\simeq 12\mkG$,
strongly suggesting that a substantial fraction of the polarized
emission is due to anisotropic random magnetic fields
(Sect.~\ref{sectAniso}).

% ----------------------------------------------
\subsection{Magnetic field strengths in the spiral arms, disc, and envelope}
\label{sectMFArms}

Substantial radio emission has been detected in the northern spiral arm of
NGC~1097, which emerges from the northern end of the bar
(around RA, DEC(J2000) = 02~46~25, $-30$~14~15, see
Figs.~\ref{n1097s6} and \ref{n1097s15}).
The magnetic field orientation suggested by the polarization angles
is along the arm. The mean degree of polarization is low
at $\lambda3.5$~cm and $\lambda6.2$~cm ($p\le5\%$).
The mean equipartition magnetic field strengths obtained from the total
synchrotron emission (after subtraction of 30\% thermal emission)
and polarized radio emission in the northern arm are
$B\tot\simeq18\mkG$ and $B\reg\simeq6\mkG$, respectively.

Bright radio emission outside of the bar of NGC~1097 is also detected in the
elongated region emerging from the end of the southern bar towards the west
(around RA, DEC(J2000) = 02~46~20, $-30$~17~50, see
Fig.~\ref{n1097s10}). Here the degree of polarization is quite high ($p\simeq10\%$).
The equipartition strengths of magnetic fields obtained from the total
synchrotron and polarized radio emissions
are $B\tot\simeq22\mkG$ and $B\reg\simeq10\mkG$, respectively.
This feature appears to be a part of
the ring-like structure discussed in Sect.~\ref{sectArms}.

The distribution of  \RM\ in NGC~1097 does not show any
systematic variation in the region of the spiral arms beyond the bar.
However, the \RM\ errors are large due to the weak polarized emission.

The spiral arms of NGC~1365 are brighter than those in NGC~1097
both in optical light and in radio continuum (Fig.~\ref{n1365s15}), while the degree
of radio polarization is similarly low ($p\le5\%$) in both galaxies.
The mean equipartition strengths of magnetic fields obtained from the total
synchrotron and polarized radio intensities
are about $22\mkG$ and $3\mkG$, respectively, for the north-western
arm of NGC~1365, and about $18\mkG$ and $5\mkG$ for the south-eastern arm.

In the low-resolution images,
both galaxies exhibit an envelope of diffuse radio emission (Figs.~\ref{n1097s15} and
\ref{n1365s25}) where the degree of polarization is high, typically 25\% at
$\lambda6.2$~cm. Assuming that the emission emerges from an extended disc with
a path length of 1~kpc, negligible thermal emission and a synchrotron spectral index of
$\alpha_\mathrm{s}=-1$, we obtain $B\tot\simeq13\mkG$ and $B\reg\simeq7\mkG$
for NGC~1097, and $B\tot\simeq9\mkG$ and $B\reg\simeq5\mkG$ for NGC~1365.
If, however, the diffuse emission emerges from a \emph{halo\/} with a path length
of 10~kpc, the fields are weaker by a factor of about 1.8.

% -----------------------------------------------
\section{The global magnetic structures}
\label{sectGlobal}

%-----------------------------------------------
\begin{figure*}[htbp]
     \label{fig:global:1097}
     \includegraphics[bb = 111 77 547 496,width=0.46\textwidth,clip=]{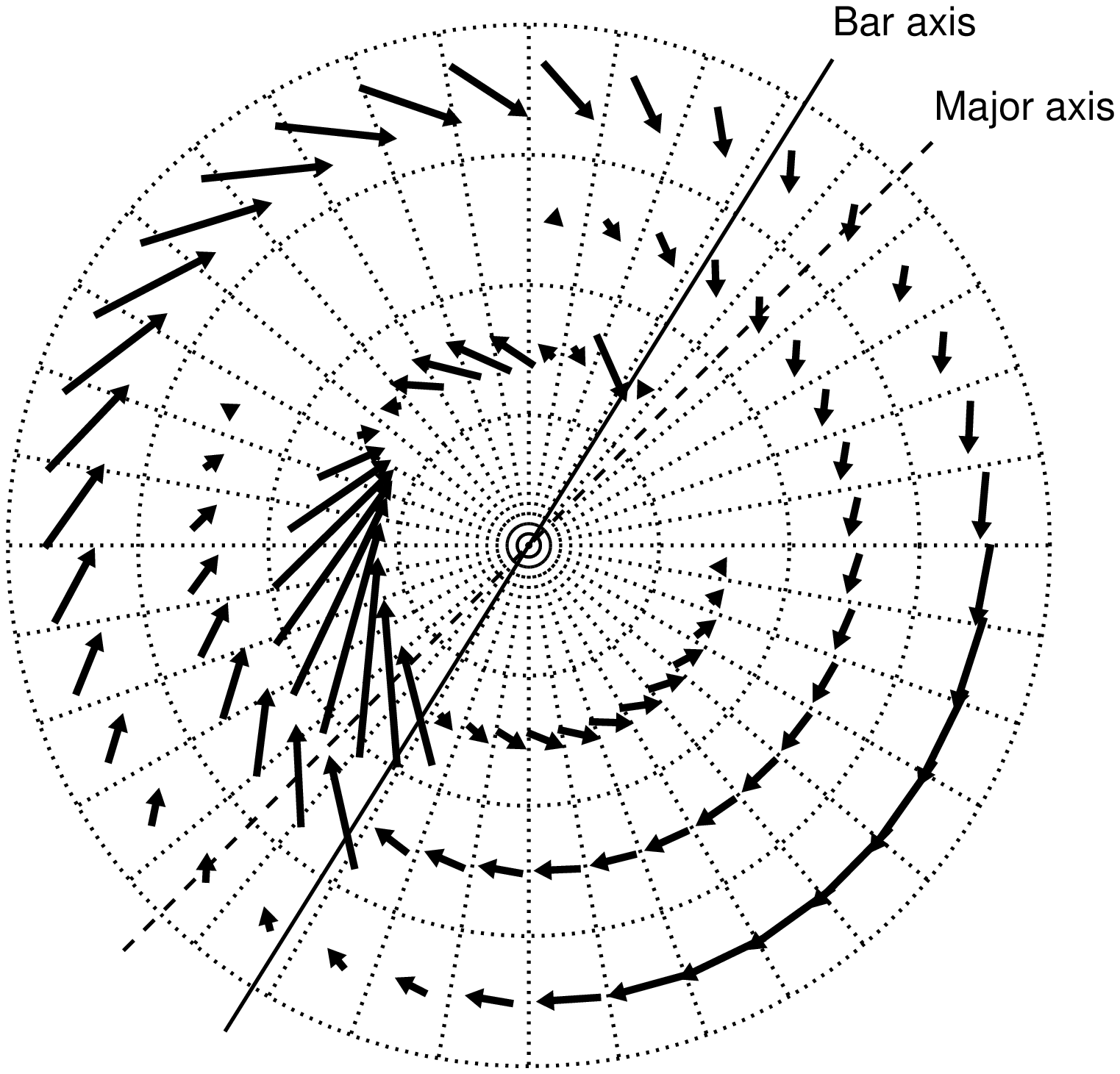}
  \hspace{-4mm}
     \label{fig:global:1365}
     \includegraphics[bb = 48 49 521 438,width=0.537\textwidth,clip=]{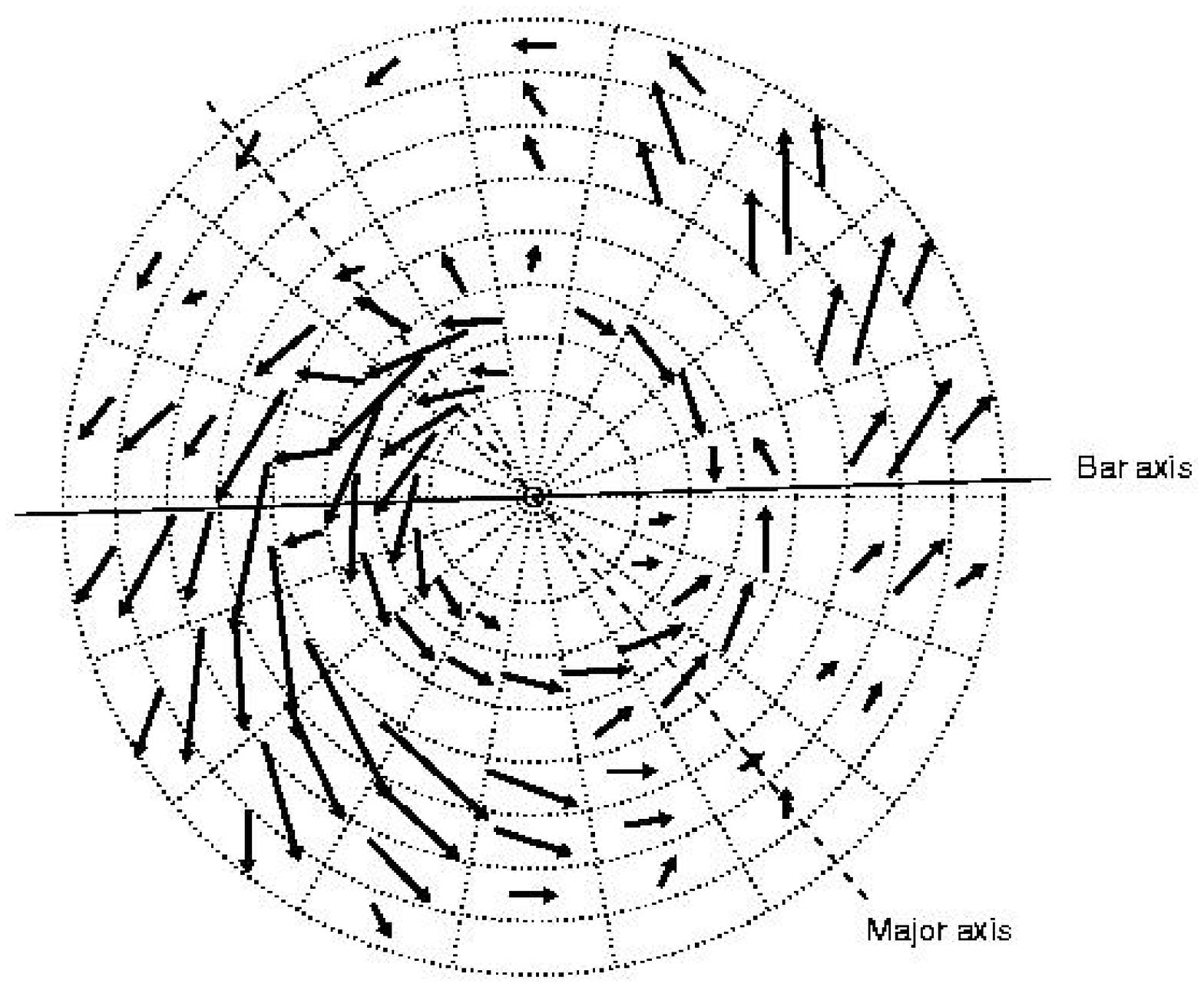}
   \caption{The global structure of the regular magnetic field in
     NGC~1097 (left) and NGC~1365 (right). Both plots are face-on views
     of the modelled galaxy, with the major axis shown by a dashed line
     and the bar axis by a solid line. The
ring boundaries for NGC~1097 are at $r=1.25,\ 2.5,$ and $3.75\kpc$,
covering the inner and
     mid-bar, where the half-length of the bar is $\simeq10$~kpc.
     The NGC~1365 rings are centred at r = 3.5,
     5.25, 7.0, 8.75, 10.5, 12.25 and 14~kpc, with a half-length of the bar
     of $r\simeq 11\kpc$. In both figures, the vector lengths are
proportional to the modelled
     Faraday rotation measure in a sector and are only shown where
     RM $>20\radm$. The vectors in the outer ring of NGC~1097
     are scaled up by a factor $1.5$ compared to those of the inner ring,
the vectors for NGC~1365 all have the same scaling.}
   \label{fig:global}
\end{figure*}
%-------------------------------------------------

The \emph{large-scale structure and direction of the regular
magnetic field\/} in a galaxy can be recovered from
polarization position angles at more than one frequency.
We have applied a method that seeks to find statistically good fits to the
polarization angles of synchrotron emission, of
a superposition of azimuthal magnetic field modes
$\exp(im\phi)$ with integer $m$, where $\phi$ is the azimuthal angle
in the galaxy's plane. We have used
\wwav{3.5}{6.2} polarization angles of NGC~1097 (10\arcsec\ resolution)
and NGC~1365 (15\arcsec\ resolution), averaged in sectors with
opening angles in azimuth of $10\degr$ and $20\degr$,
respectively. The nuclear regions, inside the innermost rings in
Fig.~\ref{fig:global}, suffer from strong Faraday depolarization
(see the right-hand panel of Fig.~\ref{rm}) and hence represent a different problem
for modelling that should be addressed elsewhere.

A three-dimensional model of regular magnetic field is fitted to the observations
of polarization angles at both wavelengths simultaneously.
The polarization angle affected by Faraday rotation is given by
$\psi=\psi_0+\RM\lambda^2$, where the intrinsic angle $\psi_0$ depends on
both the regular magnetic field and the anisotropic random field (both in the
sky plane), whereas the second term is only sensitive to the regular magnetic
field. The vector of the regular magnetic field is specified in terms of
a Fourier expansion in azimuth, and $\psi_0$ and $\RM$ are derived consistently
with each other. Then the coefficients of the Fourier series are fitted
to the observed angles using nonlinear least squares techniques, and
the quality of the fits is verified using statistical criteria.
The method, and its application to data from two spiral
galaxies, is described in more detail in Berkhuijsen et al.\
(\cite{BH97}) and Fletcher et al.\ (\cite{FB04}).
The fitted parameters of the regular magnetic fields in the two
galaxies are given in Appendix~\ref{app:fit}.
The fit parameters given there can be used to reconstruct the global
magnetic structures in the galaxies.
The resulting regular
magnetic field structures of the two barred galaxies are shown in
Fig.~\ref{fig:global}.
We applied this
method to NGC~1097 also in Moss et al.\ (\cite{MS01}), but
there used data
with lower resolution. With our new polarization data we can increase the
number of sectors, and hence the spatial sensitivity of the model, by
a factor of 2.

As discussed in Sect.~\ref{sectMFbar},
a significant part of the polarized emission can be due to a
random anisotropic magnetic field (which does not contribute to Faraday
rotation). Therefore, our fits can be a poor
representation of the observed polarization angles
where the anisotropic random field
is strong and misaligned with the regular magnetic field. The regions where
the anisotropy of the random field is strong are the radio ridges.
The anisotropic and regular fields are affected by compression
in a different way (Sect.~\ref{sectLowPI}).
However, we reasonably expect that the anisotropy produced by compression and
shear will be roughly aligned with the regular magnetic field,
as indicated by comparing Figs.~\ref{fig:global} and \ref{deproj}.

%------------------------------------------
\subsection{NGC~1097}

We were unable to achieve a statistically good fit to the data for the inner ring
($15\mbox{\arcsec}<r<30\mbox{\arcsec}$, $1.25<r<2.5\kpc$) of NGC~1097
using combinations of up to three modes selected from $m=0,1,2,3,4$
for the horizontal magnetic field, combined with either
a uniform or a $2\pi$-periodic vertical magnetic field. The reason for
this is the sharp discontinuity and sign change in the Faraday
rotation measures at the southern end of the bar major axis, shown in
Fig.~\ref{n1097rm} and discussed in Sect.~\ref{sectRM}.
However, a statistically good regular magnetic field model can be readily
obtained by treating separately the two halves of the ring
on either side of the bar
axis. Then each half can be satisfactorily described by the combination of modes
$m=0,2$. To the north-east of
the major axis (azimuth $20\degr$ to $190\degr$ measured from the
northern end of the major axis), the $m=2$ mode is twice as strong as
the $m=0$ mode. In the other half of the galaxy, the two modes have
similar amplitudes.

The motivation for splitting this ring into two halves is
that strong shear in the velocity field in the bar results in an
abrupt change in sign of the radial component of the magnetic field.
This type of sudden change, giving rise to the discontinuity and sign
change in $\RM$, cannot be well described by a superposition of
a small number of azimuthal modes.

The regular magnetic field pattern required to model the
\wwav{3.5}{6.2} polarization angles -- and hence the $\RM$ discontinuity
in Fig.~\ref{n1097rm} -- between $1.25<r<2.5\kpc$
in NGC~1097 is shown in Fig.~\ref{fig:global} (left-hand panel). At the southern
end of the bar, slightly upstream of the bar axis, the radial and azimuthal
components of the coherent regular magnetic field change sign. At the
same location the pitch angle of the field abruptly increases from
$\pa=27\degr$ to $\pa=41\degr$. (Note that the change in $\pa$ is less
than the deflection angle discussed in Sect.~\ref{sectMF} due to
averaging of the observations in sectors; we lose spatial resolution
using the model but gain the \emph{direction\/} of the regular field.)
This sharp change in the regular field may be a sign of strong
shear and compression,
and this possibility is discussed in Sect.~\ref{sectRMShear}.
Three more reversals in the regular magnetic field are also apparent
in this ring. Two of these are smoother transitions from positive to
negative $\RM$ and do not produce $\RM$ discontinuities
(Fig.~\ref{n1097rm}). At the north end of the bar there is a stronger
and sharper reversal that resembles that in the south. However, the
weak regular field in the north-west of the bar (Fig.~\ref{n1097s6})
means the global pattern is less apparent here.

The second ring ($30\mbox{\arcsec}<r<45\mbox{\arcsec}$,
$2.5<r<3.75\kpc$) was also modelled by splitting the data into two halves,
on either side of
the bar axis. In this case though, no reversals in the regular magnetic
field are required (see the left-hand panel of Fig.~\ref{fig:global}).
The variation in $\RM$ shown
in Fig.~\ref{n1097rm} (right-hand panel) is not as dramatic as in the inner
ring and the observed sign reversals of $\RM$ are consistent with
projection effects due to the galaxy's inclination.
Our fits for the whole ring are consistent with those for the two halves
reported here.
The regular field in this ring is nearly azimuthal, especially
in the half of the galaxy south east of the
major axis. In both halves of the galaxy the modelled field consists
of a strong axisymmetric component and a weaker higher azimuthal mode.
Neither the radial nor the azimuthal components of the coherent regular field
change sign in the model.
Although the observed $\RM$ seem to change sign at $\phi\simeq 190\degr$
(right-hand panel of Fig.~\ref{n1097rm}), the error bars are large, so that
the model curve is a statistically good fit.
The change in magnetic field pitch angle at the southern end of the
bar major axis is in the same sense as for the inner ring.
The change in the pitch angle from the purely azimuthal
upstream field ($\pa\simeq 0 \degr$) to the deflected field downstream
of the bar is $\simeq 40\degr$.

For the third ring ($45\mbox{\arcsec}<r<60\mbox{\arcsec}$, $3.75<r<5\kpc$,
we did not split the magnetic field into two halves; there is little
sign of an $\RM$ discontinuity in this ring. Again the regular magnetic field
is nearly azimuthal in the southern half of the galaxy and is
deflected after passing through the dust lane. The fitted regular
field in this ring comprises a strong $m=0$ mode and two
higher modes ($m=1,2$) with amplitudes which are about half as large.

%------------------------------------------------
\subsection{NGC~1365}

Polarized emission in NGC~1365 can be traced to larger radii than
in NGC~1097, so that more rings can be included in our analysis.
The polarization angles in the rings shown in
Fig.~\ref{fig:global} (right-hand panel) were fitted with a combination of $m=0, 1$
and $2$ modes, except for the third ring from the centre where only
modes $m=0,2$ were required.
The fit parameters are given in Table~\ref{tab:fit:ngc1365}.
There is a tendency for
$\pa$ to become larger (a less azimuthal, more radial field) approaching the
bar from the upstream direction, although this is not the case in all
rings. The amount of deflection in the field orientation is weaker
than in NGC~1097.

A reversal is detected in the regular magnetic field direction in the
north-west quadrant (upper right) of the second ring from the centre.
This is not seen in the $\RM$ map (bottom left panel of Fig.~\ref{rm}),
but only becomes apparent
when the data is averaged into sectors and modelled. This reversal is similar
to that seen in the ridge of NGC~1097. Since the weaker signal from NGC~1365
means we had to use only sectors with an opening angle of $20^{\circ}$ compared
to $10^{\circ}$ in NGC~1097, it is possible that further reversals will be revealed
with more sensitive polarization observations.

%--------------------------------
\section{The central regions}
\subsection{NGC~1097}
\label{sectCenter1097}

%%--------------------------------------
%FIGS NGC1097 CENTER
\begin{figure*}
\includegraphics[bb = 62 171 528 609,width=0.475\textwidth,clip=]{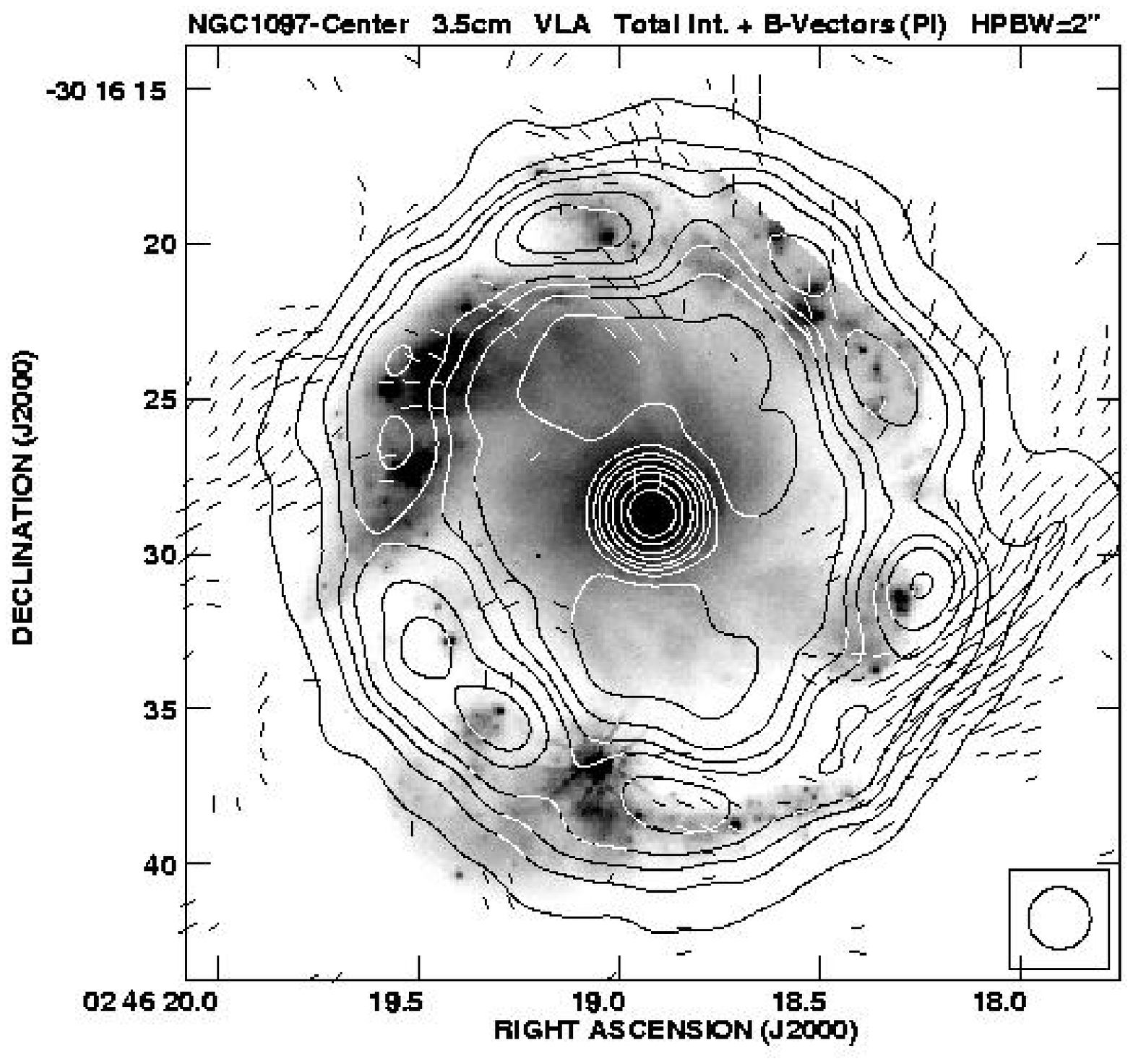}
\hfill
\includegraphics[bb = 62 171 543 609,width=0.475\textwidth,clip=]{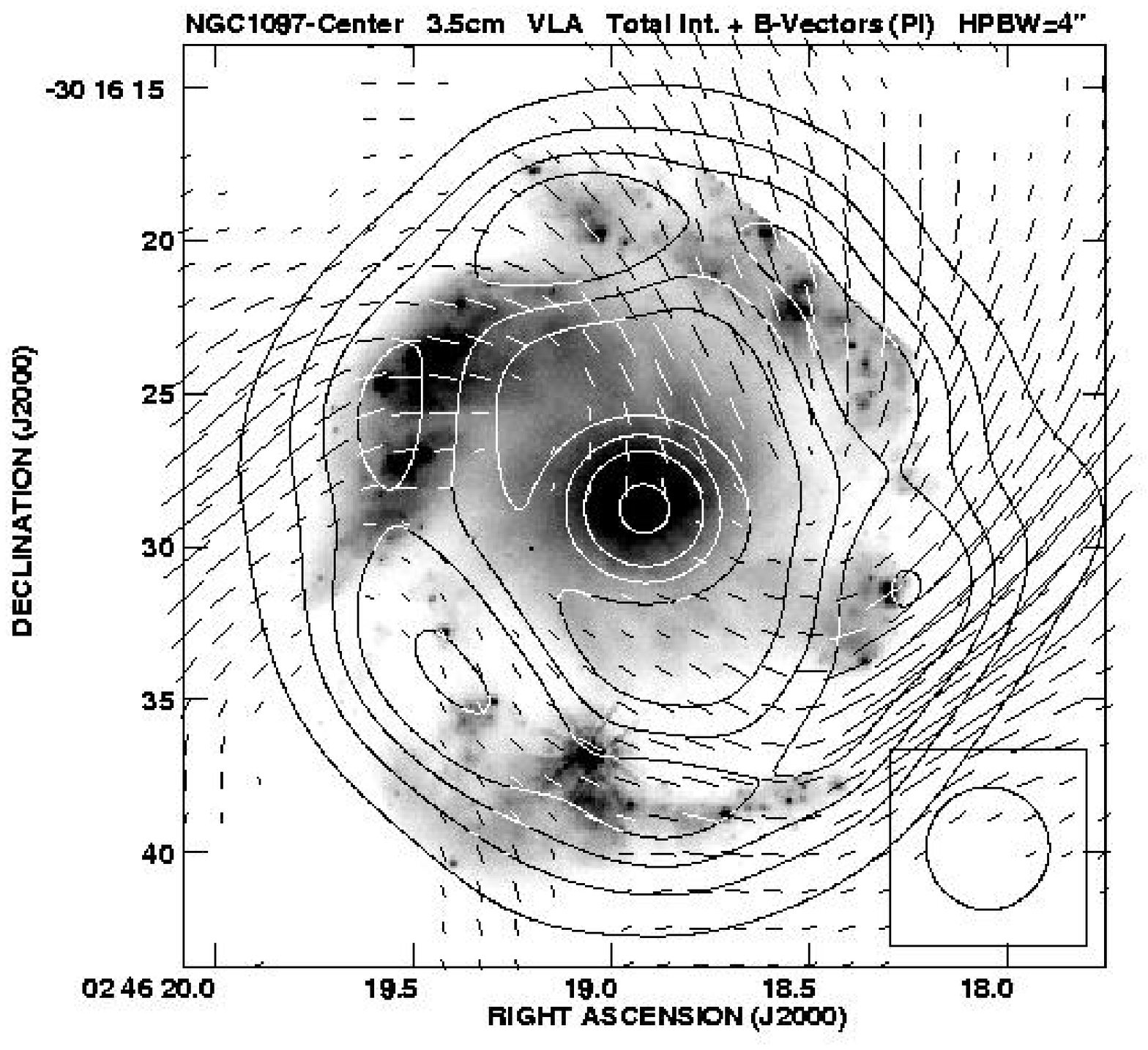}

\vspace{0.2cm}

\includegraphics[bb = 62 171 543 609,width=0.475\textwidth,clip=]{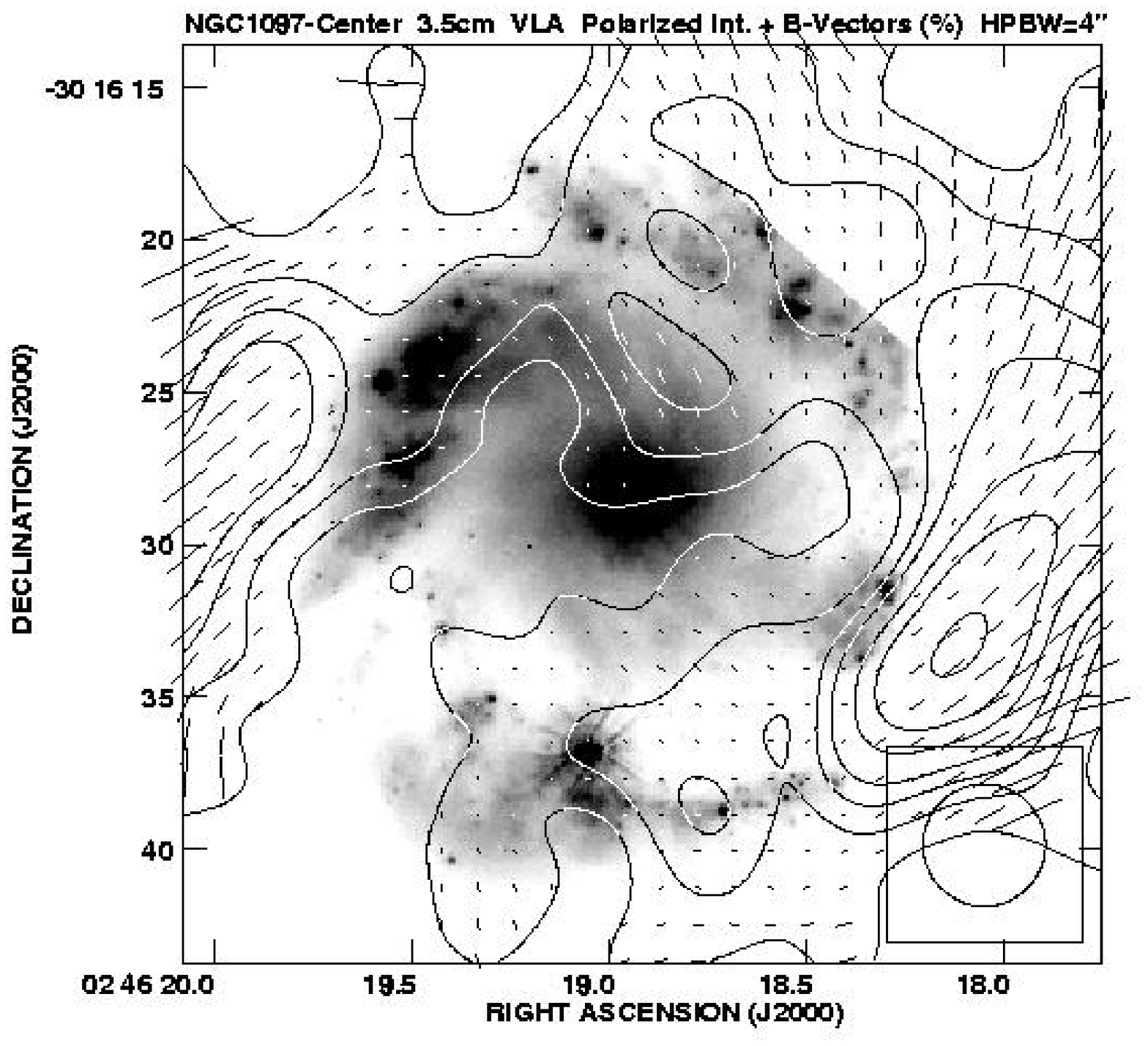}
\hfill
\includegraphics[bb = 62 171 528 609,width=0.475\textwidth,clip=]{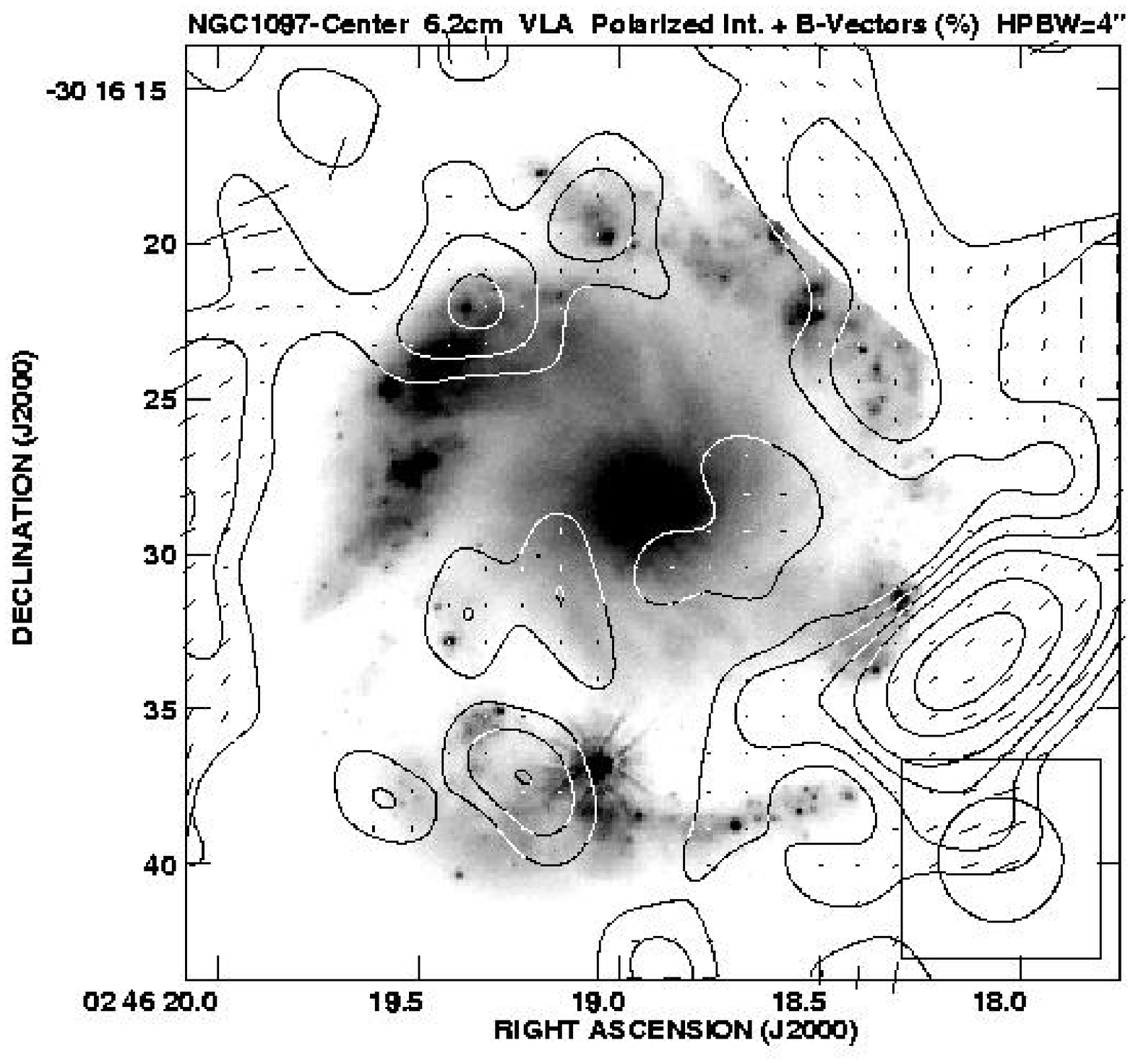}
\caption{{\it Top row:\/} Total intensity contours and observed
$B$-vectors ($E+90^\circ$) in the circumnuclear ring of NGC~1097 at
$\lambda3.5$~cm at 2\arcsec\ (left) and 4\arcsec\ (right) resolutions.
The contour levels are $1, 2, 3, 4, 6, 8, 12, 16, 32, 64, 128$ times
the basic contour level, which is 150 and 600~$\mu$Jy/beam area,
in the left- and right-hand panels, respectively.
The vector length is proportional to polarized intensity,
1\arcsec\ length corresponds to 50~$\mu$Jy/beam area.
The vector orientations are not corrected for Faraday rotation.
The background optical V-band {\it HST\/} image was kindly provided by Aaron Barth.
{\it Bottom row:\/} Polarized intensity contours and
observed $B$-vectors at $\lambda3.5$~cm (left) and
$\lambda6.2$~cm (right) at 4\arcsec\ resolution. The vector length
is proportional to fractional polarization, 1\arcsec\ length
corresponds to 17\%. The contour
levels are $1, 2, 3, 4, 6, 8, 12$ times 20~$\mu$Jy/beam area.
The vector orientations are not corrected for Faraday rotation.}
\label{n1097center}
\end{figure*}
%%--------------------------------------

%%--------------------------------------
%FIGS NGC1365 CENTER
\begin{figure*}[htbp]
\includegraphics[bb = 62 171 528 609,width=0.495\textwidth,clip=]{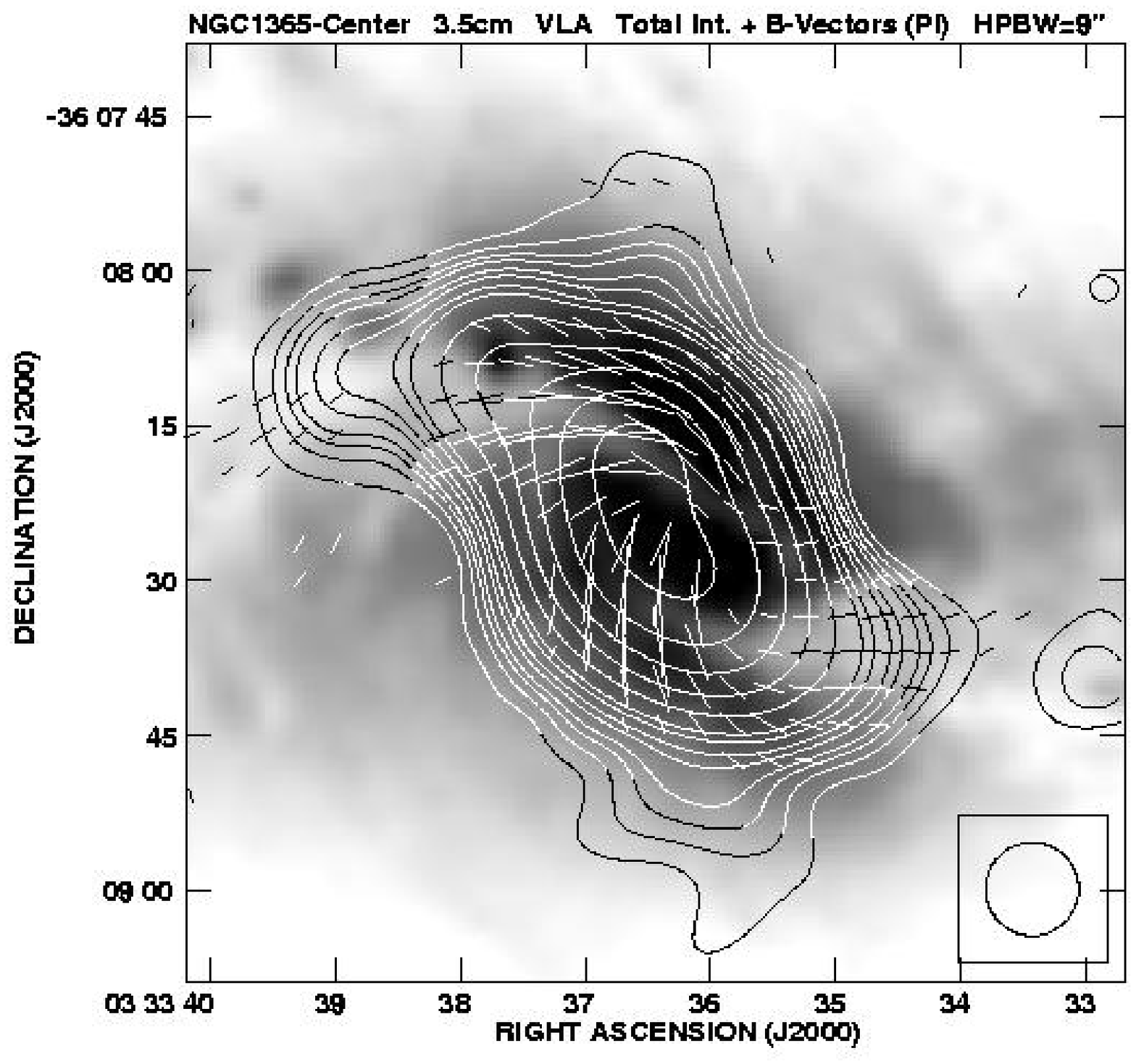}
\hfill
\includegraphics[bb = 62 171 528 609,width=0.495\textwidth,clip=]{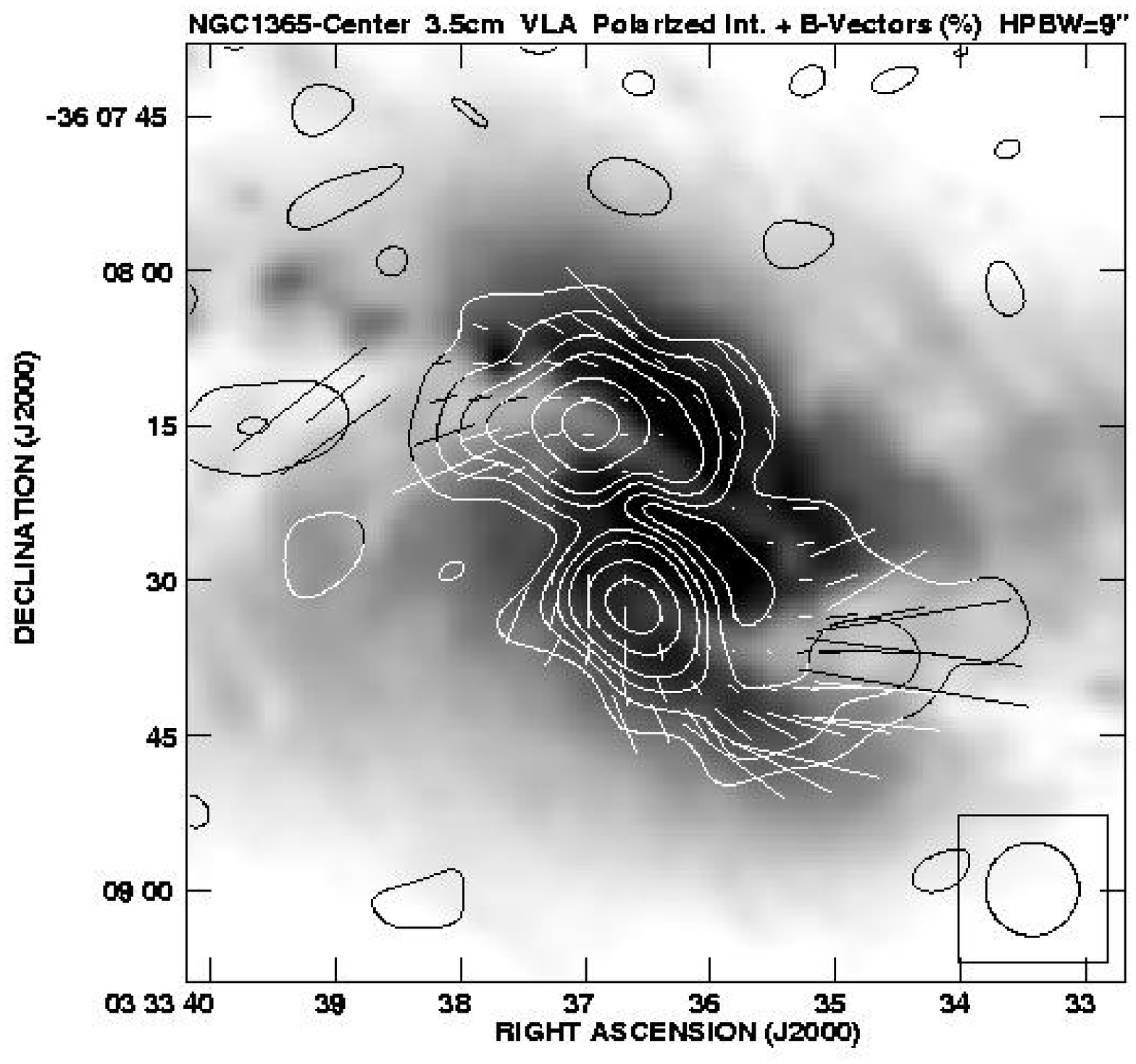}
\caption{{\it Left:\/} Total intensity contours and observed $B$-vectors
($E+90^\circ$) in the central region of NGC~1365 at $\lambda3.5$~cm
at 9\arcsec\ resolution.
The contour intervals are $1, 2, 3, 4, 6, 8, 12, 16, 32, 64, 128, 256, 512$
times 40~$\mu$Jy/beam area.
The vector length is proportional to polarized intensity,
1\arcsec\ length corresponds to 20~$\mu$Jy/beam area.
The vector orientations are not corrected for Faraday rotation.
{\it Right:\/} Polarized intensity intensity contours
and observed $B$-vectors ($E+90\degr$) of NGC~1097 at $\lambda3.5$~cm
at 9\arcsec\ resolution.
The contour intervals are $1, 2, 3, 4, 6, 8$ times 30~$\mu$Jy/beam area.
The vector length is proportional to fractional polarization,
1\arcsec\ length corresponds to 1.7\%. --
The vector orientations are not corrected for Faraday rotation.}
\label{n1365center}
\end{figure*}
%%----------------------------------

Both galaxies studied in this paper host a central starburst region. The
circumnuclear ring of NGC~1097 is a prototype example of mass inflow
and starburst ignition in a bar potential and has been studied in
various spectral ranges (Gerin et al.\ \cite{GN88}, Barth et al.\
\cite{BH95}, Quillen et al.\ \cite{QF95}, P\'erez-Olea \& Colina\ \cite{PC96},
Kohno et al.\ \cite{KI03}).
In high-resolution infrared images, the ring is resolved into a complex
network of filamentary spiral features (Prieto et al.\ \cite{P05}).

The radio continuum ring in NGC~1097 was first studied in detail by Hummel
et al.\ (\cite{HH87}). It appears as an almost perfect ring with an
average radius of
9\farcs0 (740\p) (9\farcs5 in north-south and 8\farcs5 in east-west
directions). Although it seems that the ring lies in the sky plane,
the line-of-sight gas velocity is as large as $230\kms$ at
position angle $-45\degr$ (Gerin et al.\ \cite{GN88}),
indicative of the ring actually lying in the galaxy's plane and being
intrinsically elongated along the line of sight. The radio spectrum of
the ring is nonthermal (see below) so it must contain strong magnetic
fields.  Beck et al.\ (\cite{BE99}) detected polarized emission from this ring.
Our new data have much higher resolution and also allow us to measure
Faraday effects.

The radio map of NGC~1097 at our highest resolution, shown in the top
left panel of Fig.~\ref{n1097center}, exhibits several prominent knots in total
intensity, mostly coincident with optically prominent star-forming regions.
The three brightest knots
are separated by almost precisely $120\degr$ in azimuthal angle (being located at
position angles $15\degr$, $135\degr$ and $255\degr$, measured anticlockwise
from the north).
No similar periodicity is seen in polarized intensity. Elmegreen (\cite{E94})
proposed that the gravitational instability can be
responsible for the fragmentation of inner rings into clouds.

The spectrum is significantly flatter in the
knots of the circumnuclear ring
($\alpha=-0.65\pm0.02$) than between them ($\alpha=-0.72\pm0.02$)
(see Fig.~\ref{n1097spi4}) which can be due to a higher thermal fraction, as
suggested by Hummel et al.\ (\cite{HH87}).
Assuming the same synchrotron spectral index of
$\alpha_\mathrm{s}=-1$ as in the ridges (see Sect.~\ref{sectRidge}),
the thermal fraction $f_\mathrm{th}$ at
$\lambda3.5$~cm is 45\% in the knots and 35\% in between.
If the observed spectral index is smaller than $\alpha_\mathrm{s}$,
the thermal fraction cannot be determined,
and then we assume that it is negligible.
However, the synchrotron spectral index $\alpha_\mathrm{s}$ in the knots could be
larger than $-1$, e.g.\ due to a contribution of radio
emission from individual young supernova remnants which are expected to
have an average spectral index of $\alpha_s \simeq -0.6$. Furthermore, if
bremsstrahlung loss of the cosmic-ray electrons dominates over synchrotron
loss, the synchrotron spectral index is the same as the intrinsic one
($\alpha_0=-0.5$ for strong shocks). In this case the
observed spectral indices are similar to the synchrotron spectral
index so that the thermal fractions are small.
An independent estimate of the thermal emission from the optical
H$\alpha$ line is impossible as the quality of the existing data is
insufficient.

The equipartition strength of the total magnetic field in the knots of
NGC~1097 is about $60\mkG$ (assuming a synchrotron spectral index of
$\alpha_\mathrm{s}=-1.0$
and a path length of 500\p), applying the revised equipartition
formula by Beck \& Krause (\cite{BK05}).
The average total field strength in the central ring is about $55\mkG$.
The field strength is somewhat larger than that given by Beck et al.\
(\cite{BE99})
because here we assumed a steeper synchrotron spectrum and a smaller
path length. The degree of polarization in the knots is low (1--3\%), yielding
a regular + anisotropic random field of $B\reg\simeq10\mkG$ (Table~\ref{Table4}).
Alternatively, for $\alpha_\mathrm{s}=-0.7$ and generally negligible thermal
emission, the total field strength in the knots increases to about $70\mkG$, while
$B\reg$ decreases to 4--$8\mkG$.

At the positions where the radio ridges and dust lanes enter the
central ring of NGC~1097, the degree of polarization is much higher
(note maxima in north-east and south-west in Fig.~\ref{n1097center},
bottom row panels) which yields $B\reg\simeq20\mkG$ (Table~\ref{Table4}).

%%--------------------------------------------

\begin{table*}           % Table 4.
\caption{\label{Table4}Properties of the central parts of
NGC~1097 ($r<1\kpc$) and NGC~1365 ($r<2\kpc$) from the $\lambda3.5$~cm data.
The locations of the regions are described in the text.
The thermal fractions were determined with the assumption of a
synchrotron spectral index of $\alpha_\mathrm{s}=-1.0$.}
\centering
\begin{tabular}{lcccc}
\hline\hline
\                    & Thermal  & Synchrotron & $B\tot$ & $B\reg$ \\\
                      & fraction & polarization [\%]   & [$\mu$G]  & [$\mu$G]
\\\hline
{\bf NGC~1097}\\
Bright knots         & 0.45     & 2           & 60 & 10  \\
Tangential point (W) & 0.35     & 15          & 56 & 21  \\
Tangential point (E) & $<0.05$  & 37          & 34 & 21  \\
Inside the ring (N)  & 0.45     & 17          & 40 & 15  \\
\hline
{\bf NGC~1365}\\
Inner dust lane (S)  & 0.35     & 0.5         & 63 & 6  \\
Outer dust lane (SW) & $<0.05$  & 4           & 38 & 9  \\
Outer dust lane (E)  & $<0.05$  & 22          & 21 & 12  \\
\hline
\end{tabular}
\end{table*}
%---------------------------------------------

The regular (or anisotropic random) magnetic field in the central region
of NGC~1097 is of a spiral
shape and extends well inside the ring (see the top right panel of
Fig.~\ref{n1097center}). The equipartition strength of the
magnetic field obtained from polarized emission, $B\reg$, is about $15\mkG$
inside of the ring (north of the centre) (Table~\ref{Table4}).

Faraday rotation measures are generally small in the ring of NGC~1097
(top left panel of Fig.~\ref{rm}).
The \RM\ jump in the southern ridge (see Fig.~\ref{n1097rm})
continues into the ring. With such small \RM\ values, it is impossible to
separate the coherent and anisotropic random field components
as attempted in the southern bar region (Sect.~\ref{sectMFbar}).
We suspect that, in contrast to the ridges, most of the polarized
emission from the ring originates in a coherent magnetic
field, as the conditions for
dynamo action are perfect: high star-formation rate and strong
differential rotation velocity of order
$330\kms$ (Gerin et al.\ \cite{GN88}, Kohno et al.\ \cite{KI03}).
Note that the average pitch angle of the spiral field, corrected for
Faraday rotation, is about $39\degr\pm3\degr$, which is larger than in
the discs of typical spiral
galaxies. This fact also supports strong dynamo action. However,
small \RM\ and
strong Faraday depolarization (see below) in our present data prevent
a search for global modes in the structure of the coherent regular
field.
New observations with higher sensitivity at shorter wavelengths (where
Faraday depolarization is weaker) are required.

Prieto et al.\ (\cite{P05}) interpret the spiral magnetic field
in the central region of NGC~1097 as indication of a flow of warm gas
far away from the galaxy's plane, crossing the circumnuclear ring.
Detailed measurements of Faraday rotation are necessary to clarify
the geometry of the magnetic field structure.

Faraday depolarization between $\lambda3.5$~cm and $\lambda6.2$~cm is
strong in the ring of NGC~1097 (Fig.~\ref{rm}, and compare
Fig.~\ref{n1097center}, bottom row panels), evidently
a result of strong turbulent fields and high thermal electron
density. Only the western tangential point has the same polarized
intensity at both wavelengths. The ridge emerging on the western
(near) side lies in front of the ring where it cannot be depolarized
by the medium in the ring. The hot thermal gas of the halo detected in
X-rays is too thin to cause significant
Faraday depolarization at $\lambda6.2$~cm (Sect.~\ref{sectHalo}).

The nucleus of NGC~1097 has a flat spectrum with a spectral index of
$\alpha\simeq-0.05$, typical for active nuclei. An investigation of its
properties is beyond the scope of this paper.

\subsection{NGC~1365}
\label{sectCenter1365}

NGC~1365 has a central starburst region which is even brighter in radio
continuum than the central ring in NGC~1097, and hence the total
equipartition fields are even stronger. The peak value of about $63\mkG$
in the inner part of the massive dust lane located south of the centre
is one of the largest field strengths found in any normal or barred spiral
galaxy so far.
Stronger fields were observed only in the circumnuclear rings
of the southern barred galaxies NGC~1672 and NGC~7552 (Beck et al.\
\cite{BE05}). The degree of polarization
in the inner dust lane (Table~\ref{Table4})
is lower than that
in the ring of NGC~1097 (Table~\ref{Table3}). In the outer part
of the same dust lane, south-west of the centre, and in the outer part
of the eastern dust lane, the degrees of polarization increase
significantly.  The pitch angle of the regular field in the ring
(radial range 7\farcs5--22\farcs5, or 0.7--2.0~kpc) varies strongly between
0 and $-70^\circ$.

\RM\ shows a jump in the central region of NGC~1365, south of the
nucleus (bottom left panel of Fig.~\ref{rm}). A similarity to the field
reversal in the ridges of NGC~1097 and NGC~1365 is possible, but needs
investigation with future data at higher resolution.

%--------------------------------------------------------------------------
\section{The X-ray halo of NGC~1097}
\label{sectHalo}

In order to determine the strength of regular fields from Faraday rotation
measures, independent information is needed about electron density
(Sect.~\ref{sectMFbar}), e.g. from thermal radio or H$\alpha$
emission which is significant from gas at a temperature at a few 1000~K.
Hot gas, as observed with X-rays, may also contribute to Faraday effects.

\begin{figure}
\centerline{\includegraphics[bb = 76 256 478 666,width=0.495\textwidth,clip=]{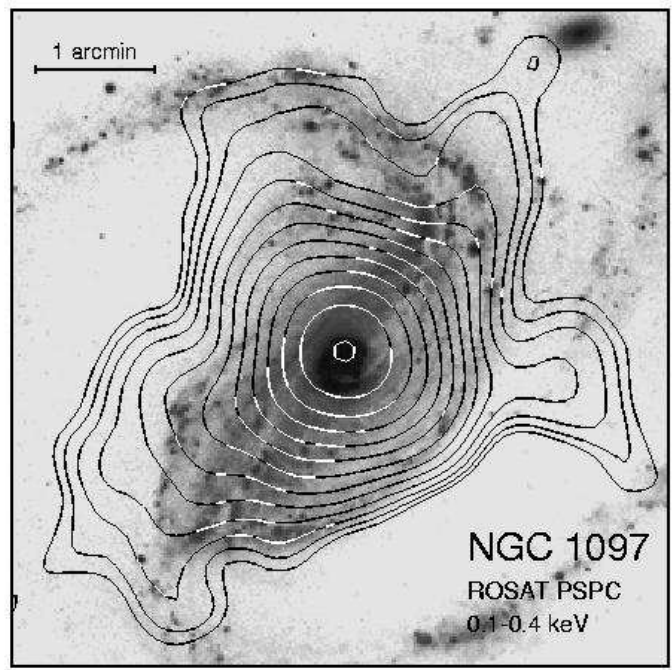}}
   \caption{Soft-band (0.1--0.4~keV) X-ray emission from NGC~1097,
     observed with the {\it ROSAT\/} PSPC detector. Contour levels are
     $2^{n/2} \times 3 \cdot 10^{-4}$ counts s$^{-1}$ arcmin$^{-2}$
     ($n=1,2,...$) above the background of $12 \cdot 10^{-4}$ counts
     s$^{-1}$ arcmin$^{-2}$.  The background optical image was kindly
     provided by Halton Arp.}
   \label{n1097x}
\end{figure}

The soft X-ray emission from NGC~1097 observed with {\it ROSAT\/}
(Fig.~\ref{n1097x}) peaks at the
galactic nucleus, which could be expected as NGC~1097 is known to
harbour an
active Seyfert nucleus. In addition to the nuclear emission (which is
smoothed out to a galactocentric radius of about 1\arcmin,
corresponding to
the point spread function of the PSPC at low energies), the soft X-ray
emission has contours elongated along the spiral arms and the
bar, and was even
detected outside the bar. This indicates the presence of extended hot gas
in a disc and/or in a halo.

The soft X-ray spur in the north-west points towards the companion
galaxy
NGC~1097A (Ondrechen et al.\ \cite{OH89}) and might be a signature of
ongoing tidal interaction between the two galaxies.

Our successful spectral fit to the dominating X-ray emission
(background-corrected) from the central region of NGC~1097
within a galactocentric radius of 1\arcmin\
used a combined emission model including a power law
component (to describe the nucleus and unresolved point-like sources)
and a Raymond--Smith plasma with `cosmic' metal abundances (Raymond \&
Smith\ \cite{RS77}), both affected by Galactic foreground absorption
(with hydrogen column density of $1.87\times10^{20}$~cm$^{-2}$). With
the fitted temperature of $9.6\times10^{6}\K$ and photon index $s$ of 1.3
(where the X-ray flux density is proportional to $E^{-\mathrm{s}}$),
the luminosity of the inner part of
NGC~1097 in the total ROSAT energy band (0.1--2.4~keV) was found to
be $11.8\times10^{40}\erg\s^{-1}$, with about
84\% of it originating from the power law component.

%% ------------------------------------------------
% FIG NGC1097 ROSAT
%% ------------------------------------------------

Unfortunately, the X-ray count rate statistics of the extended component
outside a radius of
1\arcmin\ (where we expect that the emission is no longer affected by
the Seyfert nucleus) were too poor to allow a separate fit of this
emission
component. Higher sensitivity X-ray observations of NGC~1097, e.g.,
with XMM-Newton,
are needed to further constrain this unresolved
emission. Assuming a Raymond--Smith plasma with $T=2.3\times10^6\K$
(which seems to be typical of the X-ray emitting hot gas in
spiral galaxies) and correcting for the background and the
Galactic foreground absorption, the count rate of
$2.86\times10^{-2}\,\mbox{counts}\, \s^{-1}$ in the outer area
corresponds to an X-ray luminosity of $1.65\times10^{40}\erg\s^{-1}$ in
the energy range 0.1--2.4~keV.

The plasma density $n_{\rm e}$ of the extended component
can be calculated under the assumption of radiative cooling
and ionization equilibrium (Nulsen et al.\ \cite{NS84}), where the
luminosity in soft X-rays is given by
$L_\mathrm{X}=1.11\times\Lambda(T)\,\ne^2\, V\, \eta$, with $\Lambda$
the cooling rate,
$V$ the emitting volume and $\eta$ the filling factor of the emitting
gas. For our assumed gas
temperature, Raymond et al.\ (\cite{RC76}) give a cooling coefficient
$\Lambda(T)$ of about $10^{-22}\erg\cm^3\s^{-1}$.
To address the possibility that the emitting plasma is located in the
disc of NGC~1097 we assumed a cylindrical volume (inner radius
1\arcmin, outer radius 4\arcmin\ and a disc path length of 1~kpc) to
obtain an electron volume density of $\ne \simeq 3.4 \times 10^{-3}\,
\eta^{-1/2} \cm^{-3}$.  If the thermal energy density of the
hot gas is equal to that of the magnetic field,
as is the case in other diffuse phases of the ISM,
the total field strength should be $5.2\mkG \times \eta^{-1/4}$.
The total field strength in the interarm space between the bar and the northern
spiral arm is estimated from the total radio emission as about $13\mkG$
(assuming equipartition with the cosmic rays and an average path
length of 1~kpc). Thus, either the filling factor of the hot gas is as
small as 3\%, or the magnetic field energy density is larger
than the thermal energy
density, or the magnetic field obtained from radio emission is overestimated.

Note, however, the asymmetry of the extended soft X-ray emission, which is
stronger and
falls off more slowly on the north-eastern (far) side of the galaxy than on the
south-western (near) side (Fig.~\ref{n1097x}). This may indicate that a large
fraction of the extended X-ray emission emerges from a quasi-spherical halo of hot gas
in front of an absorbing (cool) gas disc. Assuming in this case a halo radius
of 10~kpc (and excluding again the inner radius out to 1\arcmin ), we obtain an
electron volume density of $\ne \simeq 2.3\times 10^{-3}\, \eta^{-1/2} \cm^{-3}$.
A total magnetic field of $5\mkG \times \eta^{-1/4}$ would be
in equipartition with thermal energy.
If the diffuse radio emission around NGC~1097 emerges from a halo,
the total field strength is about $7\mkG$ (Sect.~\ref{sectMFArms}),
which agrees well with the above estimate from the soft X-ray emission
for a filling factor $\eta\simeq0.3$.

As the degree of polarization is about 25\% (see Sect.~\ref{sectMFArms}),
about half of the total field in the halo (or disc) is
coherent (if the anisotropic random field is negligible).
The expected Faraday rotation for a coherence length of 1~kpc is about
$5\times\eta^{-3/4}\radm$, which is much too small to explain the
observed Faraday rotation. We conclude that the hot halo gas around NGC~1097
does not contribute to Faraday effects.

% ---------------------------------------------
\section{Models of gas flow and magnetic field in the bar region}
\label{sectModels}

In this and the following sections, we interpret the radio maps in terms of
magnetic field models and discuss their compatibility with gas-dynamical
models of barred galaxies.

The most important assumption of our interpretation is the existence
of a shock. Note that shear shocks in bars behave differently from classical
shocks, and the compression region may extend deeply into the
downstream region (Syer \& Narayan\ \cite{SN93}). No 3-D models of
shear shocks including magnetic fields are available yet.

The best observational tool to identify the location of the gas shock
is the velocity field of the cold, dense gas. In NGC~1097 \HI\ line
emission from the bar is very weak so that the velocity field is known
only in the outer spiral arms (Ondrechen et al.\ \cite{OHH89}). CO line
emission from the bar is also weak, and the resolution is
insufficient to resolve the shock front (Crosthwaite\ \cite{C01}).
Velocity fields in the bar region were published for a few other galaxies.
Steep velocity gradients occur across the dust lanes in NGC~1365
(J\"ors\"ater \& van Moorsel\ \cite{JM95}, Lindblad et al.\ \cite{LL96})
and in NGC~1530 (Reynaud \& Downes\ \cite{RD98}) which indicate that the
deflection and compression regions indeed coincide.
Sensitive observations of the velocity field of NGC~1097 and NGC~1365
in the \HI\ and CO lines are required to localize the shock fronts and
to measure their compression ratio and their rate of deflection in the flow
direction. Hence there is sufficient evidence for shocks in
both galaxies of our study.

Perhaps the most notable features of gas flow
in barred galaxies are dust lanes extended along the leading edge of the bar,
which are identified with large-scale shocks in the interstellar gas
(Athanassoula\ \cite{A92b}). Strong gas compression and velocity shear
are expected to
occur in the dust lanes. Therefore, it is not surprising that they host
enhanced magnetic fields and appear as bright radio ridges in the radio
maps discussed in Sect.~\ref{sectRidge}.

Our observations have revealed several unexpected features of both
regular and random magnetic fields which are similar in both
galaxies and may be characteristic of barred
galaxies in general. The regular magnetic field does not seem to be
perfectly aligned with gas streamlines in the reference frame
corotating with the bar.  In both galaxies,
the regular magnetic field upstream of the bar is approximately
perpendicular to the dust lanes. The field then
apparently begins to turn and becomes parallel to the dust lanes at
their position (see Sect.~\ref{sectMF}, and Figs.~\ref{n1097s10}
and \ref{n1365s25}).
Hydrodynamic models of the gas flow in generic barred
galaxies (e.g.\ Athanassoula\ \cite{A92b}) have streamlines that behave
quite differently, with an acute change in the direction of the flow at the
leading side of the bar.
Two possible explanations for this effect are discussed in
Sect.~\ref{sectModel1} and \ref{sectModel2}.

In Sect.~\ref{sectCompr} we show that the
enhancement in total radio emission observed from the radio ridges is
compatible with compression and shearing of a random magnetic field in the shock
at the dust lanes.
It is striking, however, that the observed enhancement in polarized intensity
indicates that the regular magnetic field avoids any
significant amplification by
compression and shear in the shock. This behaviour is surprising indeed
as the effects of compression and shearing are clearly seen in total radio
emission, as well as in the gas distribution and kinematics. In
Sect.~\ref{sectLowPI} we
attribute this peculiar behaviour of the regular magnetic field to
the effects of conversion of atomic gas to molecular form in the shock
(i.e., the dust lane).

%------------------------------------------------
\subsection{The effect of inclination and beam smoothing}
\label{sectModel1}

%-----------------------------------------------------------------------
\begin{figure}
\includegraphics[bb = 91 349 456 730,width=0.48\textwidth,clip=]{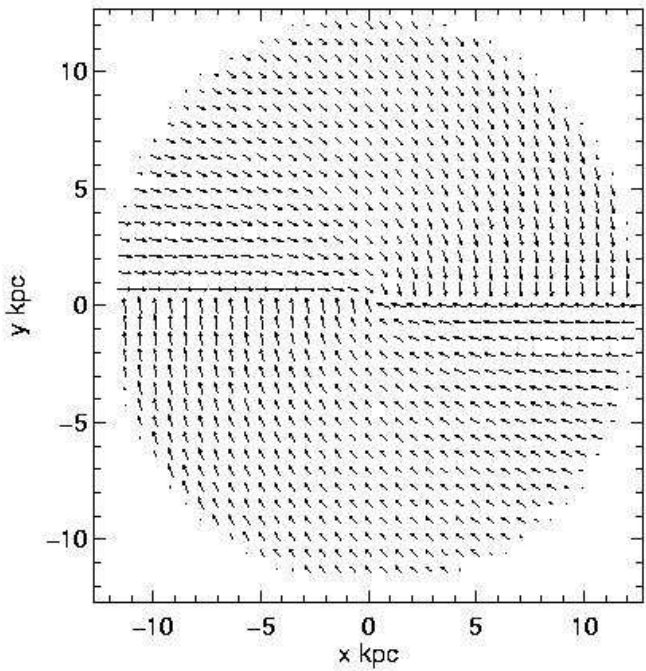}
\caption{The configuration of the magnetic field used to explore the
effects of geometry and beam smoothing.
The bar is located at $y=0$.}
\label{fig:model:field}
\end{figure}
%-------------------------------------------------------------------------
\begin{figure}
\includegraphics[bb = 104 349 456 723,width=0.48\textwidth,clip=]{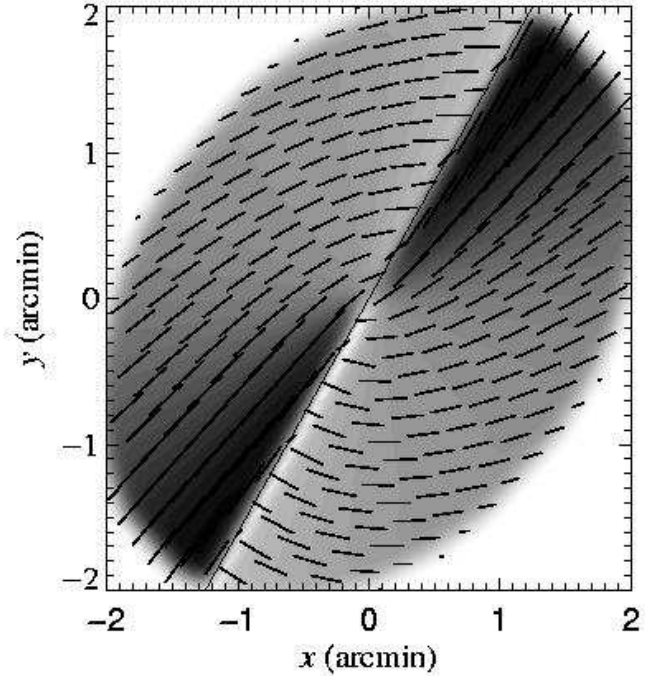}
\caption{Synthetic map obtained by rotating the field shown in
Fig.~\ref{fig:model:field}
with the vertical dependency given by Eq.~(\ref{scaleh}) to the
orientation of NGC~1097. The cosmic ray electron
distribution is assumed to be uniform and the Stokes parameters
$Q$ and $U$ are integrated along all lines of sight,
prior to smoothing with a Gaussian beam of FWHM $10''$.
The grey scale shows polarized intensity in arbitrary units
(with darker shades corresponding to larger values), and the vectors
represent the orientation of the polarization plane.
The continuous solid line shows the bar major axis.}
\label{fig:model:1097}
\end{figure}
%--------------------------------------------------------------------
\begin{figure}
\includegraphics[bb = 104 349 450 723,width=0.48\textwidth,clip=]{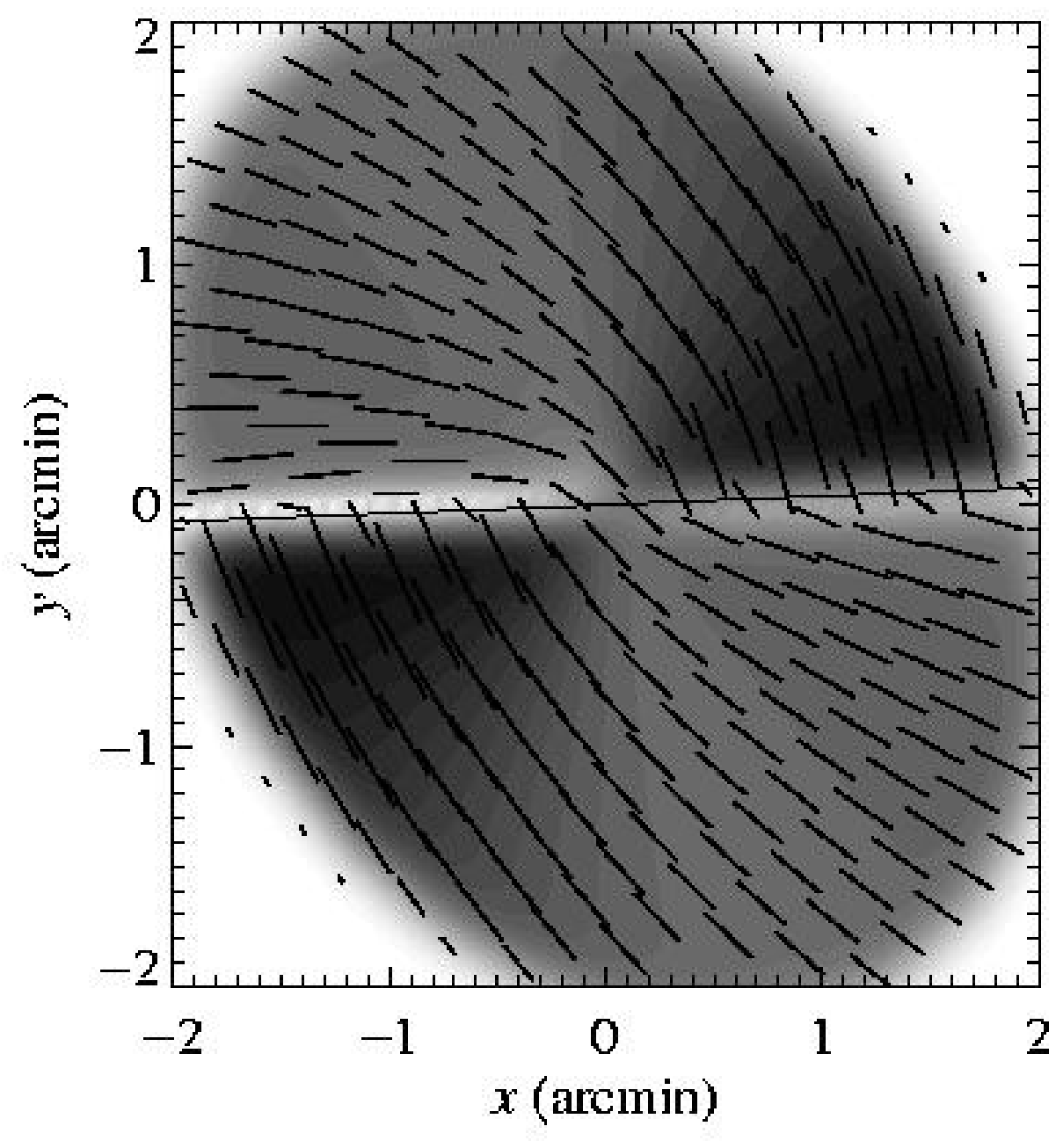}
\caption{Same as Fig.~\ref{fig:model:1097}, but for the inclination and
major axis
orientation of NGC~1365 and a Gaussian beam of FWHM $15''$.}
\label{fig:model:1365}
\end{figure}
%-----------------------------------------------------------------

The apparent offset of the depolarized strip from the dust lane and the early
onset of the magnetic field deflection could be a geometrical effect due to
the inclination of the galaxy's disc and the integration of the
emission along the line of sight through the disc. Here we demonstrate the
potential consequences of these
effects using synthetic maps produced from a simple model of the galaxy
in which the magnetic field sharply changes its direction
by 90\degr\ in the bar.

We prescribe a magnetic field in the plane of the model galaxy
in cylindrical coordinates $(r,\phi,z)$, with $\phi$ measured
in the galaxy's plane from the northern and western end
of the bar major axis to model
NGC~1097 and NGC~1365, respectively,
and the positive $z$-direction being
towards the observer. The mid-plane
horizontal magnetic field
$(\mean{B}_{r},\mean{B}_{\phi})$ is shown in
Fig.~\ref{fig:model:field} and is defined as:
\begin{displaymath}
\mean{B}_{r}=\left\{
\begin{array}{ll}
- B_{0}(z)\phi/\pi,    &  0 < \phi \leq \pi\;,    \\
B_{0}(z)(1 - \phi/\pi), &  \pi < \phi \leq 2\pi\;,
\end{array}
\right.
\end{displaymath}
\begin{displaymath}
\mean{B}_{\phi} = \left\{
\begin{array}{ll}
B_{0}(z)(\phi/\pi - 1),  &  0 < \phi \leq \pi\;,\\
B_{0}(z)(\phi/\pi - 2),  &  \pi < \phi \leq 2\pi\;,
\end{array}
\right.
\end{displaymath}
\begin{displaymath}
\mean{B}_{z}=0\;.
\end{displaymath}
The dependence of the modelled field on $z$ is described by
\begin{equation}\label{scaleh}
B_{0}(z)\propto \exp(-|z|/h_{B})
\end{equation}
with the assumed scale height $h_{B}=4\kpc.$

A $z$-dependent distribution of the form (\ref{scaleh}) is adopted
for thermal electron number density
$\ne$, with the $h_{B}$ replaced with $h_\mathrm{e} = 1\kpc$
and the radial distribution truncated
at $r = 12\kpc$. The number density of cosmic rays is
assumed to be uniform and the synchrotron spectral index is assumed to
be $\alpha_\mathrm{s}=-1$.
After rotating the major axis position angle and inclining the galaxy
to the line of sight by the same angles as in the galaxies NGC~1097 and
NGC~1365, we calculate the Stokes parameters $Q$ and $U$
by integrating emission along the line of sight
over the intervals $\pm 2h_{B}$ centred at the galaxy's midplane.
We convolve the result with
a Gaussian of FWHM $10''$ and $15''$ in the plane of the sky for
NGC~1097 and NGC~1365, respectively.
Depolarization occurs due to differential Faraday rotation along the
line of sight and due to beam smearing.
Figures \ref{fig:model:1097} and \ref{fig:model:1365} show the obtained
synthetic polarized intensity in grey scale with apparent polarization
$B$-vectors superimposed. Not surprisingly, the abrupt turn of magnetic
field by $90^\circ$ results in strips of small polarized intensity,
similar to the depolarization valleys discussed in (Sect.~\ref{sectMF}).

The combination of inclination and smoothing to the beam resolution leads to
valleys in \PI\  parallel to the bar's
major axis which are offset by $15\arcsec$ from the
ridge of maximum \PI\ , consistent with
the observations (Sect.~\ref{sectMF}). The width of the valleys of one
Gaussian FWHM is also consistent with the observations. Because of the
inclination of the galaxies, the two valleys in each model are not equally deep,
but the difference is unimportant compared with the intrinsic asymmetries
found in the observations. The $B$-vectors turn by about 90\degr in front
of the bar, much more sharply than in the observations, especially in NGC~1365.
This implies that the turn of magnetic field near the depolarization
valley in real galaxies is smoother than that in this model, and so it is
{\em partially\/} resolved in our observations.

We conclude that
the effects of disc inclination and beam smoothing
can explain the observed offset of the ``depolarization valleys''
from the ridges, but the observed turning
of the $B$-vectors is much smoother than that
predicted from our simple model.

%---------------------------------------------------------
\subsection{Misalignment between the velocity field and the magnetic field}
\label{sectModel2}

As demonstrated above, the smooth deflection of the regular
magnetic field cannot be the effect of an intrinsically sharp field reversal,
observed with limited resolution in an inclined galaxy,
and thus must have a physical reason.

The velocity field of the dense gas in the bar sharply changes its direction
in the bar which leads to a shock (Fig.~\ref{fig:vr}).
The smooth turning of the observed $B$-vectors signifies a
\emph{misalignment between the velocity field and the magnetic field orientation\/}.
This may imply that the magnetic field diffusivity is significant.
The magnetic field may also decouple from the
flow of molecular gas, as discussed in Sect.~\ref{sectLowPI}.

The dynamo model of Moss et al.\ (\cite{MS01})
predicts significant misalignment (by 20--$45^\circ$) in the bar region
(see their Fig.~7).
A similar conclusion follows from a dynamo model for NGC~1365 (Moss et al.\
2005).
The misalignment is a strong indication that the
regular magnetic field is not frozen into the flow but rather subject
to significant diffusion. Then the magnetic field can persist on
a $10^9\yr$ timescale only if
supported by dynamo action (cf.\ Moss et al.\ \cite{MS01}).

%----------------------------------------
\section{Magnetic field compression and shearing in the radio ridge of NGC~1097}
\label{MFCS}

Here we aim to interpret the observed change in the total
($I$) and polarized (\PI) radio intensities at the radio ridges
described in Sect.~\ref{sectRidge}.
In Table~\ref{Table3} we calculated that the contrast
in $I$ is $\epsilon_I\simeq 5$--10 and in \PI,
$\epsilon_P\simeq1$--$7$,
where the ranges reflect the variation with decreasing distance from the
galactic centre. First, we estimate how shock compression and shearing
of the total magnetic field (dominated by the turbulent part) is likely
to increase $I$.  Then we consider how compression and shear act to change
the regular field and hence \PI.
We shall also estimate the contribution to \PI\ of the anisotropic random field
produced from an isotropic one by compression and shearing.

%% figure
\begin{figure}[htbp]
\begin{center}
\includegraphics[width=0.485\textwidth]{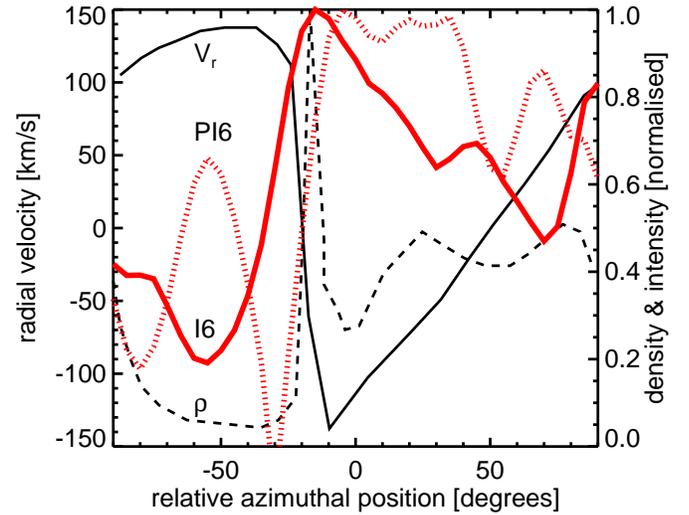}
\caption{A cut perpendicular to the bar shock front showing radial velocity
(labelled $V_r$; thin, black solid), gas density ($\rho$; thin, black dashed),
total (I6; thick, grey/red solid) and polarized (PI6; thick, grey/red dotted) radio
intensities at \wav{6}. The radial velocity and density of the gas
are from Fig.~10 of Athanassoula (\cite{A92b}) and are derived from a generic model
of a barred galaxy. The azimuthal flow is directed from left to right.
The radio data are for the southern end of the bar
in NGC~1097 and are average values for a ring at 30\arcsec\ (2.5~kpc) radius and
15\arcsec\ width. The two data sets (model and observations) have been aligned
using the peak in gas density and the peak (ridge) in total radio intensity,
defining the zero point of the $x$-axis. Note that this zero point
is different from that used in Fig.~\ref{n1097rm}.}
\label{fig:vr}
\end{center}
\end{figure}

We treat the region of the dust lanes along the bar's leading edge as a
shock front where, in addition to compression, velocity shear results
from strong flow towards the galaxy's centre behind the shock
(Fig.~\ref{shear}).
We note that the offset of  the dust lanes towards the leading edge of
the bar is a sign of a strong
shear shock in the velocity field (Athanassoula\ \cite{A92b}).
Thus, the magnetic field is compressed and sheared at the shock front.
Each effect not only amplifies
the magnetic field, but also makes its turbulent part anisotropic.

Figure~\ref{fig:vr} shows a perpendicular slice through the bar region using
radial velocity and gas density data from the model of Athanassoula
(\cite{A92b}) and our radio data from NGC~1097.

We have aligned the two datasets by the peaks in gas density and total
radio intensity. Figure~\ref{fig:vr} illustrates the general interpretation of
the radio ridges that is developed below; compression and shear produce
stronger total and polarized intensities. The depolarized valley
(at $-30^\circ$ relative azimuthal position in Fig.~\ref{fig:vr}) and
the strong upstream polarized intensity (at $-50^\circ$) are also clearly visible.

We assume that the gas speed normal to the shock immediately behind it is
comparable to the sound speed (Roberts et al.\ \cite{RHA79}, Englmaier \&
Gerhard\ \cite{EG97})
and we adopt a sound speed of $\cs=10\kms$; if the velocity normal
to the shock is subsonic (e.g.\ Roberts et al.~\cite{RHA79}), the arguments
presented below are either strengthened or unaffected.
In order to simplify our estimates, we assume
that the shock front is in the vertical plane passing through the
galactic rotation axis (thus neglecting the offset of the front from
the bar's major axis -- e.g., Athanassoula\ \cite{A92b}), so that
magnetic field components normal and tangent to it are $B_\mathrm{n}=B_\phi$ and
$\vec{B}_\mathrm{t}=(B_r,B_z)$, where $(r,\phi,z)$ are cylindrical polar
coordinates in the galaxy's frame, centred at the galaxy's centre.

%---------------------------------------------
\subsection{Contrast in total radio intensity}
\label{sectCompr}

%% figure
\begin{figure*}[htbp]
\centering
\subfigure[\mbox{Synchrotron intensity contrast:} \mbox{constant cosmic
ray density}]
{\label{fig:ei:crconst}
	\includegraphics[width=0.32\textwidth]{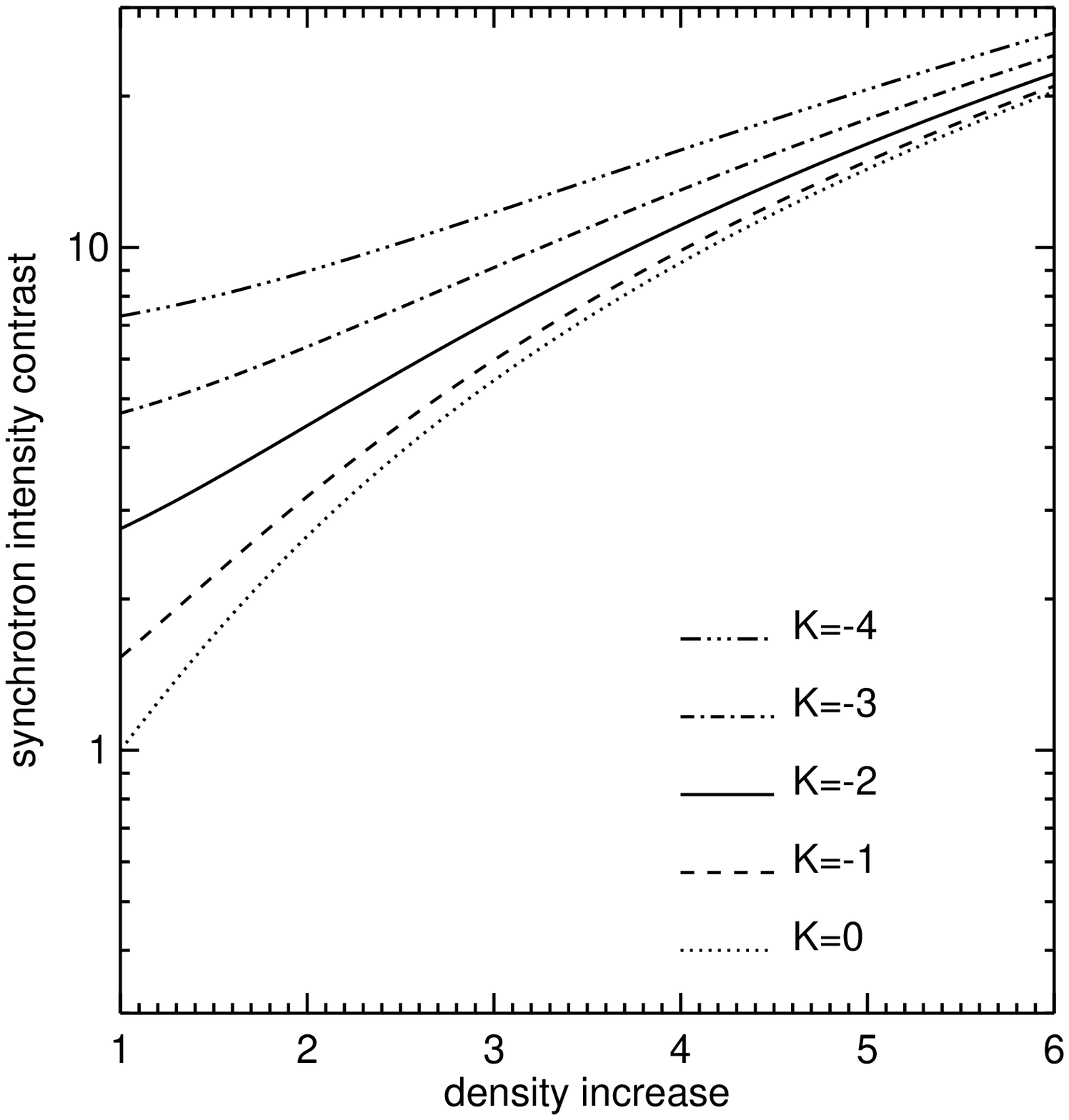}}
\subfigure[\mbox{Synchrotron intensity contrast:} \mbox{equipartition
cosmic ray density}]
{\label{fig:ei:crequi}
	\includegraphics[width=0.32\textwidth]{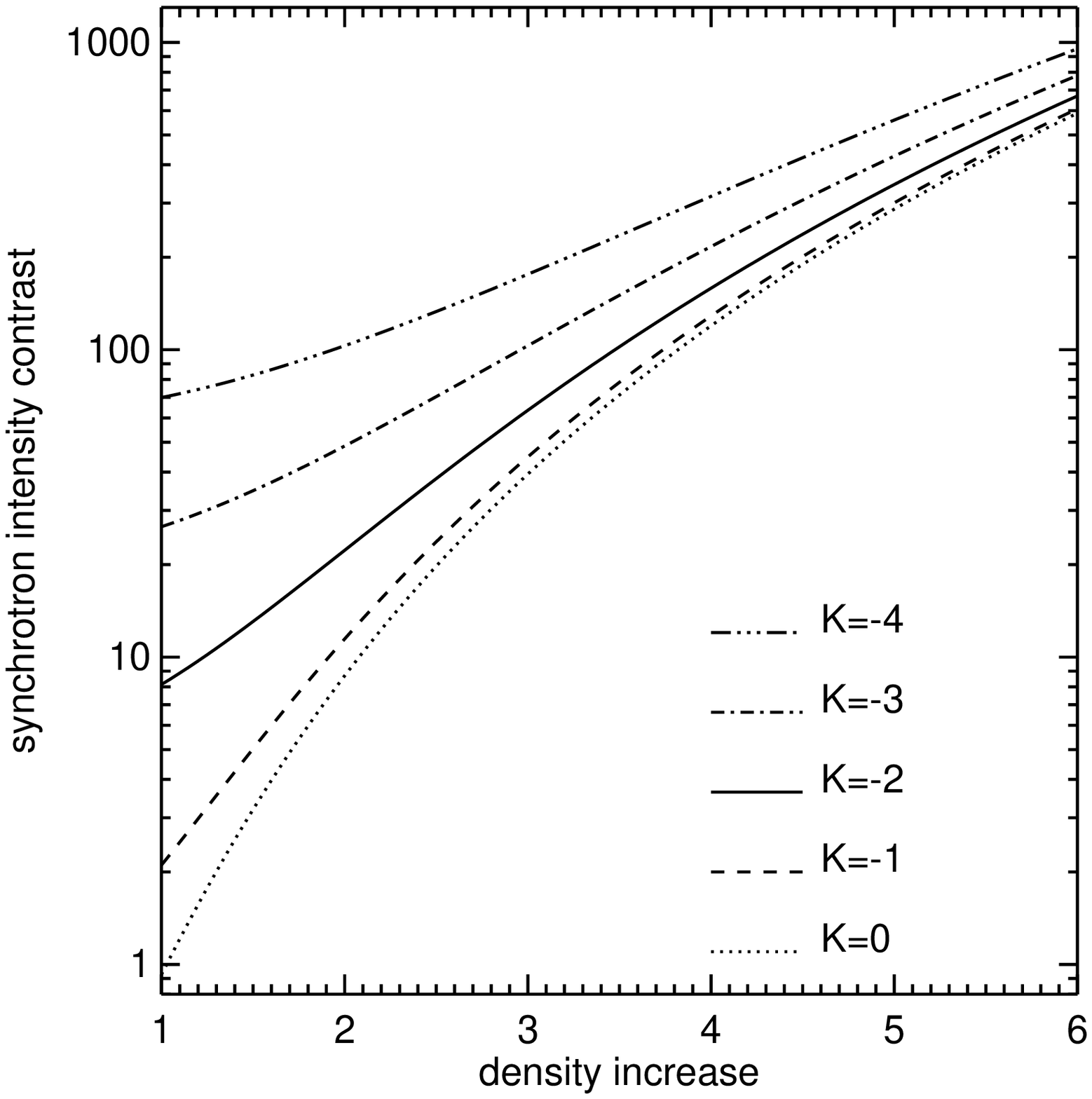}}
\subfigure[\mbox{Degree of polarization:} \mbox{constant cosmic ray
density}]
{\label{fig:epi}
	\includegraphics[width=0.32\textwidth]{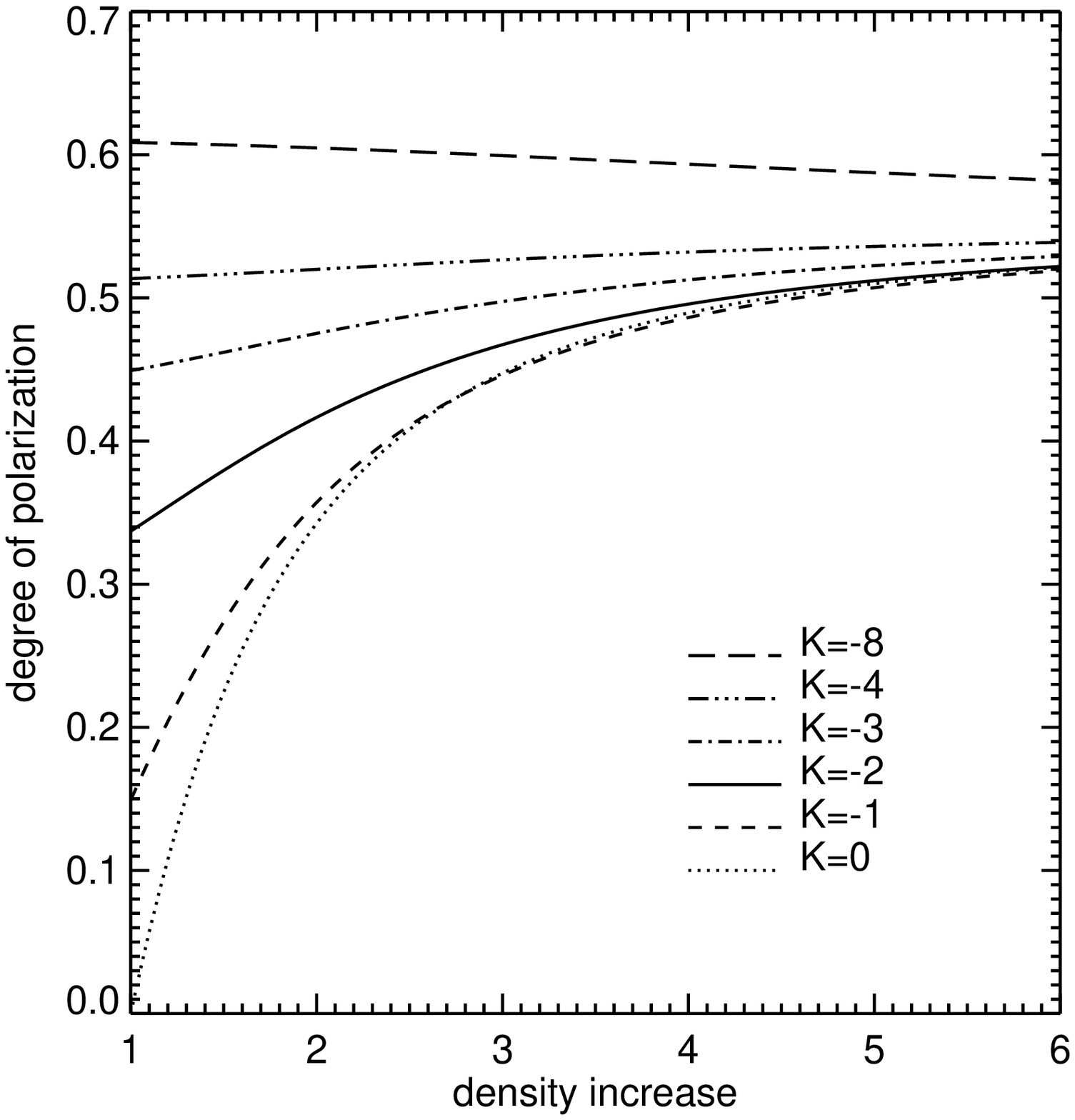}}
\caption{The changes in total intensity and degree of polarization
arising from a
sheared and compressed, initially isotropic, random magnetic field as a
function of the gas density contrast $\epsilon_\rho$ in a shock
for the geometry of the galaxy NGC~1365.
{\bf(a)}: The increase in total synchrotron intensity in the case of
constant cosmic ray
density, Eq.~(\ref{eq:ei:const}).
{\bf(b)}: The increase in
total synchrotron intensity in the case of energy density equipartition
between the magnetic field and cosmic
rays, Eq.~(\ref{eq:ei:equi}).
{\bf(c)}: The degree of polarization in the case of constant cosmic ray
density, Eq.~(\ref{eq:pol}), with $p_0=0.7$. The parameter
$K=\Delta V_r l/(c_s d)$
is the ratio of the turbulent eddy turnover timescale
$l/\cs$ to the shearing timescale $d/ \Delta V_r$
and is the appropriate measure of the strength of the shearing; high
$|K|$ means strong shear. The model inclination is $i=40^\circ$ and the
position angle of the shock is $\phi=40^\circ$, the parameters of NGC~1365.}
\label{fig:ei+epi}
\end{figure*}

In order to estimate the compression ratio (the
ratio of post- to pre-shock values)
of the total synchrotron intensity, $\epsilon_I$,
we assume that the contribution of the regular magnetic field to the
total field is negligible at the required level of accuracy, as
indicated by our observations.
In Sect.~\ref{sectMFbar} we estimated the
strength of the regular field in the ridges as about $2-4\mkG$ while the
total field is about $25\mkG$ (Table~\ref{Table3}).

In Appendix~\ref{app:shock} we derive two equations for $\epsilon_I$
when an initially isotropic, random magnetic field is subject to shock
compression
and shear. Equation~(\ref{eq:ei:const}) applies when the cosmic ray electron
density is not changed by the shock (at least, on the scale of our resolution
of about $500$\,pc) whereas Eq.~(\ref{eq:ei:equi}) includes the common assumption
that the energy density of cosmic rays is equal to the energy density of the
magnetic field -- in this case the synchrotron emission is enhanced both by
an increase in field strength and an increase in cosmic ray electron
density.  Figures~\ref{fig:ei:crconst}
and \ref{fig:ei:crequi} show how $\epsilon_I$ depends
on the compression ratio of the gas density, $\epsilon_\rho$, for different
magnitudes of the velocity shear, parameterized by the factor $K$.

The most striking difference between the constant cosmic ray density
and the equipartition cases is an order of magnitude
higher contrast in synchrotron
emission predicted by the equipartition model. For a shear strength of $K=-2$
(see Appendix~\ref{app:shock}) and the gas density contrast of an
adiabatic shock $\epsilon_\rho=4$, the model using constant cosmic ray density
(Fig.~\ref{fig:ei:crconst})
predicts $\epsilon_I\simeq10$, as observed in the inner bar of NGC~1097
and NGC~1365 (Table~\ref{Table3}). Note that, although
Fig.~\ref{fig:ei+epi} is plotted using the
geometrical parameters of NGC~1365, the different inclination and bar
position angle of NGC~1097 produce similar, but not identical,
results. The lower observed values
of $\epsilon_I$ at larger radii can be attributed to weaker shear, a
weaker shock or a combination of both. In order for the equipartition model
(Fig.~\ref{fig:ei:crequi}) to match the observed $\epsilon_I$ requires
both a much weaker velocity shear $K\simeq-1$ and a small increase in gas
density $\epsilon_\rho\simeq 2$. Since the velocity and gas density fields of
these galaxies are not known to a sufficient accuracy to allow a confident
rejection of one of the models, we can only note that with a constant cosmic ray
density, a far greater range of parameters describing the shock and shear are
compatible with the observed $\epsilon_I$, whereas equipartition requires
both weak shear and low compression.

We conclude that the observed
enhancement of total radio intensity in the radio ridges
is consistent with amplification of the random magnetic field
by shock compression and shear. Better
observational data on the gas density (CO and \HI) and velocity fields
in the bars of NGC~1097 and NGC~1365 may provide a novel way in which to test
the assumption of energy density equipartition between cosmic rays and
magnetic fields.

%------------------------------------
\subsection{Contrast in polarized intensity}
\label{sectContrast}

Polarized intensity (\PI) can be increased in the radio ridge due to
(i) shock compression and shearing of the turbulent field producing an
anisotropic turbulent magnetic field as described above and
(ii) compression and shearing of the regular magnetic field.
These effects will tend to align the magnetic
field with the shock front and so a strong increase in \PI\ can be
expected.

\subsubsection{Polarization from an anisotropic turbulent magnetic field}
\label{sectAniso}

Equation~(\ref{eq:pol}) gives the degree of polarization produced by a
compressed and sheared random magnetic field. This equation assumes that
the cosmic ray density is constant.
We have not derived a version for the case of energy density equipartition
between cosmic rays and magnetic field,
but it is sufficient for the present purposes to note that
equipartition will tend to increase the degree of
polarization as discussed in Sect.~5.2 of Sokoloff et al.\ (\cite{SB98}).

Figure~\ref{fig:epi} shows the degree of polarization, assuming
$p_0=0.7$ for the intrinsic polarization,
for the geometry of NGC~1365 as a function of $\epsilon_\rho$ for different
shear strengths. Again, the case of NGC~1097 is slightly, but not significantly
different. In the absence of compression ($\epsilon_\rho=1$), increasing the
shear produces a greater degree of polarization by producing stronger anisotropy.
With increasing compression
the degree of polarization converges to $p\simeq 0.5$, regardless of the shear.
However, for the strongest shear shown, with $K=-8$, the behaviour is
counter-intuitive: stronger compression \emph{reduces\/} the degree of
polarization. This
happens because the vertical component of the random magnetic field,
$b_z$, is increased by compression together with $b_r$. The projection of the
stronger $b_z$ into the sky plane has components parallel \emph{and orthogonal\/}
to the sheared $b_\phi$, thus reducing the anisotropy of $\vec{b}_\perp$.
We have
confirmed this explanation by re-deriving Eq.~(\ref{eq:pol})
with $b_z=0$: in this case, no matter how large $|K|$ is, the slope of $p$ is
never negative.

For $\epsilon_\rho=4$, the degree of polarization arising from anisotropic
turbulent magnetic field is $p\simeq0.5$ in NGC~1365 and $p\simeq0.4$ in
NGC~1097, with only a weak dependence on the shear strength. These are
\emph{overestimates\/} since we do not include any beam depolarization.
Nevertheless, the expected degree of anisotropy of
the turbulent magnetic field is rather high and should contribute
significantly to the observed polarized
intensity.

The above estimate is based on the anisotropy in $\vec{b}$ immediately
behind the shock, as specified in Eq.~(\ref{tn}). In fact the isotropy
of the turbulent magnetic field can be restored rather quickly
if the turbulence forcing is isotropic.  With standard
estimates of the turbulent eddy size of $l=100\p$ and
the turbulent velocity
$v=10\kms$, the magnetic field can return to isotropy in the eddy
turnover time $\tau=l/v\simeq10^7\yr$. Given that, behind the shock,
the gas speed normal to the shock is equal to the speed of sound
$\cs \simeq 10\kms$
(Roberts et al.\ \cite{RHA79}, Englmaier \&
Gerhard\ \cite{EG97}; see above)
and the width of the radio ridge is $d \simeq 500$~pc, the
residence time of the gas in the ridge is $\tau_\mathrm{r}\simeq
d/\cs\simeq5\times10^7\yr$, which exceeds the isotropization time severalfold.
Hence the degree of anisotropy estimated above only
applies to a region in the close vicinity of the shock front and may
be lower when averaged over our $6\arcsec\simeq 500\p$ telescope
beam. Thus,
whilst shock compression of the turbulent magnetic field will give
rise to a strong increase in $I$ (see Sect.~\ref{sectCompr}) it may only
have a weaker effect on \PI.

%------------------------------------------------------------------------
\subsubsection{Polarization from a sheared regular magnetic field}
\label{sectShear}

Now we consider the effect of shear in the velocity field on
the regular (coherent) magnetic field. Gas-dynamical simulations (e.g.,
Fig.~10 of Athanassoula\ \cite{A92b};
see also Fig.~\ref{fig:vr} here) confirm that the
radial velocity in the dust lane region changes very
rapidly near the ridge, from an outward to an inward direction,
and subsequently varies slowly.  If the radial
velocity changes by $\Delta V_r$ over the dust lane of a width $d$,
the radial magnetic field
$\mean{B}_r$ produced from the azimuthal one
$\mean{B}_\phi$ by the shear is
\begin{equation}\label{sc}
   \mean{B}_r\approx \mean{B}_\phi\,\frac{\Delta V_r}{d} \tau_\mathrm{r}
   \simeq \mean{B}_\phi\frac{\Delta V_r}{\cs}\simeq10 \mean{B}_\phi\;,
\end{equation}
where $\tau_\mathrm{r}\simeq d/\cs$ is the residence time in the shock region,
$\Delta V_r\simeq 100\kms$ and $\cs\simeq 10\kms$.
The enhancement of the magnetic field tangent to the shock front
by compression is
$\epsilon_B\simeq 4$ for an adiabatic shock. This is a factor
of more than two smaller than the amplification by shear, Eq.~(\ref{sc}). In addition,
the regular magnetic field is almost
perpendicular to the shock upstream of it, which reduces the effect of
compression and enhances the effect of shear. So, we neglect the compression
of the regular magnetic field.

A similar estimate of regular field amplification in the ridge follows
from the observed deflection of the polarization plane discussed in
Sect.~\ref{sectMF}, where we note that magnetic pitch angles upstream and
downstream of the shock front in NGC~1097 are ${\pa}_1\simeq15^\circ$ and
${\pa}_2\simeq75^\circ$, respectively. Given that
$\mean{B}_r/\mean{B}_\phi=\tan{\pa}$ and
$\mean{B}_{\phi1}\approx\mean{B}_{\phi2}$,
since the azimuthal field component is nearly normal to the shock front,
we obtain $\mean{B}_{r2}/\mean{B}_{r1}\approx\tan{\pa}_2/\tan{\pa}_1\simeq14$,
which is in fair agreement with Eq.~(\ref{sc}).

Such an enhancement of the regular magnetic field must result in a
very significant increase in polarized intensity. The enhancement
factor of polarized
emissivity can be estimated using Eqs.~(\ref{bpx}) and (\ref{bpy}),
but now written for the components of the transverse
(i.e. in the plane of the sky) regular magnetic
field $\vec{\mean{B}}_\perp=(\mean{B}_{x'}, \mean{B}_{y'})$.
Assuming for the sake of
simplicity that, in the galaxy's plane, the regular magnetic field is
purely azimuthal in front of the shock, $\vec{\mean{B}}_1=(0,\mean{B}_\phi,0)$, and
purely radial behind it, $\vec{\mean{B}}_2=(\mean{B}_r,0,0)$, and that the
enhancement factor in polarized synchrotron emissivity
${\varepsilon\reg}_2/{\varepsilon\reg}_1$ is equal to that in
$\mean{B}_\perp^2$ (i.e.,
the energy density of cosmic rays is constant across the shock), we obtain
\begin{eqnarray}
\label{eq:shear:B}
\epsilon\reg & = &   \frac{{\varepsilon\reg}_2}{{\varepsilon\reg}_1}\nonumber\\
   &=&
   \left(\frac{\Delta V_r}{\cs}\right)^2\,
   \frac{1-\cos^2\phi\,\sin^2 i}{1-\sin^2\phi\,\sin^2 i}\\
   &\simeq&\left\{
\begin{array}{ll}
60     &\mbox{in NGC~1097}\;,\nonumber\\
90     &\mbox{in NGC~1365}\;.\nonumber
\end{array}
\right.
\end{eqnarray}

If the region with strong shear is not resolved then the increase in
polarized
emission can be smaller than this. However, in order to achieve
$\epsilon_{P,\mathrm{obs}}\simeq 3$ (Table~\ref{Table3}) would require the
width of the sheared region to be about
$d\simeq W\,\epsilon_{P,\mathrm{obs}}/\epsilon\reg\simeq 50\p$ in NGC~1365
(where $W=1.5\kpc$ is the beamwidth) and even less in NGC~1097.
In addition, the observed $\epsilon\reg$ does not increase in our maps
with higher resolution, so the low observed values are not due to our
limited resolution.

We conclude that the observed contrast in polarized intensity is inconsistent
with the expected amount of magnetic field enhancement if it is subject to
a full amount of shearing in the dust lanes.

\subsubsection{Decoupling of the regular magnetic field from the molecular clouds}
\label{sectLowPI}

From the above arguments, it is clear that the combined effects of anisotropic
turbulent magnetic field and sheared regular magnetic field should
produce a significant increase in the polarized intensity around the shock
front/dust lane.  However, we observe
only a modest contrast of $\epsilon\reg\simeq 0.5$--$7$
(Table~\ref{Table3}).

The simplest way to resolve the conflict between the expected
and observed $\epsilon\reg$,
without having to abandon our successful description of
$\epsilon_I$ in Section~\ref{sectCompr}, is to suggest that
the regular magnetic field resists shearing by the radial velocity field.
As we argue in what follows, this can happen if \emph{the regular magnetic field
becomes decoupled from the dense gas (molecular clouds) in the shocked region}.
A typical time scale for the formation of H$_2$ molecules is of order $10^6$~yr
(Jura\ \cite{Jura75}, Bergin et al.\ \cite{Bergin04}), much less than the
residence time in the ridge $\tau_\mathrm{r} \simeq 5\times10^7$ yr.
As molecular clouds form, it is plausible that they become detached from
the regular magnetic field, carrying with them and amplifying only the random,
small-scale magnetic field. One mechanism
(Ohmic diffusion) has already been proposed for this process by Mestel
\& Strittmatter
(\cite{Mestel67}) and others are plausible, such as
ambipolar diffusion and reconnection of the
external field as a forming cloud rotates; further discussion is beyond the
scope of this paper (see Fletcher \& Shukurov \cite{FS05}).

In order for the regular field -- and hence the polarized intensity -- not to
be increased as strongly as suggested by Eq.~(\ref{eq:shear:B}), it must be
sufficiently strong to prevent the diffuse gas, to which it is coupled,
from being sheared.
Consider a shear flow $\vec{V}=(V_x(y),0,0)$ with a horizontal magnetic field
embedded into it, $\vec{\mean{B}}=(\mean{B}_x,\mean{B}_y,0)$, where the
$x$-axis is parallel to the
radial direction and the $y$-axis is directed along azimuth. The
magnetic braking of the radial shear flow is controlled by the $x$-component
of the Navier--Stokes equation,
\begin{equation}
\label{eq:dVxdt}
\deriv{V_x}{t} \simeq \frac{1}{4\pi\rho}\,\mean{B}_y\deriv{\mean{B}_x}{y}\;,
\end{equation}
where we have neglected the $x$-derivative of magnetic field in comparison with
its $y$-derivative.
For simplicity, consider the effect on the magnetic field of compression alone.
This will result in a very conservative estimate of the effect of magnetic
field on the flow since additional amplification by the shear is neglected.
Thus, $\mean{B}_x$ is compressed over a length $d_1$ by a factor $\epsilon_\rho$,
whereas $\mean{B}_y$ remains unchanged:
$\partial \mean{B}_x/\partial y\simeq(\epsilon_\rho-1)\mean{B}_x/d_1$. Since the
azimuthal speed behind the shock is close to the speed of sound $\cs$ (see above),
the increment in the radial velocity produced by magnetic stress
over a length $d_2$ follows as
\begin{eqnarray}
\label{eq:deltaV}
\Delta V_x&\simeq& \frac{\epsilon_\rho-1}{\cs}\,\frac{\mean{B}_x\mean{B}_y}{4\pi\rho}
\,\frac{d_2}{d_1}\\
&\simeq& 70\,\frac{\rm km}{\rm s}\left(\frac{\mean{B}_x\mean{B}_y}{10\mkG^2}\right)
\left(\frac{n}{0.2\cm^{-3}}\right)^{-1}
\left(\frac{\cs}{10\,{\rm km}\,{\rm s}^{-1}}\right)^{-1}\!\!, \nonumber
\end{eqnarray}
where the numerical value has been obtained for $d_1=d_2$
and $\epsilon_\rho=4$. We have the following additional constraints:
\begin{eqnarray}
\label{eq:constraints}
\mean{n}_\mathrm{e}\mean{B}_{\parallel} & \simeq & 0.3\mkG\cm^{-3}, \nonumber \\
\mean{n}_\mathrm{e}^2 L & \simeq & 100 \eta \cm^{-6} \p, \\
\mean{n}_\mathrm{e} & = & X \mean{n}, \nonumber
\end{eqnarray}
where the first is derived from Faraday rotation values in the ridge (Sect.~\ref{sectRM}),
the second from the estimated \EM\ with $\eta$ the filling factor of the diffuse ionised
gas (Sect.~\ref{sectMFbar}), and the third defines the degree of ionisation $X$ of the diffuse
gas. For the inclination of NGC~1097 and NGC~1365, $L\simeq 2\sqrt{2}h$ where $h$ is
the scale height of the diffuse gas. Combining Eq.~(\ref{eq:deltaV}),
with $\mean{B}_x=\mean{B}_y=\mean{B}/\sqrt{2}$, $\Delta V_x=100\kms$
and $\cs=10\kms$,
and Eqs.~(\ref{eq:constraints}) we obtain expressions for the regular magnetic
field strength, diffuse gas density and the
scale height of the ionized layer in terms of the least
well known parameters $\eta$ and $X$:
\begin{eqnarray}
\label{eq:BnL}
\frac{\mean{B}}{1\mkG} & \simeq & 4 X^{-1/3}, \nonumber \\
\frac{\mean{n}}{1 \cm^{-3}} & \simeq & 0.1 X^{-2/3}, \\
\frac{h}{1 \kpc} & \simeq & 3 \eta X^{-2/3}. \nonumber
\end{eqnarray}
In other words, Eq.~(\ref{eq:deltaV}) -- and hence the suppression of shear in the diffuse
gas -- is consistent with the observed \RM\ and \EM\ and the assumed disc thickness if
Eqs.~(\ref{eq:BnL}) give reasonable values of $\mean{B}$, $\mean{n}$ and $h$, using the
chosen fractional ionisation $X$ and the diffuse gas filling factor $\eta$. This is the
case for $X\gtrsim 0.5$ and $\eta\lesssim 0.2$, plausible values for both parameters.
For example, $X=0.5$ and $\eta=0.2$ gives $\mean{B}\simeq 5\mkG$, $\mean{n}\simeq 0.2 \cm^{-3}$
and $h\simeq 1\kpc$. The latter is similar to the scale height of the diffuse gas in the
Milky Way (Reynolds~\cite{Reynolds90}) and the filling factor $\eta=0.2$ is similar to
that of the diffuse thermal electrons in normal spiral galaxies
(Greenawalt et al.~\cite{Greenawalt98}).

To summarise this section: the increase in polarized intensity in
the ridge,
with respect to the upstream region, is much lower than expected because the
regular magnetic field decouples from the dense molecular gas clouds and is
sufficiently strong to prevent shearing and compression in the diffuse ionised gas.
The regular magnetic field of about $5\mkG$ in strength can be \emph{dynamically
important in the diffuse gas\/} of density $\approx0.2\cm^{-3}$. If the ionization degree
of the diffuse gas is of order 50\% and the scale height of the diffuse ionised gas
is about $1\kpc$, the resulting emission measure is compatible
with that observed in the southern ridge of NGC~1097.

%--------------------------------------------
\subsection{Variation in RM across the dust lane}
\label{sectRMShear}

\begin{figure}
\centering
\includegraphics[bb = 19 19 308 288,width=0.3\textwidth,clip=]{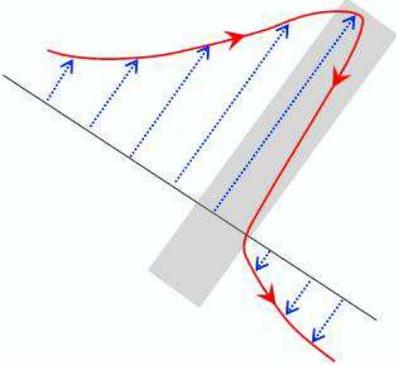}
\caption{\label{shearB}An illustration of the shearing of the
   regular magnetic field in the radio ridges. The orientation of
   the figure and the field direction roughly correspond to those in
   the southern part of NGC~1097. A magnetic line shown with
   continuous line is sheared by a velocity field component along
   the radio ridge
   (dust lane) region, whose vectors are indicated with dotted lines;
   the thin straight line shows the orientation of the original magnetic
   field; the unresolved part of the system is shown shaded.}
\label{shear}
\end{figure}

The variation of the Faraday rotation measure across the inner southern
radio ridge region of NGC~1097 (left-hand panel of Fig.~\ref{n1097rm}) is unusual.
The discontinuous change in sign of rotation measure arises because the
radial component of the regular magnetic field changes direction
across the shock front. This counter-intuitive arrangement of magnetic field
directions is confirmed
by the fitting of the large-scale polarization structure with
azimuthal Fourier modes presented in Sect.~\ref{sectGlobal}.
Figure~\ref{shear}
illustrates how such a magnetic field configuration
can be produced by a shear shock. We should remember
that the observed polarization vectors are dominated by the
anisotropic field which is strictly aligned along the shock front.
The sheared regular field sketched in Fig.~\ref{shear},
visible only in Faraday rotation, is tilted with respect to the
shock front.

%------------------------------------------
%---------------------------------------

The sharp change in magnetic field orientation at the sheared region
(Fig.~\ref{shear}) explains the origin of the depolarized valley:
If the region of strongest shearing is narrow compared to the beam
(as we argued in Sect.~\ref{sectShear}),
then the almost orthogonal field orientations will give rise to strong,
wavelength-independent beam depolarization.
A shift of the polarization ridge downstream with respect to the total
intensity ridge will also result, as observed (Sect.~\ref{sectRidge}).

%%------------------------------------------

%-------------------------------------------
\section{Mass inflow through the nuclear ring of NGC~1097}
\label{sectInflow}

The nuclear ring of NGC~1097 is a site of intense star formation with
a rate of $5M_\odot\yr^{-1}$ (Hummel et al.\ \cite{HH87}) and, presumably,
strong gas inflow to fuel the central activity. Beck et al.\
(\cite{BE99}) suggested that the stress required to drive the inflow into
the central region can be provided by the magnetic field. The total
magnetic field in the nuclear ring and its regular + anisotropic random part
are estimated, from equipartition arguments, to be $B\tot\simeq50$--$60\mkG$ and
$B\reg\simeq10$--$20\mkG$, respectively (Table~\ref{Table4}).
The regular magnetic field has the form of a trailing spiral with a
pitch angle of $\pa\simeq 40^\circ$. This field produces the
$r\phi$-component of the stress tensor
$T_{r\phi}=-\mean{u_{\mathrm{A}r}u_{\mathrm{A}\phi}}$, where
$\vec{u}_\mathrm{A}=\vec{B}/(4\pi\rho)^{1/2}$ is the Alfv\'en velocity
based on the total magnetic field.

The accretion rate driven by this
stress is $\dot M= 2\pi\sigma \Omega^{-1} T_{r\phi}$, where
$\sigma\simeq 2h\rho$ is the gas surface density, with $2h$ the
disc thickness, and $\Omega$ is the angular velocity of the gas
(e.g., Balbus \& Hawley\ \cite{BH98}). This yields
\begin{equation}
\label{eq:accrete}
\dot M\simeq-\frac{h}{\Omega}(\mean{b_r b_\phi}+\mean{B}_r \mean{B}_\phi)\;,
\end{equation}
where $\vec{b}$ and $\mean{\vec{B}}$ are the turbulent and regular
magnetic field components, respectively,
and overbar denotes averaging -- cf.\ Eq.~(\ref{eq:dVxdt}).

With $c=\mean{b_rb_\phi}/\mean{b^2}$, we obtain
\begin{eqnarray*}
\dot M & \simeq & -\frac{h}{\Omega}(c \mean{b^2}+\sfrac12 \mean{B}^2\sin 2p)\\
       & \simeq & -c \, M_\odot\yr^{-1}\;,
\end{eqnarray*}
where $c \simeq-1 $
is a numerical factor (see below) and the estimate has been
obtained for $h=0.1\kpc,\ \Omega=450\kms\kpc^{-1},\mean{b^2}^{1/2}=50\mkG,\
\mean{B}=10\mkG$ and $\pa=40^\circ$;
we have retained only the (dominant) first term in Eq.~(\ref{eq:accrete}).
Hence, we derive a mass inflow rate of a few $M_\odot\yr^{-1}$
which is sufficient to feed the active nucleus.

A correlation between the azimuthal and radial components of the
turbulent magnetic field used above arises from the shearing of
turbulent magnetic field by differential rotation. From arguments
similar to those used to obtain Eq.~(\ref{scb}) and assuming that the relevant
turbulent timescale is $h/c_\mathrm{s}$ (as in accretion discs) we obtain
\begin{equation}
b_{\phi}=r\frac{d\Omega}{dr}\,\frac{h}{c_\mathrm{s}} \, b_r
\simeq -\frac{q \Omega h}{c_\mathrm{s}} \, b_r\;,
\end{equation}
where the rotation curve is assumed to be
of the form $\Omega\propto r^{-q}$, with $q=1$ corresponding to a
flat rotation curve.
Assuming that $\mean{b_r^2}=\frac13\mean{b^2}$, we obtain
$c=\mean{b_rb_\phi}/\mean{b^2}\simeq-\frac13q\Omega h/\cs$.

The problem of mass transport from a circumnuclear ring into an active
nucleus has been addressed by many authors (see review by Knapen\ \cite{K04}).
The inflow rate due to magnetic stress is much larger than
that normally produced by graviational torques in the nuclear ring
(Maciejewski et al.\ \cite{M02}). We note that turbulent stress can result in a similar
inflow rate given that magnetic field and turbulence are in energy equipartition.
Strong magnetic fields are known to
exist in the centres of most galaxies from the observation of strong
synchrotron emission. Hence magnetically driven inflow
can be important not only in barred
galaxies, but also in
normal spiral galaxies with a magnetized inner disc
(Moss et al.\ \cite{MSS00}).

%-------------------------------------------------------------------------
\section{Summary and conclusions}

\begin{itemize}

\item We have detected polarized emission in the galaxies NGC~1097
and NGC~1365 which gives important information about the interaction of
the gas flow and the magnetic field structure in barred galaxies.

\item The total magnetic field is strong in the central star-forming
regions of NGC~1097 and NGC~1365 (about $60\mkG$) and in the radio ridges
along the galaxies' bars (20--$30\mkG$).

\item In both galaxies, the magnetic field orientation changes strongly,
but smoothly in front of the radio ridges. This causes a valley of depolarization
where the polarized emission almost vanishes.
The smooth deflection of the magnetic field is different from that of the
(likely) velocity field of the dense gas which changes its direction abruptly.

\item The enhancement of total radio intensity in the radio ridges along
the bars of NGC~1097 and NGC~1365 can be explained by shock
compression and shearing of the isotropic random field, together with
compression of the gas,
by a factor of about 4 and constant cosmic ray density across the shock.
Local equipartition conditions, where the cosmic rays are compressed together
with the magnetic field, are hardly compatible with our data as they require
both low compression and low shear.

\item The enhancement of polarized intensity is much lower than
that of the total synchrotron intensity.
\emph{The regular magnetic field resists shearing because it
is decoupled from the dense gas}, and
the Maxwell stress is sufficient to oppose efficiently the
development of a shear flow developing in the diffuse gas to which the
regular magnetic field remains coupled.
This also contributes to the misalignment between the magnetic field
and the velocity field of the dense gas in front of the ridges.

\item The enhanced polarized emission in the ridges can be
explained by the presence of
anisotropic random fields due to compression of isotropic random
fields in the shock. These anisotropic fields dominate over the
regular (coherent) fields which are visible through their Faraday effects.

\item Faraday rotation measures reveal jumps across the ridge, where the
amplitude decreases with increasing distance from the galaxy's centre.
This behaviour results from shearing of the regular field.

\item We have demonstrated that radio continuum data of the total and
polarized emission form a new and powerful tool to determine the shock
strength and the shear rate in the shearing shocks of barred galaxies.
If the gas compression
is known from independent measurements (\HI\ and CO), the validity of
energy equipartition between magnetic fields and cosmic rays can also be
tested.

\item Enhanced polarized intensity ahead of the shock fronts of
NGC~1097 and NGC~1365, at several kpc distance from the bar,
could be a signature of regular fields generated by dynamo action.
Alternatively, compression of gas and magnetic fields may occur
in the region of narrow dust filaments visible on optical images.
The magnetic field is aligned with these filaments.

\item Magnetic braking
in the circumnuclear regions of galaxies is a fundamental
process which may solve the long-standing problem of how to feed active
galactic nuclei.
We have shown that the magnetic fields in the circumnuclear ring of NGC~1097
are sufficiently strong
($\simeq60\mkG$) to drive a mass inflow of several $M_\odot\yr^{-1}$.

\item X-ray emission in NGC~1097,
detected with ROSAT, reveals  hot gas
(of a temperature about $10^7$~K) in the central region and also an extended
hot gas component.
A crude estimate of the electron density in the
extended component yields $\ne \simeq 3.4 \times 10^{-3}\eta^{-1/2} \cm^{-3}$
(where $\eta$ is the volume filling factor of the hot gas).
If the extended X-ray and the extended radio components both emerge from a halo
and $\eta\simeq0.3$, the derived
thermal energy density of the hot gas
is comparable to the energy density of the magnetic field.
A more detailed spectral and spatial X-ray analysis to
study further the hot gas in the ISM of barred galaxies such as NGC~1097
requires deep observations at high throughput observatories like
XMM-Newton.

\end{itemize}

\begin{acknowledgements}
We are grateful to Lia Athanassoula who kindly provided data from her
gas-dynamical models for barred galaxies.
We thank Elly M.\ Berkhuijsen and our referee for careful reading of the
manuscript and many useful comments.
This work was supported by the PPARC grant PPA/G/S/2000/00528, RFBR
grant 04-02-16094 and DFG/RFBR grant 96-02-00094G. We have made use of the
ROSAT Data Archive of the Max-Planck-Institut f\"ur extraterrestrische
Physik (MPE) at Garching, Germany.
\end{acknowledgements}

\appendix
\section{The effect of compression and shear on a random magnetic field}
\label{app:shock}

Here we derive equations describing the increase in total synchrotron
emission
and polarization resulting from shock compression and shearing of an
initially
isotropic, random magnetic field. For the total emission
we consider
two cases: first, and simpler, we assume that the energy
density of cosmic ray
electrons is constant (i.e.\ the density of cosmic rays is not affected
by the shock);
then we present the more complex model where the cosmic
rays are in energy
equipartition with the magnetic field. Finally, we examine how much
polarized emission
emerges from the anisotropic field produced by shock compression and
shearing.

The component of the random magnetic field
tangent to the shock front
is compressed together with the gas, whereas the
normal component remains unchanged,
\begin{equation}\label{tn}
   b_\mathrm{t2}=\epsilon_\rho b_\mathrm{t1}\;,
   \quad
   b_\mathrm{n2}=b_\mathrm{n1}\;,
\end{equation}
where $\epsilon_\rho=\rho_2/\rho_1$ is the density ratio and subscripts
1 and 2 refer to pre- and post-shock values, respectively. We assume for
simplicity that the random field is isotropic upstream of the shock,
\begin{equation}\label{tn12}
    \bra{b_\mathrm{n1}^2}=\sfrac12\bra{b_\mathrm{t1}^2}=\sfrac13\bra{b_1^2}\;,
\end{equation}
where angular brackets denote averaging.
Thus, the compressed field is anisotropic, with the dominant components
tangent to the shock front.

The anisotropy of the random magnetic field in the postshock region is
further enhanced by velocity shear.
We introduce cylindrical polar coordinates $(r,\phi,z)$ and assume
for simplicity that the shock is parallel to the $r$ direction. If the radial
velocity changes by $\Delta V_r$ across the shock region of a width $d$,
the radial magnetic field $b_{r2}$ produced from the azimuthal
$b_{\phi 1}$ by the shear is
\begin{equation}\label{scb}
b_{r2}\approx b_{\phi 1}\,\frac{\Delta V_r}{d}\tau_\mathrm{r}
=b_{\phi 1}\frac{\Delta V_r}{v_0}\frac{l}{d}\simeq -2b_{\phi 1}\;,
\end{equation}
where $\tau_\mathrm{r}=\min(d/\cs,l/v_0)$ is the relevant amplification
time (the minimum of the residence time within the sheared region,
$d/\cs$ (where $\cs$ is the sound speed), and
the eddy turnover time $l/v$). Since
the turbulent velocity $v_0\simeq\cs$
and $d>l$, we obtain the second
equality in Eq.~(\ref{scb}), and the third follows for $\Delta
V_r=-100\kms$ (Athanassoula\ \cite{A92b}), $v_0=10\kms$, $d=500\p$ and
$l=100\p$.

The radial component of the random magnetic field behind the shock thus
consists of two parts, the compressed one, denoted by $b_r^\mathrm{(c)}$
as given in Eqs~(\ref{tn}) and (\ref{tn12}), and the
sheared part $b_r^\mathrm{(s)}$ given by Eq.~(\ref{scb}). Although
$\bra{b_r b_\phi}\approx0$ ahead of the shock (if the field anisotropy
can be neglected there), the shearing produces correlated field components
behind the shock,
$\bra{b_r b_\phi}=\bra{b_r^\mathrm{(s)} b_\phi}\simeq\bra{b_\phi^2}
(\Delta V_r l)/(v_0 d)$ (where $\Delta V_r<0$). For the post-shock field
produced by compression, we have
$b_{r2}^\mathrm{(c)}=\epsilon_\rho b_{r1}$ and likewise for the
$z$-component,
whereas the $\phi$-component remains unchanged as it is normal to the
shock
front. The compression enhances an isotropic random magnetic field by a
factor of
$\left[\sfrac13(1+2\epsilon_\rho^2)\right]^{1/2}$.
For a strong shock ($\epsilon_\rho=4$) the compression factor is
approximately $3$.

Since synchrotron intensity depends on the magnetic field component in
the plane of the sky, its compression factor $\epsilon_I$ depends on
the orientation of the shock front with respect to the observer.
Appropriate expressions for the magnetic field projection to the plane
of the sky, $\vec{b}_\perp=(b_{x'},b_{y'})$, in terms of its
cylindrical components $\vec{b}=(b_r,b_\phi,b_z)$ in the galaxy's
reference frame can be found in Appendix~A of Berkhuijsen et al.\
(\cite{BH97}):
\begin{eqnarray}
b_{x'}&=&b_r\cos\phi-b_\phi\sin\phi\;,\label{bpx}\\
b_{y'}&=&(b_r\sin\phi+b_\phi\cos\phi)\cos i+b_z\sin i\;,\label{bpy}
\end{eqnarray}
where $i$ is the inclination angle of the galactic disc ($i=0$
corresponds to the face-on view) and $\phi$ is the azimuthal angle in
the galaxy's plane measured anticlockwise from the northern end of the
major axis, related to that in the sky plane, $\phi'$, by $\tan\phi
\cos i=\tan\phi'$. The
$x'$-axis in the sky plane points to the northern end of the galaxy's
major axis; here and below, coordinates in the plane of the sky are
denoted with primes, $(x',y')$.

In order to calculate the contrast in synchrotron intensity,
we consider the two models described below.

%--------------------------------------------------------------------------
\subsection{Total intensity contrast for uniform cosmic ray distribution}
If the energy density of cosmic rays does not vary across the
radio ridge (i.e., at a scale of 1 kpc), the increase in synchrotron
emission depends solely on the change in
$b_\perp$, the magnetic field component parallel to the plane of the
sky, and any variation in the disc thickness between the shocked and
upstream region. Then
Eqs.~(\ref{tn}), (\ref{tn12}) and (\ref{scb}) yield the contrast in
total
synchrotron intensity between the shocked and upstream regions:
\begin{eqnarray}
\label{eq:ei:const}
   \epsilon_I&=&\frac{\bra{b_{\perp 2}^2}}{\bra{b_{\perp
1}^2}}\,\frac{L_2}{L_1} \nonumber\\
   &=&\sfrac12\,\frac{L_2}{L_1}
         \left[2+(\epsilon_\rho^2+K^2-1)(1+\sin^2i\cos^2\phi) \right.\\
     & &\mbox{}\left. - K(K+\sin 2\phi)\sin^2i\right], \nonumber
\end{eqnarray}
where $K=(\Delta V_r l)/(v_0 d)$ is a measure of the strength of the
shear and for the canonical values given above $K\simeq -2$. We have
assumed
that different components of the compressed part of $\vec{b}$ remain
statistically
independent, $\bra{b_r^\mathrm{(c)}b_\phi}=0$ and
similarly for the other components. For $L_1=L_2$,
$\epsilon_\rho=1$ and $\Delta V_r=0$,
we obtain, as it should be, $\epsilon_I=1$; for $i=0$ (the face-on
view), the
contrast in $I$ becomes independent of azimuth.

\subsection{Total intensity contrast under energy equipartition
between cosmic rays and magnetic fields}

Here we consider the case where the energy density of cosmic rays
is proportional to the energy density of the total magnetic field.
Then the synchrotron emissivity behind the shock will be increased both
because of the field amplification by compression and shear and,
additonally, due
to the accompanying increase in the cosmic ray electron density where
$n_\mathrm{cr}\propto b^2$. Now we obtain, for $L_1=L_2$:
\begin{eqnarray}
\label{eq:ei:equi}
\epsilon_I&=&\frac{\bra{b_{\perp 2}^2 b^2}}{\bra{b_{\perp 1}^2 b^2}}\,
                \nonumber\\
&=& 1+\sfrac{3}{10}(\epsilon_\rho^2+K^2-1)^2(\cos^2i+\cos^2\phi\,\sin^2i)\nonumber\\
&&\mbox{}+\sfrac{1}{10}(\epsilon_\rho^2+K^2-1)(4+9\cos^2i+7\cos^2\phi\,\sin^2i)\nonumber \\
&&\mbox{}+\sfrac{1}{10}(\epsilon_\rho^2-1)(\epsilon_\rho^2+K^2-1)(1+\cos^2\phi\,\sin^2i)\nonumber\\
&&\mbox{}+\sfrac{1}{10}(\epsilon_\rho^2-1)(3\epsilon_\rho^2+4)\sin^2i\nonumber\\
&&\mbox{}-\sfrac{1}{10}K(1+\epsilon_\rho^2+K^2)\sin2\phi\,\sin^2i\;,
\end{eqnarray}
following careful treatment of the averages of the components of the
field and their cross terms ($\bra{b_r^4}$, $\bra{b_r^2 b_\phi^2}$
etc.) and some algebra.

\subsection{Polarization from an anisotropic random magnetic field}

For a uniform cosmic ray distibution, we will now derive the
 degree of polarization
from the anisotropic random magnetic field created by the combined
effects of compression and velocity shear in the shock.

An anisotropic random field in the plane of the sky,
$\vec{b}_\mathrm{t}=(b_{r'},b_{z'})$, produces a degree of
polarization (Sokoloff et al.\ \cite{SB98}):
\begin{equation}
\label{eq:pol}
   p=p_0\left|
      \frac{\bra{b_\mathrm{t2'}^2}-
\bra{b_\mathrm{n2'}^2}}{\bra{b_{\perp2}^2}}\right|
        \equiv p_0\left|\frac{C_1}{C_2}\right|,
\end{equation}
where
\begin{eqnarray*}
  C_1&=&\sfrac14 (\epsilon_\rho^2+K^2-1)\\
   & & \!\!\!\times\left[(1-\cos i)^2\cos 4\phi
        - 2\cos2\phi\, \sin^2i+(1+\cos i)^2\right] \\
   & & \mbox{}+ K^2\cos 2\phi\,\sin^2i - \sfrac 1 2 K\sin4\phi\,(1-\cos i)^2\;,\\
C_2&=&2 + (\epsilon_\rho^2+K^2-1)(1+\sin^2i\,\cos^2\phi)\nonumber\\
   & & \mbox{}-K(K+\sin 2\phi)\sin^2i\;,
\end{eqnarray*}
and $p_0\approx0.7$ is the maximum degree of polarization.

The observed degree of polarization depends on the position in the
galaxy because of projection effects. As expected, the observed degree of
anisotropy of the magnetic field components in the sky plane behind the
shock is independent of azimuth $\phi$
(together with $p$) for face-on view $i=0$. The observed anisotropy is
maximum
when the shock occurs at the minor axis $\phi=90\degr$ and zero at the
major axis $\phi=0$ when the disc is viewed edge-on, $i=90^\circ$.

%---------------------------------------------------------------
\section{Parameters of the fitted regular magnetic fields}
\label{app:fit}

In Tables~\ref{tab:fit:ngc1097} and \ref{tab:fit:ngc1365} we give the
parameters of the fitted regular magnetic field models discussed in
Section~\ref{sectGlobal}.
Although a component of the regular field perpendicular to the
disc plane ($B_z$ in Eq.~(\ref{eq:model})) is allowed in the model and
we searched for fits using this component, a vertical field was not
required to obtain a good fit in any of the rings in either galaxy.
Polarization angles at $\lambda3.5$~cm and $\lambda 6.2$~cm were averaged
in sectors with an opening angle of $10^{\circ}$ in
NGC~1097 and $20^{\circ}$ in NGC~1365. The greater of the standard deviation
and the noise within a sector was taken as the error in polarization angle.

The regular magnetic field is modelled as
\begin{eqnarray}
  B_r & = & R_0\sin p_0 + R_1\sin p_1\cos(\phi-\beta_1)\nonumber \\
      & & +\, R_2\sin p_2\cos(2\phi-\beta_2),\nonumber \\
  B_{\phi} & = & R_0\cos p_0 + R_1\cos p_1\cos(\phi-\beta_1)\label{Bmod} \\
      & & +\, R_2\cos p_2\cos(2\phi-\beta_2),\nonumber\\
  B_z & = & R_{z0} + R_{z1}\cos(\phi-\beta_{z1}) + R_{z2}\nonumber
  \cos(2\phi-\beta_{z2}),\nonumber
  \label{eq:model}
\end{eqnarray}
where $R_i$ is the amplitude of the $i$'th mode in units of $\!\radm$,
$p_i$ is its pitch angle and $\beta_i$ determines the
azimuth where the corresponding non-axisymmetric mode is maximum.
The magnetic field in each non-axisymmetric mode of this model is
approximated by a logarithmic spiral, $p_i=\mbox{const}$, within a given ring.
However, the superposition of such modes is not restricted to represent a
logarithmic spiral. For further details of the method, see
Berkhuijsen et al.\ (\cite{BH97}) and Fletcher et al.\ (\cite{FB04}).

The foreground Faraday rotation due to the magnetic
field of the Milky Way was neglected in all cases,
since both galaxies
lie far from the Galactic plane; the expected (but unknown) contribution
of the Milky Way to \RM\ in the direction of the two galaxies is about $10\radm$
and will not significantly affect the fits. $S$ is the residual of the fit and
the appropriate $\chi^2$ value, at the $95$\% confidence level, is shown for
the number of fit parameters and data points.
A fit is statistically acceptable if $S\leq\chi^2$.
The $\chi^2$ values vary
from ring to ring (even when the same number of fit parameters are used)
as some sectors are excluded from the model, either because the average
measured signal is too weak or because the sector represents a clear
outlier from the global pattern.

% TABLE******
\begin{table*}
\caption{Parameters of fitted model for NGC 1097,
with notation as in Eq.~\ref{Bmod}}.
\label{tab:fit:ngc1097}
\begin{center}
\begin{tabular}{lllllll} \hline \noalign{\smallskip}
  & Ring & $1.25<r<2.5\kpc$ & $1.25<r<2.5\kpc$ & $2.5<r<3.75\kpc$ & $2.5<r<3.75\kpc$ & $3.75<r<5\kpc$ \\ \noalign{\smallskip}
  & Azimuth & $20^{\circ}\le\phi\le190^{\circ}$ & $200^{\circ}\le\phi\le370^{\circ}$ &
  $20^{\circ}\le\phi\le190^{\circ}$ & $200^{\circ}\le\phi\le370^{\circ}$  & $0^{\circ}\le\phi\le350^{\circ}$ \\ \noalign{\smallskip}
\hline
\noalign{\medskip}
$R_0$ & $\radm$ &
  $-169$\,\scriptsize{$\pm30$} & $55$\,\scriptsize{$^{+44}_{-7}$} &
  $-126$\,\scriptsize{$^{+33}_{-44}$} & $-57$\,\scriptsize{$\pm14$} &
  $-155$\,\scriptsize{$\pm8$} \\ \noalign{\smallskip}
$p_0$ & deg &
  $41$\,\scriptsize{$\pm2$} & $27$\,\scriptsize{$\pm2$} &
  $36$\,\scriptsize{$\pm4$} & $35$\,\scriptsize{$\pm6$} &
  $23$\,\scriptsize{$\pm2$} \\ \noalign{\smallskip}
$R_1$ &  $\radm$ &
        & & & $43$\,\scriptsize{$\pm12$} & $-51$\,\scriptsize{$\pm7$} \\ \noalign{\smallskip}
$p_1$ & deg &
        & & & $131$\,\scriptsize{$\pm22$} & $82$\,\scriptsize{$\pm16$} \\ \noalign{\smallskip}
$\beta_1$ & deg &
        & & & $-105$\,\scriptsize{$\pm10$} & $82$\,\scriptsize{$\pm5$} \\ \noalign{\smallskip}
$R_2$ &  $\radm$ &
  $-339$\,\scriptsize{$^{+44}_{-52}$} & $48$\,\scriptsize{$^{+7}_{-61}$} &
  $45$\,\scriptsize{$\pm24$} & & $75$\,\scriptsize{$\pm7$} \\ \noalign{\smallskip}
$p_2$ & deg &
  $40$\,\scriptsize{$\pm2$} & $-153$\,\scriptsize{$\pm2$} &
  $13$\,\scriptsize{$^{+16}_{-24}$} & & $3$\,\scriptsize{$\pm4$} \\ \noalign{\smallskip}
$\beta_2$ & deg &
  $-25$\,\scriptsize{$\pm3$} & $-21$\,\scriptsize{$\pm41$} &
  $12$\,\scriptsize{$\pm15$} & & $5$\,\scriptsize{$\pm5$} \\ \noalign{\smallskip}
\hline \noalign{\medskip}
$S$&& $38$ & $37$ & $40$ & $33$ & $74$ \\
\noalign{\smallskip}
$\chi^2$ && $44$ & $45$ & $47$ & $45$ & $78$ \\ \noalign{\smallskip}
\hline \noalign{\medskip}
\end{tabular}
\end{center}
\end{table*}
%************

% TABLE******
\begin{table*}
\caption{Parameters of fitted model for NGC 1365,
with notation as in Eq.~\ref{Bmod} and ring width $1.75\kpc$.}
\label{tab:fit:ngc1365}
\begin{center}
\begin{tabular}{lllllllll} \hline \noalign{\smallskip}
  & & \multicolumn{7}{c}{Ring centre (kpc)}  \\
  & & 3.50 & 5.25 & 7.00 & 8.75 & 10.50 & 12.25 & 14.00 \\
\hline
\noalign{\medskip}
$R_0$ & $\radm$ &
  $55$\,\scriptsize{$\pm3$} & $65$\,\scriptsize{$\pm2$} & $52$\,\scriptsize{$\pm6$} & $100$\,\scriptsize{$\pm1$} & $90$\,\scriptsize{$^{+18}_{-1}$} & $56$\,\scriptsize{$\pm14$} & $32$\,\scriptsize{$\pm6$} \\ \noalign{\smallskip}
$p_0$ & deg &
  $34$\,\scriptsize{$\pm2$} & $17$\,\scriptsize{$\pm1$} & $31$\,\scriptsize{$\pm1$} & $22$\,\scriptsize{$\pm1$} & $37$\,\scriptsize{$^{+1}_{-6}$} & $29$\,\scriptsize{$\pm11$} & $33$\,\scriptsize{$\pm6$} \\ \noalign{\smallskip}
$R_1$ &  $\radm$ &
  $-69$\,\scriptsize{$\pm4$} & $-112$\,\scriptsize{$\pm2$} &  & $-142$\,\scriptsize{$\pm3$} & $-56$\,\scriptsize{$^{+13}_{-1}$} & $13$\,\scriptsize{$\pm17$} & $8$\,\scriptsize{$\pm4$} \\ \noalign{\smallskip}
$p_1$ & deg &
  $23$\,\scriptsize{$\pm3$} & $11$\,\scriptsize{$\pm1$} &  & $21$\,\scriptsize{$\pm1$} & $-21$\,\scriptsize{$^{+1}_{-31}$} & $74$\,\scriptsize{$^{+133}_{-65}$} & $-90$\,\scriptsize{$^{+42}_{-52}$} \\ \noalign{\smallskip}
$\beta_1$ & deg &
  $-118$\,\scriptsize{$^{+8}_{-1}$} & $-108$\,\scriptsize{$\pm1$} & & $266$\,\scriptsize{$\pm1$} & $-58$\,\scriptsize{$^{+1}_{-13}$} & $-290$\,\scriptsize{$\pm50$} & $-38$\,\scriptsize{$^{+43}_{-30}$} \\ \noalign{\smallskip}
$R_2$ &  $\radm$ &
  $32$\,\scriptsize{$\pm6$} & $85$\,\scriptsize{$\pm4$} & $-53$\,\scriptsize{$\pm6$} & $-64$\,\scriptsize{$\pm2$} & $62$\,\scriptsize{$^{+1}_{-30}$} & $-65$\,\scriptsize{$\pm16$} & $32$\,\scriptsize{$\pm5$} \\ \noalign{\smallskip}
$p_2$ & deg &
  $41$\,\scriptsize{$\pm4$} & $18$\,\scriptsize{$\pm2$} & $-148$\,\scriptsize{$\pm2$} & $25$\,\scriptsize{$\pm2$} & $-131$\,\scriptsize{$^{+9}_{-1}$} & $56$\,\scriptsize{$\pm22$} & $-114$\,\scriptsize{$\pm14$} \\ \noalign{\smallskip}
$\beta_2$ & deg &
  $39$\,\scriptsize{$\pm4$} & $11$\,\scriptsize{$\pm1$} & $21$\,\scriptsize{$\pm3$} & $-9$\,\scriptsize{$\pm2$} & $0$\,\scriptsize{$^{+11}_{-1}$} & $-14$\,\scriptsize{$\pm6$} & $-11$\,\scriptsize{$\pm5$} \\ \noalign{\smallskip}
\hline \noalign{\medskip}
$S$&& $29$ & $32$ & $35$ & $34$ & $18$ & $23$ & $20$ \\
\noalign{\smallskip}
$\chi^2$ && $34$ & $33$ & $43$ & $36$ & $22$ & $33$ & $22$\\ \noalign{\smallskip}
\hline \noalign{\medskip}
\end{tabular}
\end{center}
\end{table*}

%-----------------------------------------------------------------------

\end{document}